\documentclass[traditabstract,longauth]{aa}
\pdfoutput=1 
\usepackage[varg]{txfonts}
\usepackage{graphicx,amsmath,amssymb, natbib, wasysym}
\usepackage{booktabs}
\usepackage{multirow}

\bibpunct{(}{)}{;}{a}{}{,} 

\usepackage{colortbl,xcolor}

\usepackage{hyperref}
\hypersetup{
bookmarks=true,     
unicode=true,       
colorlinks=true,    
linkcolor=red,      
citecolor=blue,     
filecolor=magenta,  
urlcolor=black      
}

\newcommand{\swift}{\textit{Swift}}
\newcommand{\konus}{Konus-\emph{Wind}}

\def\gr{\hbox{ \raisebox{-1.0mm}{$\stackrel{>}{\sim}$} }}
\def\kr{\hbox{ \raisebox{-1.0mm}{$\stackrel{<}{\sim}$} }}

\begin{document}

\title{Four GRB-Supernovae at Redshifts between 0.4 and 0.8}
\subtitle{The bursts GRB 071112C, 111228A, 120714B, and 130831A\thanks{{ 
Based on observations collected with GROND 
at the MPG 2.2m telescope at ESO La Silla Observatory (PI: J. Greiner), 
the Very Large Telescope of the European Southern
Observatory, Paranal Observatory, Chile (ESO programme
092.A-0231B, PI: T. Kr\"uhler), Keck LRIS and MOSFIRE
(PI: D. A. Perley), Spitzer (PI: D. A. Perley), and publicly available
data obtained from the Gemini, Hubble Space Telescope (HST), and 
Telescopio Nazionale Galileo (TNG) data archives.}}}

\author{
S.~Klose\inst{1},
S.~Schmidl\inst{1},
D.~A.~Kann\inst{1,2}, 
A.~Nicuesa Guelbenzu\inst{1},
S.~Schulze\inst{3,4,5}, 
J.~Greiner\inst{6}, 
F.~Olivares\inst{5,7}, 
T.~Kr\"uhler\inst{6}, 
P.~Schady\inst{6}, 
P.~M.~J.~Afonso\inst{8},
R.~Filgas\inst{9},
J.~P.~U.~Fynbo\inst{10},
D.~A.~Perley\inst{11},
A.~Rau\inst{6},
A.~Rossi\inst{12}, 
K.~Takats\inst{5},
M.~Tanga\inst{6},
A.~C.~Updike\inst{13},
\and
K.~Varela\inst{6}
}

\institute{
Th\"uringer Landessternwarte Tautenburg, Sternwarte 5, 07778 Tautenburg, 
Germany %
\and 
Instituto de Astrof\'isica de Andaluc\'ia (IAA-CSIC), Glorieta de la
Astronom\'ia s/n, E-18008 Granada, Spain 
\and 
Department of Particle Physics and Astrophysics, Weizmann Institute of Science,
Rehovot 76100, Israel 
\and 
Instituto de Astrof\'isica, Facultad de F\'isica, Pontificia Universidad 
Cat\'olica de Chile, 306, Santiago 22, Chile
\and 
Millennium Institute of Astrophysics (MAS), Nuncio Monse\~{n}or S\'otero Sanz
100, Providencia, Santiago, Chile
\and 
Max-Planck-Institut f\"ur Extraterrestrische Physik, Giessenbachstra\ss{}e 1, 85748 Garching, Germany
\and 
Departamento de Astronom\'ia, Universidad de Chile, 
Camino el Observatorio 1515, Santiago, Chile 
\and 
American River College, Department of Physics and Astronomy, 4700 College Oak Drive,
Sacramento, CA 95841, USA 
\and 
Institute of Experimental and Applied Physics, Czech Technical University in
Prague, Horska 3a/22, 128 00, Prague 2, Czech Republic 
\and 
Astrophysics Research Institute, Liverpool John Moores University,
IC2, Liverpool Science Park, 146 Brownlow Hill, Liverpool L3 5RF, UK
\and 
The Cosmic Dawn Center, Niels Bohr Institute, University of
Copenhagen, Juliane Maries Vej 30, 2100 Copenhagen, Denmark 
\and 
INAF-OAS Bologna, Via Gobetti 93/3, I-40129 Bologna, Italy 
\and 
Department of Chemistry and Physics, Roger Williams University,
One Old Ferry Road, Bristol, RI 02809, USA} 

\date{Received 29 January 2018}
\titlerunning{Four GRB-SNe}
\authorrunning{Klose et al.}


\abstract{Twenty years ago, GRB 980425/SN~1998bw revealed that long Gamma-Ray
  Bursts  (GRBs) are physically associated with 
  broad-lined type Ic  supernovae. Since
  then more than 1000 long GRBs have been localized to high angular precision,
  but only in $\sim50$ cases the underlying supernova (SN) component was
  identified. 
  Using the multi-channel imager GROND (Gamma-Ray Burst Optical Near-Infrared
  Detector) at ESO/La Silla, during the last ten years we have devoted a
  substantial amount of observing time to reveal and to study SN components
  in long-GRB afterglows. Here we report on four more GRB-SNe (associated with
  GRBs 071112C, 111228A, 120714B, and 130831A) which were 
  discovered and/or followed-up with
  GROND and whose redshifts lie between $z=0.4$ and 0.8. We study their
  afterglow light curves, follow the associated SN bumps 
  over several weeks, and characterize their host galaxies. 
  Using SN~1998bw as a template, the derived
  SN explosion parameters are fully consistent with the  corresponding
  properties of the so-far known GRB-SN ensemble, with no evidence for an
  evolution of their properties as a function of redshift. In two cases 
  (GRB 120714B/SN~2012eb at $z=0.398$ and GRB 130831A/SN~2013fu at
  $z=0.479$) additional Very Large Telescope (VLT) spectroscopy of the
  associated SNe revealed a photospheric expansion velocity at maximum light of
  about 40\,000 and 20\,000 km s$^{-1}$, respectively. For GRB 120714B, which
  was an intermediate-luminosity burst, we find additional evidence for a
  blackbody component in the light of the optical transient at early times,
  similar to what has been detected in some GRB-SNe at lower redshifts.}

\keywords{(stars:) gamma-ray burst: individual:
GRB 071112C, GRB 111228A, GRB 120714B, GRB 130831A - (stars:)
supernovae: individual: SN 2012eb, SN 2013fu}

\maketitle

\section{Introduction}

The association of SN~1998bw in the spiral galaxy ESO 184-G82
($z=0.0085$, \citealt{Tinney1998IAUC}) with the long GRB 980425
(\citealt{Galama1998Natur}) provided the first clue that long-duration GRBs
are associated with the deaths of massive stars. Twenty years after SN~1998bw
there is mounting observational and theoretical evidence that long GRBs have
their origin in a subclass of broad-lined  type Ic supernovae (SNe), which
spectroscopically reveal high expansion velocities
(for reviews see, e.g., \citealt{Cobb2012IAUS,Hjorth2012grbu,Hjorth2013RSPTA,
Schulze2014,Olivares2015a,Kann2016,Cano2016}). 
Since long GRBs signal the explosions of massive stars, they potentially 
allow for a zoom-in into the 
high-$z$ universe at times when Population III stars formed and exploded
(e.g., \citealt{Mesler2014ApJ...787...91M}).

Currently, about 50 GRB-SNe have been discovered
photometrically as a late-time bump in GRB afterglows, but only
50\%  of these have a
spectroscopic confirmation (\citealt{Cano2016}). At least some
might be linked to the formation of a magnetar 
\citep{Mazzali2014MNRAS.443...67,Greiner2015Natur,Kann2016,Wang2017ApJ,
Wang2017ApJ...850..148W,Lu2018ApJ}.
Supernova bumps  have been
detected up to a redshift of $z=1.06$ (GRB 000911; \citealt{Masetti2005})
and spectroscopically studied up to $z=1.01$ (GRB
021211/SN~2002lt;
\citealt{DellaValle2003IAUC8197,DellaValle2003}). For two decades GRB
980425/SN~1998bw has remained the closest GRB-SN detected. As such it is also 
the best-studied GRB-SN event and, therefore, used as the standard
template for basically all GRB-SN studies in the optical bands
\citep{Zeh2004a,Cano2013a}. 

In the pre-\swift~ (Neil Gehrels \emph{Swift} Observatory)
satellite (\citealt{Gehrels2004}) era (1997-2004) the annual
discovery rate of long-GRB afterglows with a redshift $z<1$ ($<0.5$),
i.e., those potentially suited for GRB-SN detections, was on average  about
2--3 (1--2) events per year. In the
\swift~ era (2005+) this rate increased  to about 6--8 (3--4) per year 
(no $z<0.5$ burst in 2007, and only 1 burst 
in 2008 and 2014), while the discovery
rate of the accompanying GRB-SNe  settled at on average 1 to 2 events
per year. For $0.5 < z < 1.0$ visibility constraints and
substantial observational efforts for a required photometric long-term follow
up might be the main reasons why most GRB-SNe were missed, though
in some cases host-galaxy extinction 
(e.g., \citealt{Soderberg2006ApJ...636..391S})
or an intrinsically faint SN (e.g., \citealt{Niino2012PASJ...64..115N}, but see \citealt{Postigo2018arXiv180704281D})
might have played a role too. 

One could also speculate that a long-lasting  bright optical afterglow
could hide a rising SN component (thanks to the referee for pointing this
out). Indeed,  this was basically the case for GRB 030329; here the $R_C$-band
light curve did not show a bump since the transition between afterglow light
and SN light was very smooth (see figure 3
in \citealt{Zeh2005NCimC..28..617Z}).  Though a detailed investigation of this
possibility remains to be done, {\it ad hoc} it appears to be a less likely
situation.  On the one hand, at least the GROND data archive always includes
multi-color data. This might strongly reduce the probability to miss  a rising
SN component. On the other hand, once a redshift information was known and
$z\kr0.5$ found, very likely spectroscopic observations  were triggered by the
GRB community.

At redshifts $z\lesssim0.1$ observational efforts to monitor an
expected/accompanying SNe were usually high and led to the
discovery of thermal components in early GRB afterglows
(e.g., \citealt{Campana2006,Waxman2007,Olivares2012,
Starling2012MNRAS.427.2950S,Schulze2014}) and
allowed for detailed studies of the SN explosion parameters
(for a review see \citealt{Cano2016}, and references therein).
Moreover, it led to the discovery of three events where no
underlying SN component was found down to deep flux limits
(GRB 060505 at $z=0.089$ and GRB 060614 at 
$z=0.125$: 
\citealt{DellaValle2006Natur};
\citealt{Fynbo2006Natur};
\citealt{Gal-Yam2006Natur};
\citealt{Xu2009ApJ...696..971X};
\citealt{McBreen2008ApJ...677L..85M};
GRB 111005A at $z=0.01326$, \citealt{Michalowski2016}; 
\citealt{Tanga2017arXiv170806270T}).
This has raised the question
whether some long bursts could have their origin in 
failed supernovae which immediately collapse into a black hole.
The low redshift of these well-studied SNe also allowed for 
detailed studies of their host galaxies (e.g., 
\citealt{Wiersema2007,
Christensen2008A&A...490...45C, 
Thone2008ApJ...676.1151T,
Levesque2011ApJ...739...23L,
Levesque2012ApJ...758...92L,
Leloudas2011A&A...530A..95L,
Fynbo2012grb..book..269F,
Michalowski2012ApJ...755...85M,
Michalowski2016,
Schulze2014, Thone2014MNRAS.441.2034T, 
Izzo2017MNRAS.472.4480I,
Kruhler2017A&A...602A..85K,
Tanga2017arXiv170806270T}). Though, 
even for events at higher redshifts detailed host-galaxy studies have been
performed (\citealt{Postigo2018arXiv180704281D}).

The relatively small annual discovery rate of GRB-SNe calls for detailed
follow-up observations of each event. While spectroscopic observations usually
need the biggest telescopes in order to get a reasonable signal-to-noise
ratio, photometric studies are less demanding and can be
performed using smaller telescopes as well.

Here we report on observations of a further set of four GRB-SNe  observed with
GROND (MPG 2.2m, ESO/La Silla; \citealt{Greiner2007Msngr,Greiner2008}) in the
optical/NIR bands in the years between 2007 and 2013.  Previous results of
follow-up observations of GRB-SNe  with GROND were presented
in \cite{Olivares2012} (GRB/XRF 100316D/SN~2010bh),
\cite{Olivares2015a} (GRBs 081007/SN~2008hw, 091127/SN~2009nz, 
101219B/SN~2010ma) as well as \cite{Greiner2015Natur} and \cite{Kann2016} 
(GRB 111209A/SN~2011kl). Three of the events discussed here 
are studied for the first time (GRBs 071112C, 111228A, 120714B), while 
GRB 130831A/SN~2013fu
was also explored by \cite{Cano2014a} using an independent data set. Two of the
events we study could also be investigated based on spectroscopic follow-up 
campaigns with the 
Very Large Telescope (GRB 120714B/SN~2012eb, GRB 130831A/SN~2013fu; 
\citealt{Klose2012GCN13613,Klose2012a,Klose2013GCN15320,Klose2013b}).

The paper is organized as follows.  We start with a brief overview concerning
the observational details (Sect.~\ref{SecObs}) and then focus on the SN light
curves (Sect.~\ref{Photometry}). Thereafter, we report (i) on the results of
our early-time  VLT/X-shooter spectroscopy of the optical transient that
followed GRB 120714B (Sect.~\ref{Shocking}) and (ii) on the results of the
VLT/FORS2 (FOcal Reducer and low dispersion Spectrograph)
spectroscopy around SN maximum (GRB 120714B/SN~2012eb and GRB
130831A/SN~2013fu; Sect.~\ref{SN.Spectr}).  In Sect.~\ref{Discussion} we
derive the relevant explosion parameters of the SNe and put
the properties of the four GRB-SNe in the context of the present world-sample
of well-observed GRB-SNe. In addition, we summarize the properties of the
corresponding afterglows and GRB host galaxies.

In the following, we use the convention $F_\nu(t)\sim t^{-\alpha}\nu^{-\beta}$
to describe the temporal and spectral evolution of the flux density $F_\nu(t)$
of an afterglow. We use a $\Lambda$CDM cosmology with $H_0=71$~km~s$^{-1}$
Mpc$^{-1}$, $\Omega_M=0.27$, and $\Omega_\Lambda=0.73$ \citep{Spergel2003a}.

\section{Observations and data reduction \label{SecObs} }
\subsection{GROND multi-color imaging}

The multi-color GROND data were reduced in a standard fashion (bias
subtraction, flat fielding, co-adding) with a customized pipeline \citep[for
  details see][]{Kruehler2008a, Yoldas2008a} which is based on standard
routines in \texttt{IRAF} \citep[Image Reduction and Analysis Facility;
][]{Tody1986}. To measure the brightness of the optical/near-infrared  (NIR)
transient {we employed aperture photometry 
as well as} point-spread-function (PSF) 
photometry using the DAOPHOT
and ALLSTAR packages in \texttt{IRAF}  \citep{Tody1993a}, similar to the
procedure described in \cite{Kruehler2008a} and \cite{Yoldas2008a}.
Once an
instrumental magnitude was established, it was photometrically calibrated
against the brightness of a number of field stars measured in a similar
manner. 

{PSF photometry was used when  
the afterglow dominated the light of the optical transient at early times
(GRBs 071112C, 111228A, and 130831A), while aperture
photometry was always applied for the fainter supernova and host-galaxy 
component.
In the case of the GRB 120714B the afterglow, the supernova, 
and the host were 
of similar brightness so that aperture photometry was used.  
Since all four host galaxies turned out to be either
compact on our images (GRBs 071112C, 111228A, and 120714B) or very faint
(GRBs 071112C, 111228A, and 130831A)
and since in all cases the afterglow/SN was 
situated well within the host's light, there is no mismatch  
between PSF and aperture photometry even though we did not 
perform image subtraction against a late-time reference image.
}

Photometry was tied to the Sloan Digital Sky Survey (SDSS) DR7
catalog \citep{Abazajian2009a} in the optical filters ($g^\prime r^\prime
i^\prime z^\prime $) and  the Two Micron All Sky Survey (2MASS;
\citealt{Skrutskie2006a})  in the NIR bands ($JHK_s$).  Details on the GROND
extinction corrections and Vega--AB conversions  can be found in, e.g.,
\cite{Rossi2012}, central wavelengths of the GROND filter bands are listed in,
e.g., \cite{Rossi2011}. 

\subsection{Late-time host-galaxy imaging using other telescopes}

{Late-epoch, host-galaxy 
imaging data using other instruments than GROND were obtained
with the High Acuity Wide field $K$-band Imager (HAWK-I) at ESO/VLT 
on ESO/Paranal,  DoLoRes (Device Optimized for the LOw
RESolution) mounted at the Telescopio Nazionale Galileo (TNG),  La Palma, the
Gemini Multi-Object Spectrographs (GMOS) at the  Frederick
C. Gillett Gemini North telescope on Mauna Kea, the Low Resolution Imaging
Spectrometer (LRIS) mounted at Keck I on Mauna Kea, and the Infrared Array
Camera (IRAC) at the Spitzer Space Telescope.

Gemini raw data were downloaded from the Gemini 
archive \footnote{\texttt{\url{https://archive.gemini.edu/}}},  
TNG raw data from the 
TNG archive\footnote{\texttt{\url{http://ia2.oats.inaf.it/archives/tng}}}.
Keck data stem from  
SHOALS (\citealt{Perley2016ApJ...817....7P,Perley2016ApJ...817....8P}),
Spitzer/IRAC data from the Spitzer 
GRB host galaxy data base (\citealt{Perley2016sptz.prop13104P}).

All data were reduced in a standard fashion using  IRAF (\citet{Tody1986}).
In all cases host-galaxy magnitudes were obtained by aperture
photometry. Optical data were calibrated against field stars from the SDSS DR
12 catalog (\citealt{Alam2015}), near-infrared data  were calibrated against
the Two Micron All Sky Survey (\citealt{Skrutskie2006a}). The transformations
proposed by \cite{Lupton2005} were used to transform 
SDSS into $R_C$ and $I_C$-band magnitudes. }

\subsection{UVOT imaging of GRB 111228A}

We expand our photometric database of GRB 111228A by adding the  Ultraviolet
Optical Telescope (UVOT;  \citealt{Roming2005SSRv..120}) observations from
the Neil Gehrels \emph{Swift} Observatory. 
Photometry with UVOT was carried out on pipeline-processed sky
images downloaded from the \emph{Swift} data
center\footnote{\url{www.swift.ac.uk/swift_portal}} following the standard UVOT
procedure \citep{Poole2008MNRAS}. Source photometric measurements were
extracted from the UVOT early-time event data and later imaging data files
using the tool {\sc uvotmaghist} (v1.1) with a circular source extraction
region of 3\farcs5 to maximize the signal-to-noise. In order to remain
compatible with the effective area calibrations, which are based on $5\arcsec$
aperture photometry \citep{Poole2008MNRAS}, an aperture correction was
applied. At late times, in each filter, we stacked all observations that had
yielded upper limits only, to achieve deeper constraints on the SN/host
galaxy. No detections were made in these deep stacks.

\subsection{Light-curve analysis}

GRB-SN parameters were
extracted using an analytical ansatz which expresses the light of the optical
transient (OT) as the sum of afterglow light (AG), host-galaxy light, and
light from an underlying SN component. The flux density, $F_\nu$,  in a given
photometric band that is characterized by its frequency $\nu$ is then given by
(\citealt{Zeh2004a,Zeh2005texas}) 
\begin{equation}
 F_\nu^{\rm OT}(t) = F_\nu^{\rm AG}(t) + k\,F_\nu^{\rm SN}(t/s)
 + F_\nu^{\rm host}\,.
\label{ot}
\end{equation}
Here, the parameter $k$ describes the observed luminosity ratio between the
GRB supernova at peak time, and the SN template in the considered band (in the
observer frame). The parameter $s$ is a stretch factor with respect to the
used template. If $s<1$ ($s>1$) the SN is developing faster (slower)
than the template GRB-SN. Following basically all GRB-SN studies,  the
template used here is SN~1998bw (\citealt{Galama1998Natur}), though other
template SNe can be constructed too (\citealt{Ferrero2006a}). The fit equation
for the SN component has the form
\begin{equation}
F_\nu^{\rm SN}(x=\frac{t}{s}) = 
q_1 \, \exp \left\{ -\left( \frac{(x - q_2)^2}{q_3}\right)\right\} \, x^{q_4} + 
q_5\,x^{q_6} \, e^{-q_7\,x}\,,
\label{eq:SNtheo}
\end{equation}
where the first term models the SN rise and peak and the second term the
exponential decay.  The coefficients $q_1$ to $q_7$ are determined based on a
fit to the  numerical SN~1998bw light curves shifted to the redshift under
consideration  as described in detail in \cite{Zeh2004a}. 

Practically, at first we shift the SN template to the redshift under
consideration (including the cosmological $k$-corrections). Then we
approximate this SN light curve by Eq.~\ref{eq:SNtheo} and finally
include it in the numerical fit (Eq.~\ref{ot}). When shifting the template
light curves to a certain redshift, we have to integrate over the filter under
consideration. In doing so we take into account that 
the $(k,s)$ values are different from band to band. 
In order to take this wavelength-dependence into
account, we build a sample of monochromatic $(k,s)$ values by 
interpolating between the corresponding broad-band values of
$k$ and $s$ of the template SN. Finally, when performing the light-curve fits
(Eq.~\ref{ot}), the (broad-band) $k$ and $s$ values were allowed to be
different from band to band as it is the case for SN~1998bw,
while the afterglow parameters were not. In other words,
afterglows were considered to evolve achromatically.

We always performed a joint fit, i.e.,
we fit all bands simultaneously. We require an achromatic  evolution
of the afterglow (spectral index $\beta=\textnormal{const.}$).  When fitting the (Galactic
extinction-corrected; \citealt{Schlafly2011a}) 
spectral energy distribution (SED)  of an afterglow, we
assume a power-law shape and in addition allow for a contribution from dust in
the GRB host galaxy. In doing so,  we use the analytic expressions from
\cite{Pei1992} to model extinction light curves based on Small Magellanic
Cloud (SMC),  Large Magellanic Cloud (LMC), or  Milky Way (MW) dust, following
\cite{Kann2006ApJ}. The slope $\beta$ of the SED of the afterglow is a direct
output of the joint fit of the multi-color light curves.  When plotting the
SED, we refer to the flux density at $t=1$ day in case of single power-law
fits, and $t=t_b$ assuming $n=\infty$ in case of broken power-law fits. In the
latter case $t_b$ is the break time and  $n$ the smoothness of the
break. Here,  $n=\infty$ describes a sharp break (see \citealt{Zeh2006a}
for more details).  Similarly, we assume that the break frequency $\nu_c$ is
either below or above the optical/NIR bands for the entire time span of the data from 
which the SED was constructed.

Afterglow light curves were fitted following the standard model according to
which the observer lies in the  cone of a jetted outflow and the observed
break time in the  optical light curve is a measure for the actual degree of
collimation (\citealt{Rhoads1999,Sari1999ApJ}). Jet half-opening angles are
calculated following \cite{Frail2001ApJ562}  for an 
interstellar medium (ISM) and
\cite{Bloom2003ApJ} for a wind medium, 
\begin{eqnarray}
\Theta_{\rm ISM}&=&0.057 \, \left(\frac{2\,t_b}{1+z}\right)^{3/8} \,E_{\rm iso, 53}^{-1/8}\, \left(\frac{\eta}{0.2}\right)^{1/8}\left(\frac{n}{0.1}\right)^{1/8}\, \nonumber \\
\Theta_{\rm wind}&=&0.169 \, \left(\frac{2\,t_b}{1+z}\right)^{1/4}\left(\frac{E_{\rm iso, 52}}{\eta \,A_\star}\right)^{-1/4}\,,
\label{theta}
\end{eqnarray}
where $\Theta$ is measured in units of radians,  $n$ (cm$^{-3}$) is the gas
density, and $\eta$ is the efficiency of the shock in converting the energy in
the ejecta into gamma radiation. The break time $t_b$ is measured in days and
the isotropic equivalent energy  $E_{\rm iso, x}$ in units of $10^{x}$ erg. We
assumed $n=1$ and $\eta=0.2$. The parameter $A_\star$ describes the wind mass
loss rate. Following \cite{Chevalier2000ApJ536}, we set $A_\star$ = 1.0.

\subsection{VLT spectroscopy} 

Spectroscopy with FORS2 mounted at the VLT  was performed to reveal the SN
features in the case of the two cosmologically nearest events, GRB 120714B and
GRB 130831A. In the case of GRB 120714B  spectroscopy of the optical transient
was performed on the night of Aug. 1/2, 2012, 18.3 days after the burst (13.2
days in the GRB rest frame; \citealt{Klose2012GCN13613,Klose2012a}),  which
was about 2 days (rest frame) after the SN maximum in the $r^\prime $ band
(about 3.5 to 5 days before the peak in the $i^\prime,z^\prime$ bands). The
exposure time was $4\,\times\,1450$~s.   Spectroscopy of the optical transient
following GRB 130831A was executed with FORS2 on Sep~29 and Sep~30, 2013
(exposure time  $2\,\times\,1378$~s in each run), 28.5 and 29.5~days after the
burst (about 19.5 days in the rest frame;
\citealt{Klose2013GCN15320,Klose2013b}). This was about 6 days (rest frame)
after the SN maximum in the $r^\prime,i^\prime,z^\prime$ bands. Observations
were performed under excellent sky conditions with a mean seeing of 0\farcs6.

\begin{table}[t!]
\renewcommand{\tabcolsep}{3pt}
\caption{Summary of the four GRB events studied here.}
\begin{center}
\begin{tabular}{l | rrrr}
\toprule
GRB                     & 071112C          & 111228A         & 120714B         & 130831A \\
SN                      &                  &                 & SN~2012eb   & SN~2013fu\\
\midrule
redshift $z$            &  0.823           & 0.7163          &  0.3984         & 0.4791 \\
$T_{90}$ (s)            & $15\pm2$         & $101.2\pm5.4$   & $159\pm34$     & $32.5\pm2.5$  \\
$\log E_{\rm \gamma, iso}$   & $52.28^{+0.11}_{-0.12}$ & $52.61^{+0.05}_{-0.06}$ & $50.76^{+0.12}_{-0.14}$ & $51.86^{+0.04}_{-0.04}$ \\[1mm]
log $L_{\rm \gamma, iso}$  & $51.10^{+0.15}_{-0.22}$ & $50.60^{+0.07}_{-0.09}$ & $48.56^{+0.18}_{-0.31}$ & $50.35^{+0.07}_{-0.08}$ \\
\bottomrule
\end{tabular}
\end{center}
\tablefoot{Burst durations are given in the observer frame and were taken
from \url{http://swift.gsfc.nasa.gov/archive/grb_table/}. The
isotropic-equivalent energies 
(in units of erg) were
derived from the observed high-energy properties \citep{Briggs12744,Golenetskii2013a}, 
and using the statistically inferred high-energy properties of GRBs
(\citealt{Butler2007a}). $L_{\rm \gamma, iso}$ is the time-averaged luminosity  (erg s$^{-1}$) of the
burst, here defined as $E_{\rm iso}/T_{90}$. References for the redshifts are
given in appendix~\ref{appgrbs}.}
\label{Tab:summary0}
\end{table}

\begin{figure*}[t!]
\hspace*{-10mm}
\includegraphics[width=20.3cm]{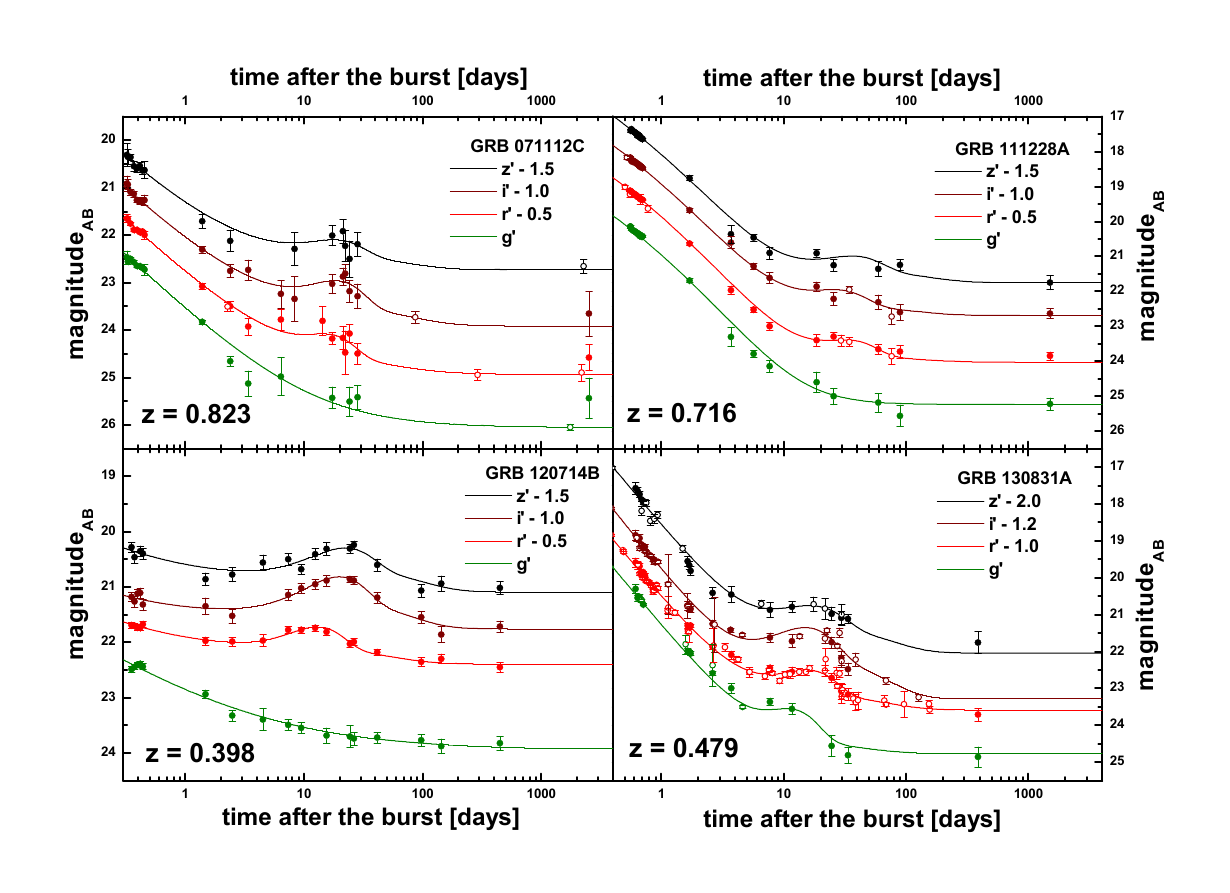}
\caption{
Afterglow and SN $g^\prime r^\prime i^\prime z^\prime$ light curves of the four SNe in our sample
(apparent AB magnitudes are not corrected for Galactic extinction, time refers
to the observer frame). Filled circles represent GROND data, 
open circles data taken from the literature.
\emph{GRB 071112C}:\ The $r^\prime$-band data point
at $t\sim300$ days was taken from \cite{Vergani2015a}, other late-time data
are listed in Table~\ref{noGRONDmag}. \emph{GRB 111228A}:\ The light curves
include GROND data (Table~\ref{071112Cmagdata}) as well as late-time data
points from the TNG (Table~\ref{noGRONDmag}). \emph{GRB 120714B}:\  For the
GROND data see  Table~\ref{120714Bmagdata}. \emph{GRB 130831A}:\ Data points
stem from GROND observations (Table~\ref{130831Amagdata}) as well as  (in
$r^\prime i^\prime z^\prime $) from \cite{Cano2014a}.}
\label{fig:SNe}
\end{figure*}

In all observing runs we used the 300V+10 grism, which covers the
wavelength range from 4450 to 8650~\AA\, (dispersion 112~\AA~ mm$^{-1}$,
1.68~\AA\, pixel$^{-1}$ in case of no binning, resolving power $R=440$ at
5900~\AA)\footnote{see the FORS2 User Manual dated 26/02/2013, page 12}
together with the GG435 order sorting filter. Data reduction was performed
applying the standard cosmic ray, flatfield, and bias correction using
\texttt{IRAF}.  Wavelength calibration was performed relative to HgCdHe+Ar
calibration lamps. The spectrophotometric standard stars LTT~1020  and
LTT~1788 were observed to flux-calibrate  the spectra for the SNe associated
with GRB 120714B and GRB 130831A, respectively.

In the case of GRB 120714B additional VLT spectroscopy of the optical
transient was obtained with X-shooter 0.35 days after the burst
(program ID 089.A-0067; PI: J. P. U. Fynbo)
and published by \cite{Kruehler2015AA1}. Data reduction was performed in a 
standard manner (for details, see \citealt{Kruehler2015AA1}).

\section{Results}

Here we focus on the phenomenological GRB-SN parameters based on the observed
GROND multi-color light curves,  i.e., the stretch factor $s$ and the
luminosity factor $k$ in the different photometric bands. The results of
our VLT spectroscopic follow-up campaigns for two of the four GRBs are
outlined thereafter. More details on the  derived individual afterglow and
host-galaxy  parameters are provided in the appendix. Table~\ref{Tab:summary0}
provides a summary of the properties of the four GRBs studied here.

\subsection{Supernova bumps \label{Photometry} }

Following the phenomenological classification scheme as summarized in,
e.g., \cite{Hjorth2013RSPTA} and
\cite{Cano2016}, based on the observed isotropic equivalent luminosity
in the  gamma-ray band (Table~\ref{Tab:summary0}), GRB 120714B is an
intermediate-luminosity GRB ($48.5 < \log L_{\gamma, \rm iso} < 49.5$ erg
s$^{-1}$), while the other three are high-luminosity bursts  ($\log
L_{\gamma, \rm iso} > 49.5$ erg s$^{-1}$). 

The GROND light curves of the four optical transients (as well as a few
host-galaxy magnitudes from other sources and telescopes which were used in
the fits) are shown in Fig.~\ref{fig:SNe}. These data demonstrate how well
GRB-SN light curves can be monitored with 2-m class optical telescopes even at
redshifts beyond $z=0.5$. In all cases we performed late-epoch observations
several hundred to several thousand days after the event in order to obtain a
reliable host-galaxy magnitude, which should guarantee a proper subtraction of
the host-galaxy flux when the SN fit was performed. Shown in
Fig.~\ref{fig:SNe} are only the GROND
$g^{\prime}r^{\prime}i^{\prime}z^{\prime}$ band data since in no case were we
able to detect the SN component in $JHK_s$\footnote{Note that
early NIR observations of SN~1998bw do not exist; there is no
template for these bands.}; successful
NIR observations of GRB-SN light curves remain a challenge,\footnote{e.g., 
GRB 031203/SN~2003lw (\citealt{Cobb2004ApJ...608L..93C,Gal-Yam2004ApJ...609L..59G});
GRB 060218/SN~2006aj (\citealt{Cobb2006ApJ...645L.113C,Kocevski2007ApJ...663.1180K});
GRB 100316D/SN~2010bh (\citealt{Olivares2012});
GRB 130702A/SN~2013dx (\citealt{Toy2016ApJ})}
not to mention the fact that early NIR observations 
of SN~1998bw do not exist. The best-sampled cases are GRBs
120714B and 130831A, 
and the error bars of their derived SN light-curve parameters
(luminosity factor $k$, stretch factor $s$) are correspondingly small
(Table~\ref{Tab:summary2}). 

The light-curve fit revealed that all four SNe were less luminous than the
template SN~1998bw ($k<1$),  though in some cases within the $1\sigma$  error
bar a $k>1$ in a certain photometric band is not ruled out. Within the
statistical uncertainties, only the SN associated with GRB 111228A might have
developed slower than SN~1998bw (i.e., $s>1$), all others developed faster
($s<1$). 

All four events are characterized by a rather small magnitude difference 
($\kr1$ mag) between the peak of the SN bump (as the sum of
SN, afterglow and host-galaxy flux) and the magnitude of the underlying host
galaxy.  All four events do however notably differ in their 
magnitude difference 
between the peak of the emerging SN component and the flux of the
optical transient (OT) at $t_{\rm obs}=1$ day.  The light curve of the OT that
followed GRB 111228A was rather typical for GRB-SN events at moderate
redshifts (see \citealt{Zeh2004a,Zeh2005texas}). At early times the afterglow
flux clearly dominated the light,  while the SN bump was rather modest, with
its peak around $2-3$ weeks after the burst.  In the case of GRB 071112C and
GRB 130831A this difference in magnitude at $t_{\rm obs}=1$ day
was only about 1 mag, while it was even
negative for GRB 120714B, indicating for the latter case a relatively
low-luminosity afterglow (see also Sect. \ref{AGLum}).  In this respect  the
light curve of the OT that followed GRB 120714B is phenomenologically closer
to nearby GRB-SN events such as GRB
031203/SN~2003lw \citep[e.g.,][]{Malesani2004a} and GRB/XRF
060218/SN~2006aj \citep[e.g.,][]{Ferrero2006a}. 

What is immediately apparent in the light curves is that only for GRB 130831A
our data reveal an indisputeable SN bump in the $g^\prime$ band. In the case of GRB 071112C
an inspection of the GROND $g^\prime$-band light curves seems to suggest the
existence of a break around 2 days and as such also a rising SN
component. However, when following this ansatz further we encountered some
shortcomings with the corresponding fit results (see
appendix \ref{app071112C}). Also, for this event no contemporaneous break is
apparent in the X-ray light curve (see the \swift/XRT
repository, \citealt{Evans2007a}) which one would expect to find if this
feature is due to a collimated explosion and if X-ray and optical flux have a
common origin. Therefore, also for this burst we focused on a model with no
break, which actually provides a better reduced $\chi^2$ for the joint fit
than the approach based on a broken power-law.

The lack of a clearly detectable SN component in the $g^\prime$ band  for GRBs
071112C, 111228A, and 120714B is worth noting. It is probably 
primarily due to metal-line absorption in the UV bands
(e.g., \citealt{Olivares2015a}) and not due to
substantial host-galaxy visual extinction along the line of sight; the
corresponding values we found for $A_V^{\rm host}$  based on the afterglow
light curve fits lie between 0 and $\sim$0.15 mag
(appendices~\ref{app071112C}, \ref{App.Sect.111228A}, \ref{Appendix:120714B}).
Possibly, in these three cases  the visibility of the SN bump in
the $g^\prime$ band is also affected by the comparably bright afterglow and/or
underlying host galaxy. Otherwise, for GRB 120714B the VLT spectrum
of the accompanying supernova shows that the $g^\prime$-band window is
dominated by metal-line absorption, which reduces the flux in this
band correspondingly (Fig.~\ref{fig:spectra}).  
Last but not least it has to
be taken into account that our light curve fits are based on published $UBVRI$
light curves of SN~1998bw (\citealt{Galama1998Natur}).  Consequently, for the
given redshifts  in particular our $g'$-band templates represent an
extrapolation into the ultraviolet domain, where in our
calculations we assume that the flux density of the SN falls $\propto \nu^{-3}$.

We finally note that with respect to their redshifts GRBs 071112C and
111228A resemble the long GRBs 040924 ($z=$0.86) and 041006 ($z=$0.71), which 
also revealed evidence for a SN bump
(\citealt{Stanek2005ApJ...626L...5S,Soderberg2006ApJ...636..391S}). 
Both SNe were probably dimmed by
host-galaxy extinction by 1.5 and 0.3 mag, respectively, but could
be observed with HST until about 120 days post burst
(\citealt{Soderberg2006ApJ...636..391S}). In both cases the SN~1998bw
light curve template fit the observations well.

\subsection{Spectroscopic identification of SN light \label{SN.Spectr}}

\begin{figure}[t!]
\includegraphics[width=\columnwidth,angle=0]{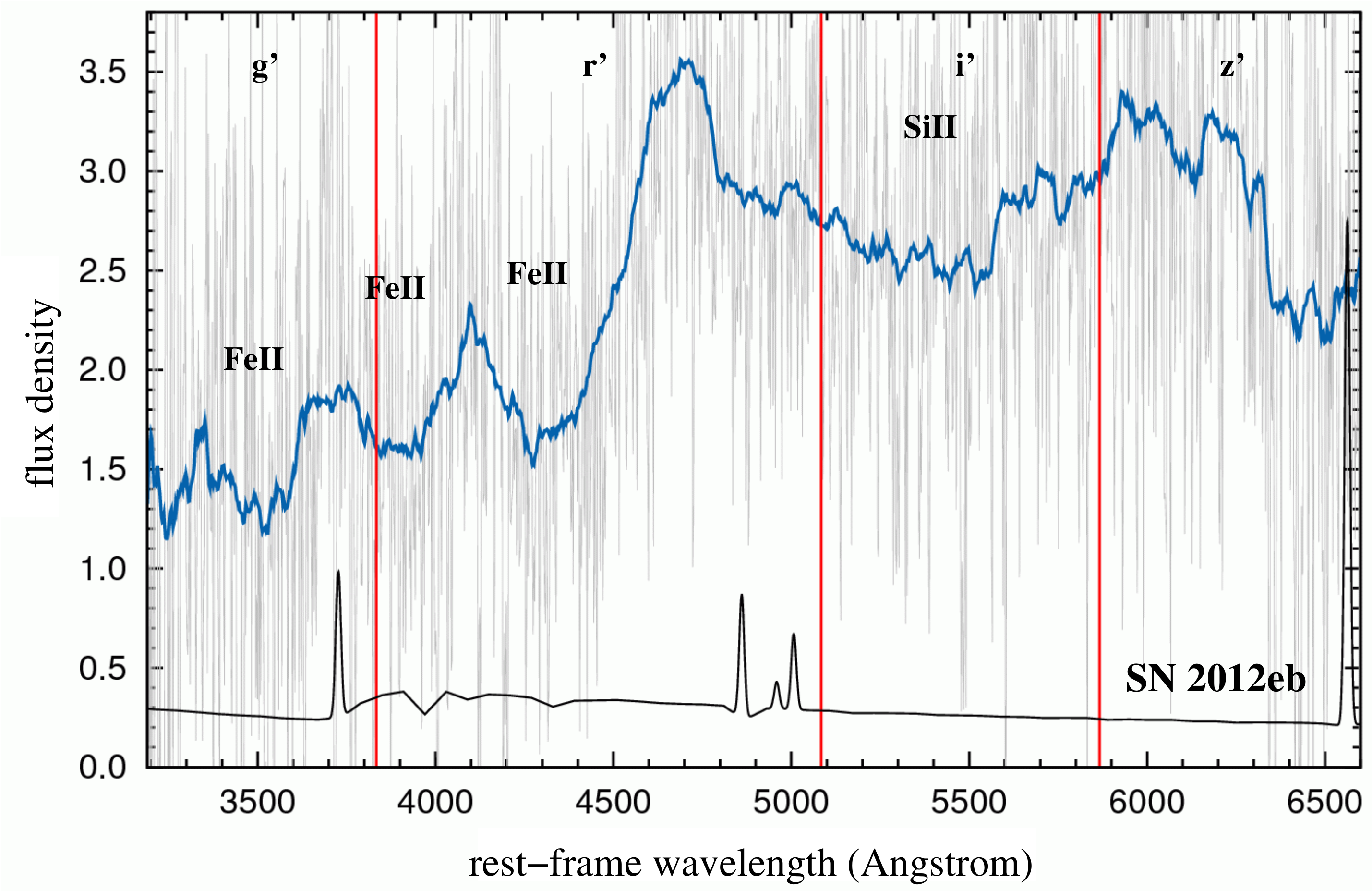}
\includegraphics[width=\columnwidth,angle=0]{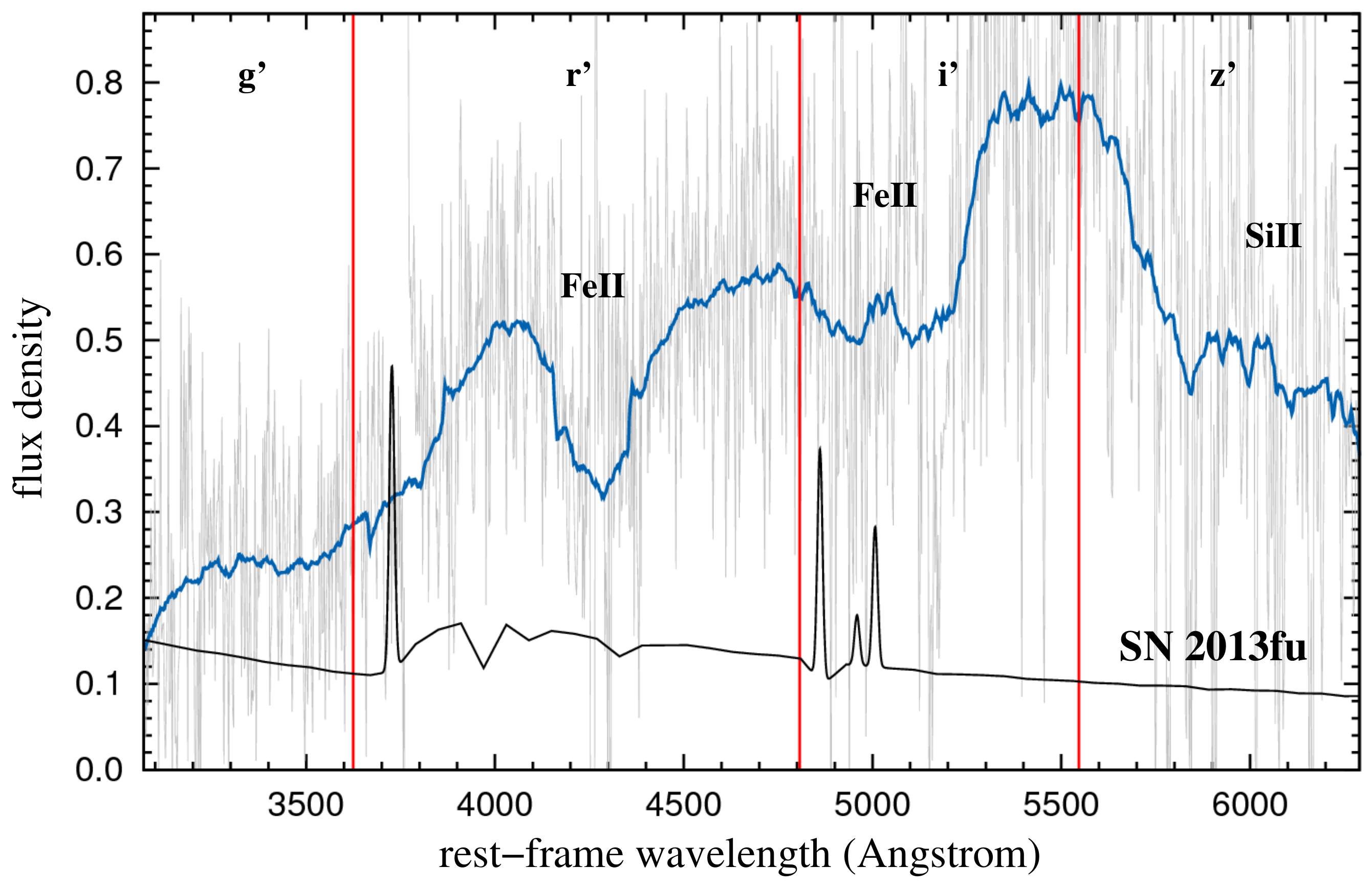}
\caption{
FORS2 spectra of SN~2012eb and SN~2013fu (plus underlying host-galaxy emission). These
spectra are not host-galaxy subtracted, since no spectra of the hosts could be
taken.  In gray is shown the original spectrum and in blue the spectrum
smoothed with a boxcar filter 250~\AA\ wide. The spectra are shifted into the
GRB rest frames but the flux density (in units of $10^{-18}$ erg s$^{-1}$
cm$^{-2}$ \AA$^{-1}$) is measured  in the observer frame. Also drawn are the
positions and widths of the GROND filter bands in the GRB rest frame. Note
that in the case of the $g^\prime$ and the $z^\prime$ bands the (redshifted)
filter width is slightly larger than shown
here \citep{Greiner2007Msngr,Greiner2008}.  \emph{Top:} \ SN~2012eb, taken on
day 18.3 post-burst (observer frame), about 2 days (rest frame) after the SN
maximum  in the $r^\prime$
band \citep{Klose2012GCN13613,Klose2012a}. \emph{Bottom}:\ SN~2013fu,
obtained at a mean time of $t=29$ days post-burst (observer frame), about 6
days (rest frame) after the SN maximum in the $r^\prime i^\prime z^\prime$
bands \citep{Klose2013GCN15320,Klose2013b}.  Also shown are  the host-galaxy
template SEDs (black lines; appendices~\ref{Appendix:120714B}, 
\ref{Sect:130831A})
mainly to illustrate where emission lines from the hosts might
exist. (The template for the host of GRB 130831A 
obviously overestimates the flux in the emission lines.)}
\label{fig:spectra}
\end{figure}

For two of the four events discussed here (GRBs 120714B and 130831A)
additional VLT spectroscopy was obtained which revealed the underlying SNe
based on their characteristic broad-band features (now designated
SN~2012eb, \citealt{Klose2012a}, and SN~2013fu, \citealt{Klose2013GCN15320},
respectively). In addition, the VLT spectroscopy allowed for measurements
of the photospheric expansion velocities close to maximum light of the two SNe. Since these
spectra are relatively noisy, a boxcar filter was applied in order to recover
their overall shape.  Depending on the chosen boxcar smoothing parameter, this
not only  reveals several prominent absorption troughs. It also shows  the
tight morphological similarity between both SN spectra even though the
corresponding cosmological redshift  as well as relativistic Doppler blueshift
are  different from each other (Fig.~\ref{fig:spectra}). 

Since for both events no pure host-galaxy spectra could be obtained with the
VLT, we use the template spectra that followed from the 
\texttt{Le PHARE} fits (Appendix~\ref{appgrbs}) to reveal where prominent
emission lines might lie. These are lines due to star-forming activity from 
the [\ion{O}{II}] $\lambda\lambda3727,3729$ doublet,
H$\gamma$ (4341~\AA), H$\beta$ (4861~\AA), [\ion{O}{III}] $\lambda\lambda4959,5007$, 
and H$\alpha$ (6563~\AA) (Fig.~\ref{fig:spectra}). {However, none of these
lines are detected in our spectra. This suggests that 
\texttt{Le PHARE} may overestimate the SFR for both hosts.}

\begin{figure}[t!]
\includegraphics[width=\columnwidth,angle=0]{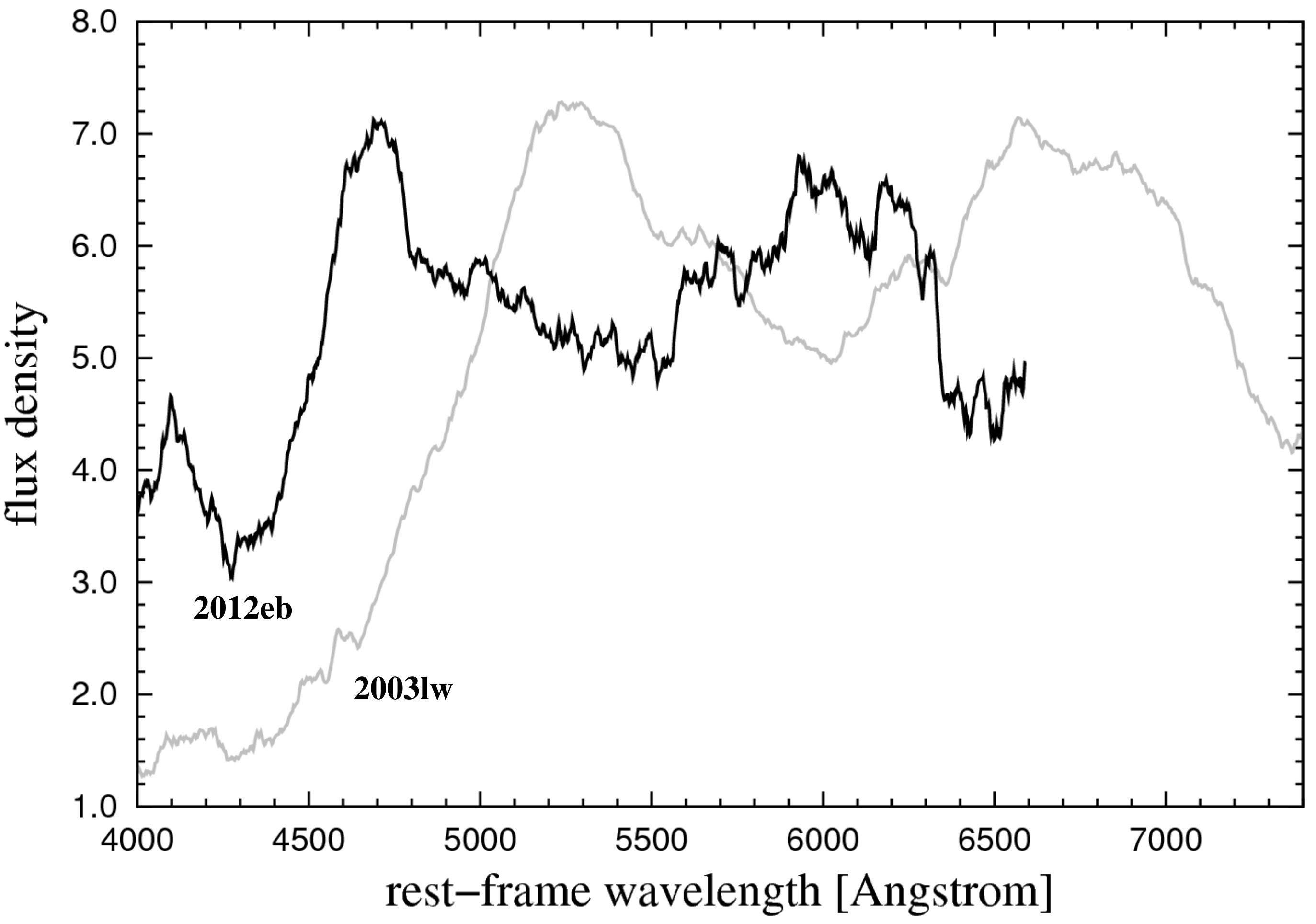}
\caption{Comparison of the spectrum of SN~2003lw taken on Dec. 20, 2003,
  $\sim15$ days (rest frame) after the burst and about 3 days before the SN
  maximum in the $R_C$ band (in light gray, \citealt{Malesani2004a}) with the
  spectrum of SN~2012eb taken on Aug. 1/2, 2012, 13.2 days after the burst
  (black line; shifted to the rest frames). The spectrum of SN~2003lw was
  taken from the {\it Open Supernova Catalog} data base
  \citep{Guillochon2016}. According to \cite{Mazzali2006a}, it implies
  that on Dec. 20 the expansion velocity was $17\,000\pm1\,000$ km
  s$^{-1}$.  The SN~2012eb spectrum shows a substantially higher
  blueshift. Both spectra are smoothed with a boxcar filter of 250~\AA\
  width. Flux densities are arbitrarily scaled.}  
\label{spectra.comp}
\end{figure}

In the case of SN~2012eb the most prominent absorption trough is  located at
$5500\pm100$~\AA\ rest frame.\footnote{The  relatively large error has its
origin in the application of the boxcar filtering procedure to the
spectra. Depending on the chosen boxcar smoothing parameter (number of
pixels), the shape and position (wavelength) of the broad-band absorption
feature  varies.} The shape of this feature is very much reminiscent of the
one seen in SN~2003lw/GRB~031203 ($z=0.1055$) around maximum light
(Fig.~\ref{spectra.comp}). It is generally attributed to \ion{Si}{II}
$\lambda$ 6355, while the broad absorption
features bluewards of \ion{Si}{II} are  believed to be due to blended lines of
single-ionized iron
(\citealt{Nakamura2001,Malesani2004a,Mazzali2006a,Mazzali2013}).\footnote{
\ion{Si}{II} $\lambda6355$: doublet $\lambda\lambda6347,6371$; 
\ion{Fe}{II} $\lambda4100$: multiple blends; 
\ion{Fe}{II} $\lambda4570$: potentially blended lines of \ion{Mg}{II} 
     $\lambda4481$, \ion{He}{I} $\lambda4471$,
     \ion{Ti}{II} $\lambda4550$; 
\ion{Fe}{II} $\lambda5100$: triplet $\lambda\lambda4924,5018,5169$}

Interpreted in this way, the \ion{Si}{II} feature corresponds to a blueshifted
velocity of $43\,000\pm5\,300$  km s$^{-1}$. The large error is a
direct consequence of the aforementioned boxcar filtering procedure and the
related uncertainty ($\pm$100~\AA)  of the central wavelength of the
absorption trough.  Applying the SYNOW code \citep{Fisher1997} to an
individual smoothed spectrum provided smaller errors. It also attributed the
absorption trough at around 4300~\AA\ to broadened lines of single-ionized Fe
$\lambda\lambda5100$  at basically the same high expansion velocity, while the
identification  of the other broad-band features is less secure. 

In the case of SN~2013fu the spectrum is relatively noisy.  The prominent
absorption trough redwards of 5500~\AA \ (rest frame) is probably again due
to \ion{Si}{II}. This feature is partly redshifted out of our spectral window,
however; the location  of its central wavelength cannot be determined.  The
shape of the absorption trough centered at around 5000~\AA\ appears to be
affected by emission lines from the underlying host galaxy, so that it might
not be a reliable tracer of the expansion velocity.  If the absorption feature
centered at $4270\pm50$~\AA\ (rest frame) is due to (dominated
by) \ion{Fe}{II} $\lambda4570$,  then the expansion velocity is
$20\,400\pm3\,500$ km s$^{-1}$  (Fig.~\ref{fig:spectra}).

\begin{figure}[t!]
\includegraphics[width=\columnwidth,angle=0]{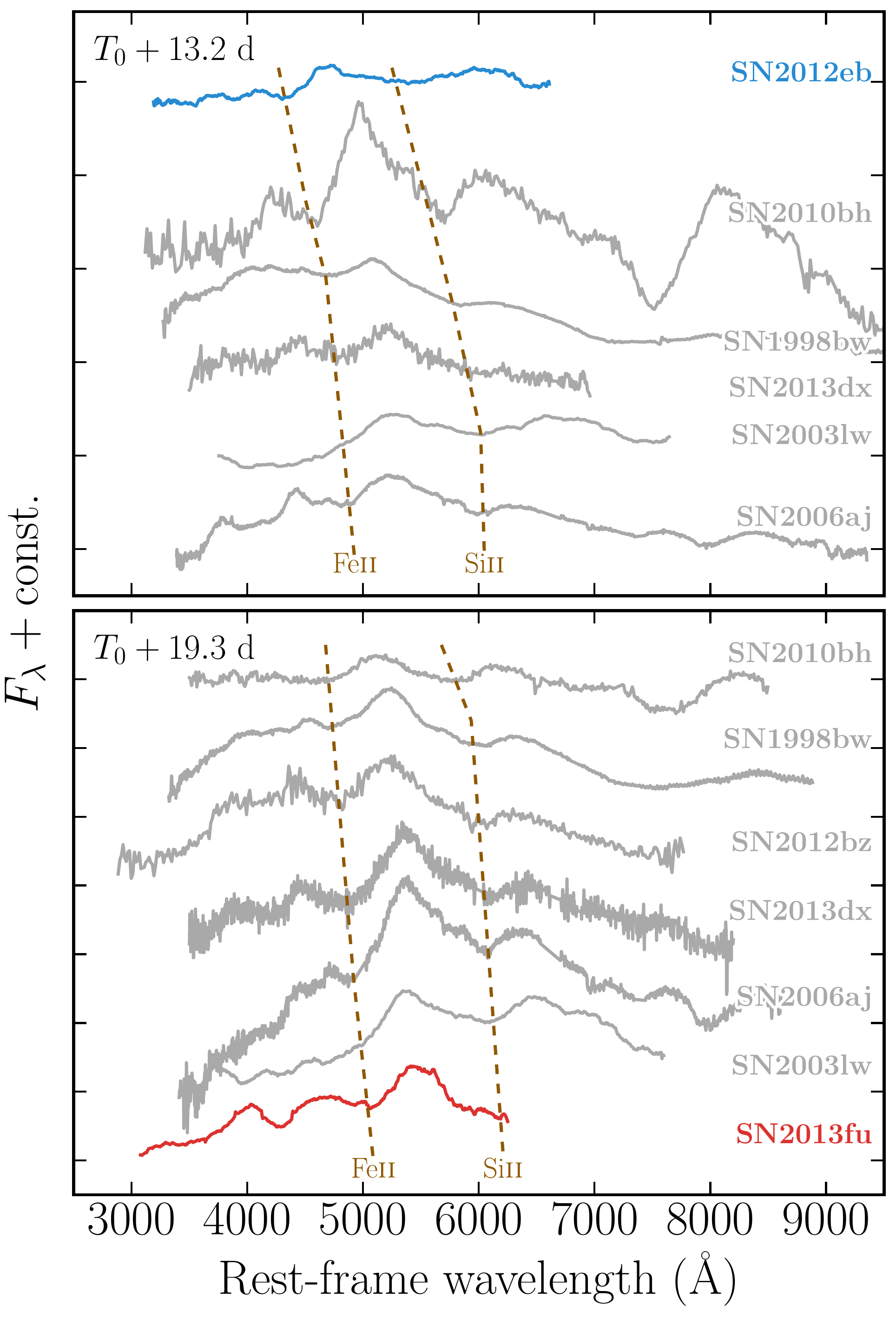}
\caption{Comparison of SNe~2012eb (blue) and 2013fu (red) to spectra of other
GRB-SNe (gray) at two different phases, 13.2 and 19.3 days past explosion,
respectively. All comparisons are made in the rest frame. The dashed lines
connect the approximate minima for the Fe\textsc{ii} and Si\textsc{ii}
features, and the spectra are shown in an expansion velocity sequence from the
fastest (SN~2010bh) to the slowest (SN~2006aj). 
All data are corrected for Galactic extinction along the line of sight. Data references:
SN~1998bw: \cite{Patat2001ApJ...555..900P},
SN~2003lw: \cite{Malesani2004a},
SN~2006aj: \cite{Pian2006Nature},
SN~2010bh: \cite{Bufano2012ApJ},
SN~2013dx: \cite{Elia2015A&A...577A.116D},
SN~2012bz: \cite{Schulze2014}. Further data were taken from the 
Interactive Supernova Data Repository (WISeREP; \citealt{
Yaron2012PASP..124..668Y}).
All spectra were corrected for host reddening, except for that of SN~2003lw.}
\label{SNspectra.comp}
\end{figure}

\begin{figure}[t!]
\includegraphics[width=\columnwidth,angle=0]{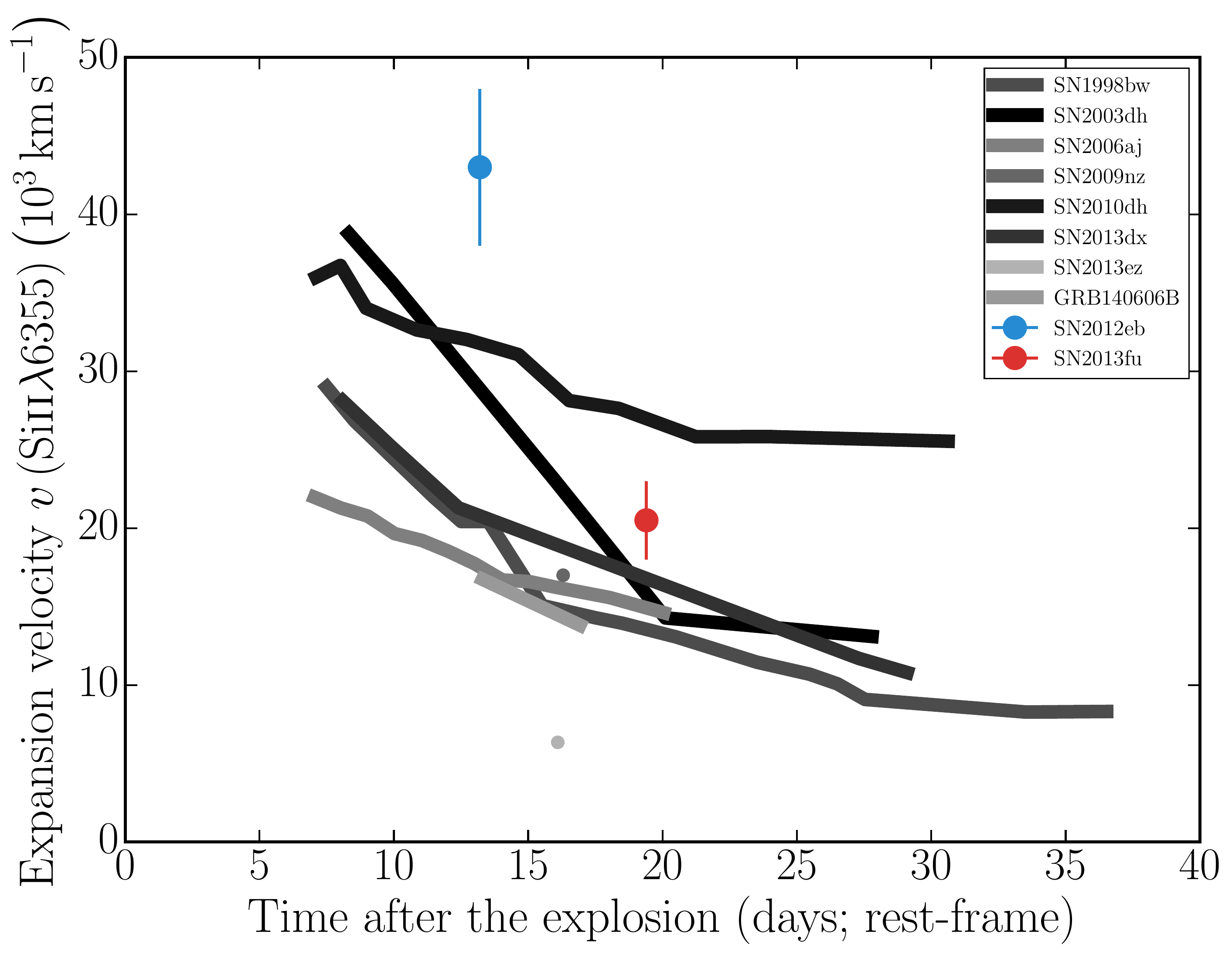}
\caption{Evolution of the expansion velocities measured from Si\textsc{ii}$\lambda$6355
for SN~2012eb and SN~2013fu and a comparison sample 
(for references see Fig.~\ref{SNspectra.comp}) of GRB-SNe with good spectroscopic data.}
\label{SNspectra.vel.comp}
\end{figure}

Compared to other GRB-SNe where the photospheric expansion velocity  was
measured via the  \ion{Si}{II} $\lambda6355$  or the \ion{Fe}{II} feature(s)
the inferred velocity for SN~2012eb is the highest so far measured close to
the time of maximum light 
(e.g., \citealt{Ben-Ami2012ApJ,Bufano2012ApJ,Schulze2014,
Elia2015A&A...577A.116D,Olivares2015a,Toy2016ApJ,Cano2016}).
Even higher velocities, though, close to 50\,000 km s$^{-1}$ 
have been reported for GRB 100316D/SN~2010bh during the SN
rise time (\citealt{Bufano2012ApJ}).
On the other hand, SN~2013fu lies rather at the
lower end of the expansion velocities for GRB-SNe close to maximum
light and measured via the \ion{Fe}{II} features (Figs.~\ref{SNspectra.comp},
\ref{SNspectra.vel.comp}).

Last but not least it is worth stressing that the isotropic equivalent energy
of GRB 120714B in the gamma-ray band
(Table~\ref{Tab:summary0}) was more than 1000 times less than
the isotropic equivalent energy of GRB 130427A/SN~2013cq 
at $z=0.34$  (log $E_{\rm iso}$
[erg] = 53.98; \citealt{Xu2013ApJ...776...98X}) 
and ten times higher than the one of GRB 100316D/SN~2010bh at $z=0.059$
(log $E_{\rm iso}$ [erg] = 49.6; \citealt{Starling2011MNRAS.411.2792S}); 
the corresponding SN explosion as well as light curve
parameters however are not substantially different from each other (see also \citealt{Melandri2014AA}). 

\subsection{Excess of blue light in the optical transient that
followed GRB 120714B \label{Shocking}}

\begin{figure}[t!]
\includegraphics[width=\columnwidth]{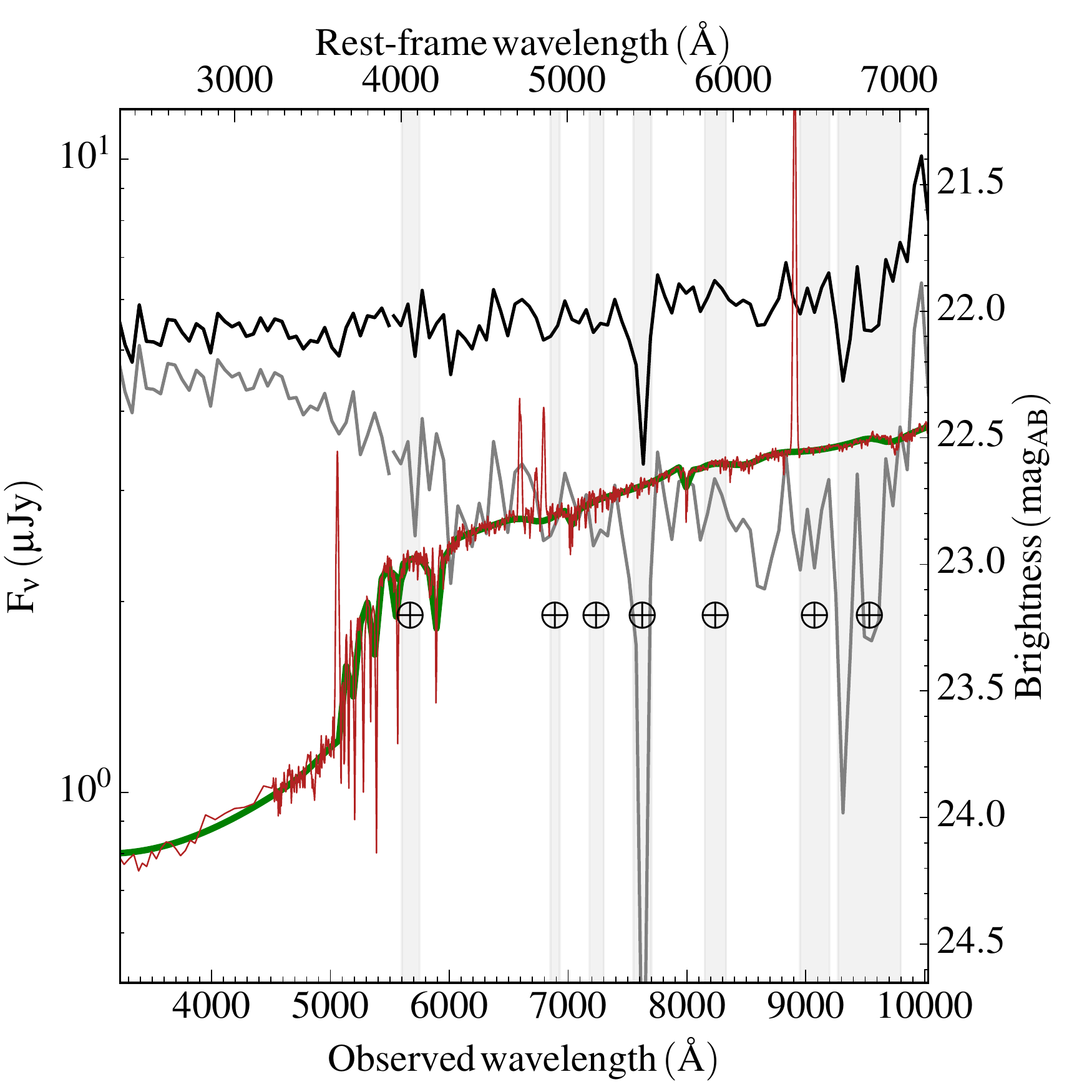}
\caption{VLT/X-shooter spectrum of the optical transient that followed GRB
  120714B, taken at $t_{\rm obs}=0.353$ days after the burst.  In black is
  shown the binned original spectrum, in red the host-galaxy SED template as
  obtained from the  \texttt{Le PHARE} fit (appendix~\ref{Appendix:120714B}).
  The green curve shows the smoothed host-galaxy template that was subtracted
  from the original spectrum. The host-subtracted SED  is drawn in light
  gray. Flux densities refer to the observer frame.  Telluric features are
  indicated. } 
\label{fig:120714B.shock}
\end{figure}

During the GROND first-epoch observations of the optical transient that
followed GRB 120714B,  at $t_{\rm obs}=0.265$ days, the seeing was about $3''$,
resulting in a relatively large magnitude error in $i^\prime$ and
$z^\prime$ (Table~\ref{120714Bmagdata}). Therefore, we analyzed the early SED
of the optical transient  (all data are corrected for Galactic reddening along
the line of sight; $E_{\rm (B-V)}^{\rm Gal}$ = 0.01 mag \citealt{Schlafly2011a})
based on the GROND data  taken at $t_{\rm obs}=0.354$ days in combination with
the VLT/X-shooter spectrum taken at basically the same time (at $t_{\rm
obs}=0.353$ days post burst; Fig.~\ref{fig:120714B.shock}). After subtracting
the (smoothed) broad-band SED of the underlying host-galaxy flux (based on the
\texttt{Le PHARE} fit; appendix~\ref{Appendix:120714B}), an inspection of the
X-shooter data shows that bluewards of $\sim600$~nm  the flux density is
increasing with increasing frequency ($(r^\prime -g^\prime )>0.5$ mag). This
is opposite to what is expected for the SED of a GRB afterglow. This excess of
blue light can be modeled as the low-energy tail of soft thermal X-ray
emission, similar to what was observed in other low-redshift GRB-SNe (GRB/XRF
060218/SN~2006aj,
\citealt{Campana2006}; GRB/XRF 100316D/SN~2010bh, \citealt{Cano2011,Olivares2012};
GRB 120422A/SN~2012bz, \citealt{Schulze2014}).\footnote{In the case of GRB
120422A/SN~2012bz the blackbody component had a low luminosity compared to the
afterglow and a too low temperature to show up in the X-ray band.}

Support for the interpretation of the excess of blue light as an underlying
additional blackbody component comes from the spectral slope in the wavelength
range between about $\lambda_{\rm rest}=320$~nm and 450~nm. Here, the
host-subtracted SED can be fit with a power-law and the  best-fit spectral
slope ($\beta=-1.78\pm0.19$) lies close to the
value expected for the Rayleigh-Jeans tail of a blackbody ($\beta=-2$). For
the given wavelength region this requires a temperature $T_{\rm bb}^{\rm
rest} \gr10$~eV (rest frame).


Since the first discovery of a thermal component in the afterglow of GRB/XRF
060218/SN~2006aj \citep{Campana2006} the preferred interpretation in the
literature is that this is emission from the cooling 
photosphere after SN shock breakout
(\citealt{Waxman2007,Nakar2010}; but see also
\citealt{Irwin2016MNRAS.460.1680I,Ruffini2017A&A...600A.131R}). 
Substantial observational efforts have
been put forward to search for such a signal in the X-ray data of
long-duration GRBs \cite[e.g.,][]{Olivares2012,Starling2012MNRAS.427.2950S,
  Sparre2012MNRAS.427.2965S} as well as in the optical data of  SNe in
general \citep[e.g.,][]{Forster2016ApJ...832..155F}, mostly with negative
results. As such, adding another positive detection to the database would be
of great interest.

\begin{figure}[t!]
\includegraphics[width=\columnwidth]{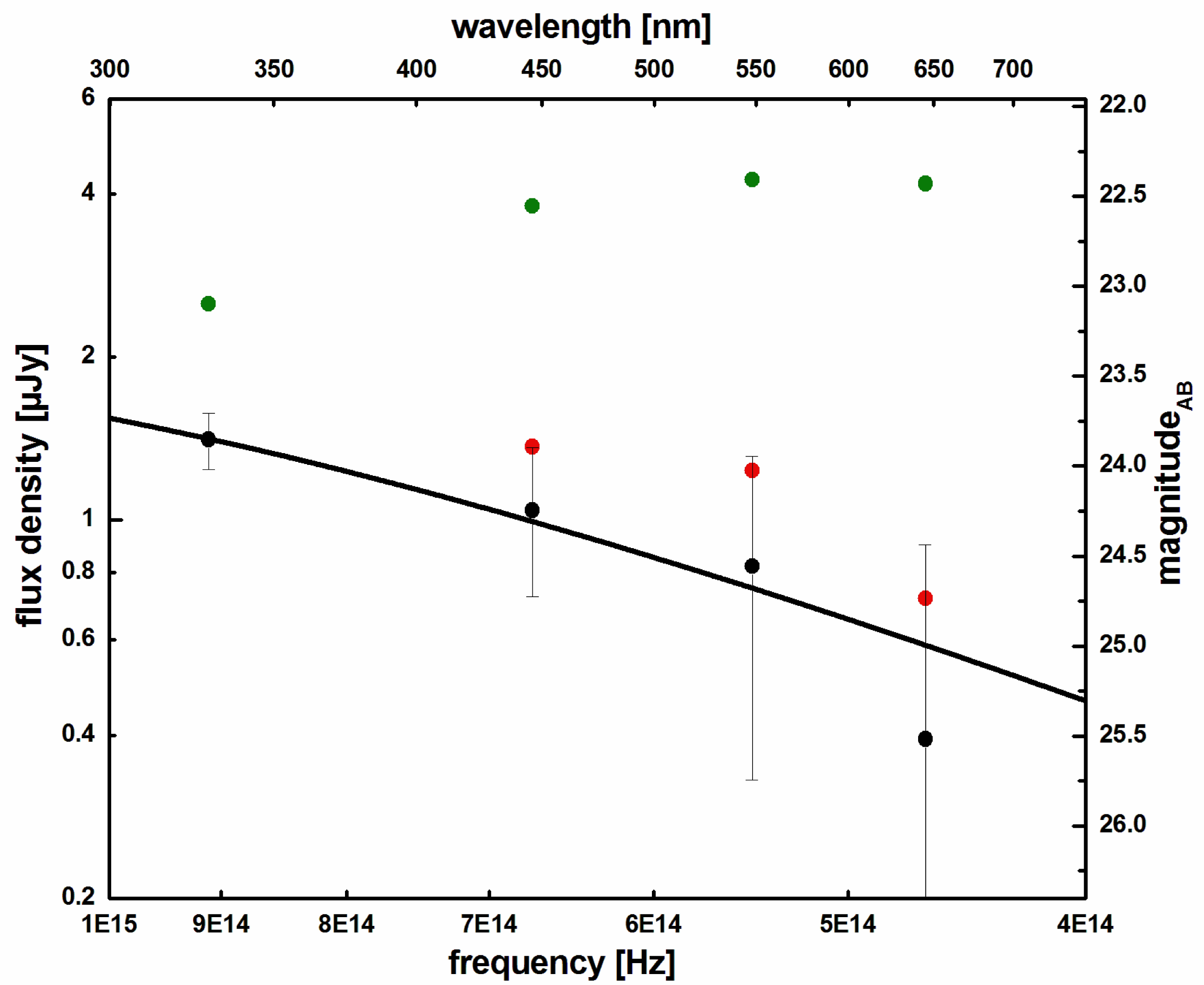}
\caption{Rest-frame SED of the optical transient 
  following GRB 120714B at $t_{\rm obs}=1.48$ days based on the GROND
  $g^\prime r^\prime i^\prime z^\prime $-band data (Table~\ref{120714Bmagdata}). Green-colored data points
  refer to observed, red-colored to host-subtracted, and black-colored to
  (host+SN)-subtracted magnitudes.  The thick black line shows the best fit,
  the SED of  a blackbody with $T_{\rm bb}^{\rm rest}=2.4\pm0.7$ eV.  The data
  imply that the peak of the blackbody flux still lies bluewards of the
  $g^\prime$ band. Wavelengths  and frequencies are given  in the host-galaxy
  frame.}
\label{fig:120714B_SED_DAY2}
\end{figure}

The lack of X-ray data does not allow us to determine $T_{\rm bb}^{\rm rest}$
at $t_{\rm obs}=0.353$ days in a more precise way.  However, we can estimate
its value by taking into account the optical GROND multi-color data at $t_{\rm
obs}=1.48$ days. At that time the emerging SN component cannot be neglected
anymore.  After the subtraction of host and SN light, the $(r^\prime -g^\prime
)$ color is still positive ($+0.3$ mag), corresponding to a spectral slope of
$\beta\sim-1$.  Compared to the first-epoch observations the SED has clearly
flattened.  A fit of our host- and SN-subtracted  (host-frame) data with a
blackbody component now yields $T_{\rm bb}^{\rm rest}=2.4\pm0.7$ eV, which we
adopt as a lower limit  for $T_{\rm bb}^{\rm rest}$ at 1.48 days (observer
frame; Fig.~\ref{fig:120714B_SED_DAY2}).   Smaller temperatures are excluded
as they would change the shape of the SED in a way that the data become
incompatible with the model.  Higher temperatures are formally not constrained
as the spectral slope would be at most $\beta=-2$, which is qualitatively
still compatible with the observed blue color.  If we adopt the hypothesis
that we see radiation from a cooling blackbody, then the lower bound on
$T_{\rm bb}^{\rm rest}$ implies that at 1.48 days the peak of the  blackbody
component had not yet passed the $g^\prime$ band.
 
The lower limit we can place on $T_{\rm bb}^{\rm rest}$ at  $t_{\rm obs}=1.48$
(in the following $t_2$) days can now be used to constrain its value  at
$t_{\rm obs}=0.353$ days (in the following $t_1$).  Following the procedure
outlined in \cite{Olivares2012},  for the time  evolution of $T_{\rm bb}$ we
make the ansatz $T(t)=T_{\rm init}-a\,t^b$,  where  $T_{\rm init}$, $a$, and
$b$ are constants.  Consequently, for two different times
$T(t_1)-T(t_2)=a\,(t_2^b-t_1^b)$, where we measure  $T$ in keV and $t$ in
seconds (in the host frame).  Using  $T_{\rm bb}^{\rm rest}(t_2=1.48$
days$)=2$~eV, $a=0.0036$ and  $b=0.3\pm0.2$ (as it was found
by \citealt{Olivares2012} for GRB/XRF 100316D/SN 2010bh), we obtain  $T_{\rm
bb}^{\rm rest} (t_1=0.353$ days$)\gr40$~eV. Similar results are obtained when
we use the parameters that describe  the temperature evolution of GRB/XRF
060218/SN 2006aj \citep{Campana2006}. Using their data, we find $a=0.0006$,
$b=0.47$ and obtain $T_{\rm bb}^{\rm rest} (t_1=0.353$ days$)\sim65$~eV. 

In principle, a cooling blackbody should progressively shift the peak flux of
the thermal emission redwards into the optical bands, potentially leading to a
detectable time evolution of the observed broad-band optical/NIR SED, and
perhaps  even a rebrightening as it was observed for GRB/XRF 060218/SN~2006aj
\citep{Campana2006}. Unfortunately, for GRB 120714B/SN~2012eb the rapid
rise of the supernova component in the $r^\prime i^\prime z^\prime$  bands 
prevents any detection of such an effect at later times.

\begin{figure}[t!]
\hspace*{-6mm}
\includegraphics[width=\columnwidth]{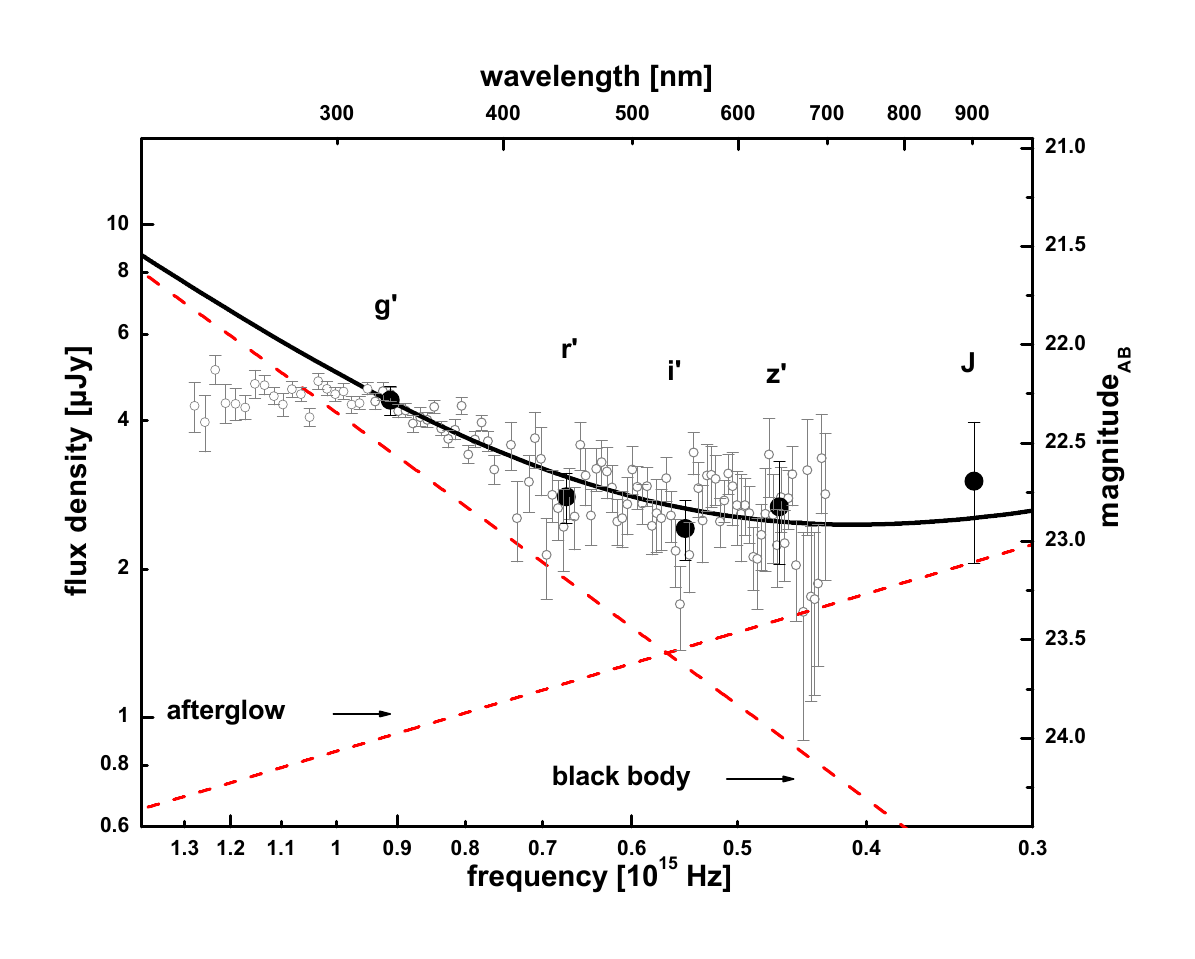}
\caption{Host-subtracted SED of the optical transient 
  following GRB 120714B at $t=0.353$ days
  (observer frame), consisting of GROND (black) and binned X-shooter (grey)
  data (step size 60~nm). 
  The thick black line shows the best fit for a two-component model
  (indicated by the red broken lines)
  consisting of afterglow light with  a spectral slope of  $\beta=0.7$ and a
  blackbody with a temperature of $T_{\rm bb}^{\rm rest}=50$~eV.  
  Wavelengths  and frequencies are given in the host-galaxy frame.}
\label{fig:120714B_SED_bb.ag}
\end{figure}

Once the blackbody component has been identified in the host-subtracted data
we can derive the spectral slope of the pure afterglow light at $t_{\rm
obs}=0.353$ days. In doing so, the scatter of the X-shooter data at larger
wavelengths and the uncertainty of the $J$-band magnitude do not constrain the
spectral slope  $\beta$ tightly enough; adopting for the blackbody component
$T_{\rm bb}^{\rm rest}$ = 50 eV, the  best fit to the $g^\prime r^\prime
i^\prime z^\prime J$-band data provides $\beta=0.7\pm0.4$
(Fig.~\ref{fig:120714B_SED_bb.ag}). We note that in this figure all data are
corrected for host-galaxy flux. Here, the  $g^\prime r^\prime i^\prime
z^\prime$-band data points are based on GROND observations of the transient at
$t_{\rm obs}=0.354$ days. The $J$-band data point, however, is based on GROND
observations at  $t_{\rm obs}=0.396$ days since the transient was not detected
in this band at earlier times (Table~\ref{120714Bmagdata}). Therefore, this
data point was shifted to $t_{\rm obs}=0.353$ days before the host-galaxy flux
was subtracted.  In doing so, we assumed a time evolution of the $J$-band flux
according to a power law with $\alpha=0.58$, as follows from our light-curve 
fits (Sect.~\ref{Photometry}; Table~\ref{Tab:summary1}). This  procedure
is justified since the data suggest that around 0.35 to 0.40 days the
host-galaxy corrected flux in the $J$ band  was dominated by afterglow light.

We finally note that  bluewards of the $g^\prime$ band the flux density of the
host-subtracted SED clearly deviates from a combined blackbody and afterglow
fit, there is a lack of flux with increasing frequency
(Fig.~\ref{fig:120714B_SED_bb.ag}). Milky Way-like dust  along the line of
sight with $A_V^{\rm host}\sim0.5$ mag could explain this feature.   The
deduced SN peak luminosities (parameter $k$; Table~\ref{Tab:summary2})  would
then scale correspondingly, which could potentially also explain the
non-detection of the SN in the $g^\prime$ band.  However, we cannot rule out
that this feature is (at least partly)  an artifact introduced by the
subtraction of the host-galaxy synthetic spectrum (green curve in
Fig.~\ref{fig:120714B.shock}), which is not well known bluewards of the
$g^\prime$ band.  Therefore, we followed Occam's razor and adopted for this
burst $A_V^{\rm host}=0$ mag.

\section{Discussion \label{Discussion}}
\subsection{Bolometric SN light curves \label{Sect:bols}}

In order to compare our sample of GRB-SNe with  the current data base of
GRB-SNe, we need to construct  bolometric light curves. In doing so, we are
confronted with the problem that these targets are at moderate redshifts. Even
if there is a good set of multi-color data in the observer frame, in the SN
rest frame the situation can look much less favorable. For instance, for the
cases discussed here we lack any GRB-SN detection redwards of the $z^\prime$
band.  In the worst case (GRB 071112C; $z=0.823$) the $z^\prime$ band is
centered at 490~nm rest-frame wavelength, in the best case (GRB 120714B,
$z=0.398$) at 639~nm. 

To correct for, e.g., the unknown NIR part of a  supernova SED, it is
usually assumed  that the SED can be reconstructed based on its snapshot in
the optical bands  \citep[e.g.,][]{Tomita2006ApJ,Takaki2013,Olivares2015a}.
We follow this approach in a more rigorous way by adopting the following two
assumptions: (i) The bolometric light curve of a GRB-SN can be described by a
stretch and a luminosity factor as well, and these two factors can be
normalized to the  bolometric light curve of SN~1998bw.  (ii) The bolometric
stretch and luminosity factor (in the following $s_{\rm bol}$ and $k_{\rm
  bol}$) can be constructed from the  stretch and luminosity factors in the
optical/NIR bands. In other words, we assume that the in some way 
wavelength-averaged SN light curve parameters {in the optical
bands} are
a very good proxy for the bolometric light curve parameters 
(light curve shape). {In order to avoid the introduction of too many 
mathematical symbols, and in order to underline that the procedure
which follows relies on the aforementioned basic ideas, 
in the following we 
will always refer to $s_{\rm bol}$ and $k_{\rm bol}$. Though 
the reader should keep in mind that these calculated numbers 
are approximate (quasi-bolometric) values for 
the true (and basically unknown) numbers of $s_{\rm bol}$ and $k_{\rm bol}$. 

In doing so, 
{\it we define} a quasi-bolometric} 
light-curve parameter $k_{\rm bol}$ as
\begin{equation}
k_{\rm bol} = \left( \frac{1}{n}\,\sum_{i=1}^n \ k_i^3\right)^{1/3}\,. 
\label{kmean}
\end{equation} 
We include in Eq.~(\ref{kmean}) the $r^\prime i^\prime z^\prime$ bands, but do
not use the $g^\prime$ band, i.e., we have $n=3$.\footnote{{In order to
be consistent, also in the case of GRB 130831A we did not use 
the $k$ value for the $g^\prime$ band (Table~\ref{Tab:summary2}).}}  

Our omission of the
$g^\prime$ band is immediately justified by the non-detection of the GRB-SNe
in $g^\prime$ in three of the four cases discussed here.  For SN~1998bw it is
$k_i=1$ for all $i$ so that Eq.~(\ref{kmean}) delivers $k_{\rm bol}=1$,
whatever $i$ (photometric bands) are selected.  

Similar to Eq.~(\ref{kmean}), we {\it define} a 
quasi-bolometric stretch factor 
\begin{equation}
s_{\rm bol} = \left( \frac{1}{n}\,\sum_{i=1}^n \ s_i^3\right)^{1/3}\,. 
\label{smean}
\end{equation} 
{
It should be stressed that there is {\it ad hoc} no 
physical reason for the use of a 
cubic averaging of the light-curve parameters. Also, 
the obvious weakness of  our definition of $k_{\rm bol}$ is that it ignores
any redshift effect. For different SNe at different redshifts the
corresponding $k_i$-parameters are measured in the  observer frame and hence
in the SNe host frames they can trace different wavelength regimes. As such we
caution that we build the 
quasi-bolometric light curve based on snapshots of the
light curves in different parts of a supernova SED. We adopt here the
assumption that $k_{\rm bol}$ closely follows a weighted ensemble average of
the $k_i$ values. Our way of averaging gives photometric bands with a large $k$
factor a higher weight than bands with a smaller $k$ factor. The philosophy
behind this approach is that it is probably more likely to underestimate the
flux in a certain band (potentially affected by unknown absorption) than to
overestimate it. Clearly, this argument cannot be exported to 
the averaging of the $s_i$ values. Nevertheless, here we use the same 
kind of averaging again in order to achieve some 
``internal mathematical consistency''
in our approach. Much more justified equations to construct 
quasi-bolometric parameters might be
found, but for the present discussion this is not strongly relevant.}

For the further discussion we need the SN expansion velocities during maximum
light, $v_{\rm ph; peak}$. To determine this, 
we follow \cite{Sanders2012} and make
the ansatz
\begin{equation}
v_{\rm ph}(t)=v_{\rm ph; peak} \ \left(\frac{t}{t_{\rm peak}}\right)^{-\alpha}\,.
\label{dv}
\end{equation}
In addition, we define $t_{\rm peak, bol} = s_{\rm bol} \,\times\,t_{\rm peak,
  bol, bw}$. Based on the analysis of \cite{Sanders2012} we assume
$\alpha=0.8\pm0.2$ for GRB-SNe.

In the case of GRB 120714B/SN~2012eb the bolometric peak time was at  $t_{\rm
  peak, bol} =13.6\pm0.7$ days (rest frame; Table~\ref{Tab:summary2}),  while
the SN spectrum was taken at $t=13.2$ days (rest frame).  Using
Eq.~(\ref{dv}), we obtain $v_{\rm ph; peak}=42\,000\pm5\,500$ km s$^{-1}$.  In
the case of GRB 130831A/SN~2013fu, we have $t_{\rm peak, bol}=11.9\pm0.3$ days
(Table~\ref{Tab:summary2}),  while the SN spectrum was taken at $t=19.5$
days. This leads to  $v_{\rm ph; peak}=30\,300\pm6\,000$ km s$^{-1}$.

In order to construct a comparison sample, we used the homogeneous multi-color
data set of $(s,k)$ pairs published by \cite{Kann2016} for GRB-SNe from the
years between 1999 and 2014 and applied Eqs.~(\ref{kmean}) and (\ref{smean}).
In addition, we included the GROND multi-color fit results for GRB
101219B/SN~2010ma \citep{Olivares2015a}. Finally, we selected only GRB-SNe
with  known $(k,s)$ data in the $r^\prime i^\prime z^\prime$ or $VR_cI_c$
bands. Altogether the entire sample contains 17 events, including the four
GRB-SNe studied here (but excluding the template GRB 980425/SN~1998bw;
Table~\ref{skall}). 

\begin{table}[!t]
\caption[]{Bolometric $(k,s)$ values and redshifts for 17 GRB-SNe.}
\renewcommand{\tabcolsep}{6.0pt}
\begin{tabular}{llccl}
\toprule
GRB     & SN     &    $k_{\rm bol}$   &  $s_{\rm bol}$ & $z$ \\
\midrule
990712  &        &   0.68 $\pm$ 0.11 & 0.67 $\pm$ 0.05 & 0.433\\  
020405  &        &   0.76 $\pm$ 0.09 & 0.85 $\pm$ 0.05 & 0.691 \\  
030329  &2003dh  &   1.46 $\pm$ 0.16 & 0.87 $\pm$ 0.01 & 0.1685\\  
031203  &2003lw  &   1.47 $\pm$ 0.26 & 1.08 $\pm$ 0.15 & 0.1055\\[2mm] 
\bf   071112C   &&   0.60 $\pm$ 0.10 & 0.75 $\pm$ 0.15 & 0.823\\
080319B &        &   1.03 $\pm$ 0.10 & 0.80 $\pm$ 0.05 & 0.938\\  
081007  &2008hw  &   0.93 $\pm$ 0.09 & 1.03 $\pm$ 0.06 & 0.530\\  
091127  &2009nz  &   1.30 $\pm$ 0.15 & 0.97 $\pm$ 0.04 & 0.490\\  
100316D &2010bh  &   0.48 $\pm$ 0.01 & 0.62 $\pm$ 0.01 & 0.0591\\  
101219B &2010ma  &   1.41 $\pm$ 0.09 & 0.76 $\pm$ 0.06 & 0.552\\
111209A &2011kl  &   2.41 $\pm$ 0.16 & 1.10 $\pm$ 0.03 & 0.677 \\
\bf   111228A   &&   0.75 $\pm$ 0.13 & 1.43 $\pm$ 0.14 & 0.716\\
120422A &2012bz  &   1.13 $\pm$ 0.01 & 0.91 $\pm$ 0.01 & 0.283\\ 
\bf   120714B & 2012eb &  0.69 $\pm$ 0.04 & 0.86 $\pm$ 0.04 & 0.3984\\  
130702A &2013dx  &   1.27 $\pm$ 0.03 & 0.86 $\pm$ 0.01 & 0.145 \\
\bf   130831A &2013fu & 0.77 $\pm$ 0.04 & 0.75 $\pm$ 0.02 & 0.4791\\ 
140606B &        &  0.70 $\pm$ 0.02 & 0.96 $\pm$ 0.06 & 0.384\\  
\bottomrule
\end{tabular}
\tablefoot{For the first four events the $k_{\rm bol}, s_{\rm bol}$ 
values are based on $VR_cI_c$-band data. For all other 
SNe these values are based on $r'i'z'$-band data. 
The four GRB-SNe discussed here are highlighted.
Bolometric values were calculated using Eqs.~(\ref{kmean}) and (\ref{smean}).}
\label{skall}
\end{table}

What is apparent in the data is that low- and medium-$z$ GRB-SNe
favor the case $s_{\rm bol}<1$. Only 4 events have $s_{\rm bol}>1$ (neglecting
the error bars). The spread in the individual  $k_{\rm bol}$ factors is much
larger, 9 events have $k_{\rm bol}<1.0$ and 8 have $k_{\rm bol}>1.0$. This
might reflect the fact that the determination of $k_{\rm bol}$ depends on the
(adopted) host-galaxy extinction along the line of sight as well as the
(adopted) internal SED of the corresponding SN while the determination of
$s_{\rm bol}$ does not. Noteworthy,
there is no evidence for a redshift-dependence of these parameters.

\begin{figure}[t!]
\includegraphics[width=1\columnwidth,angle=0]{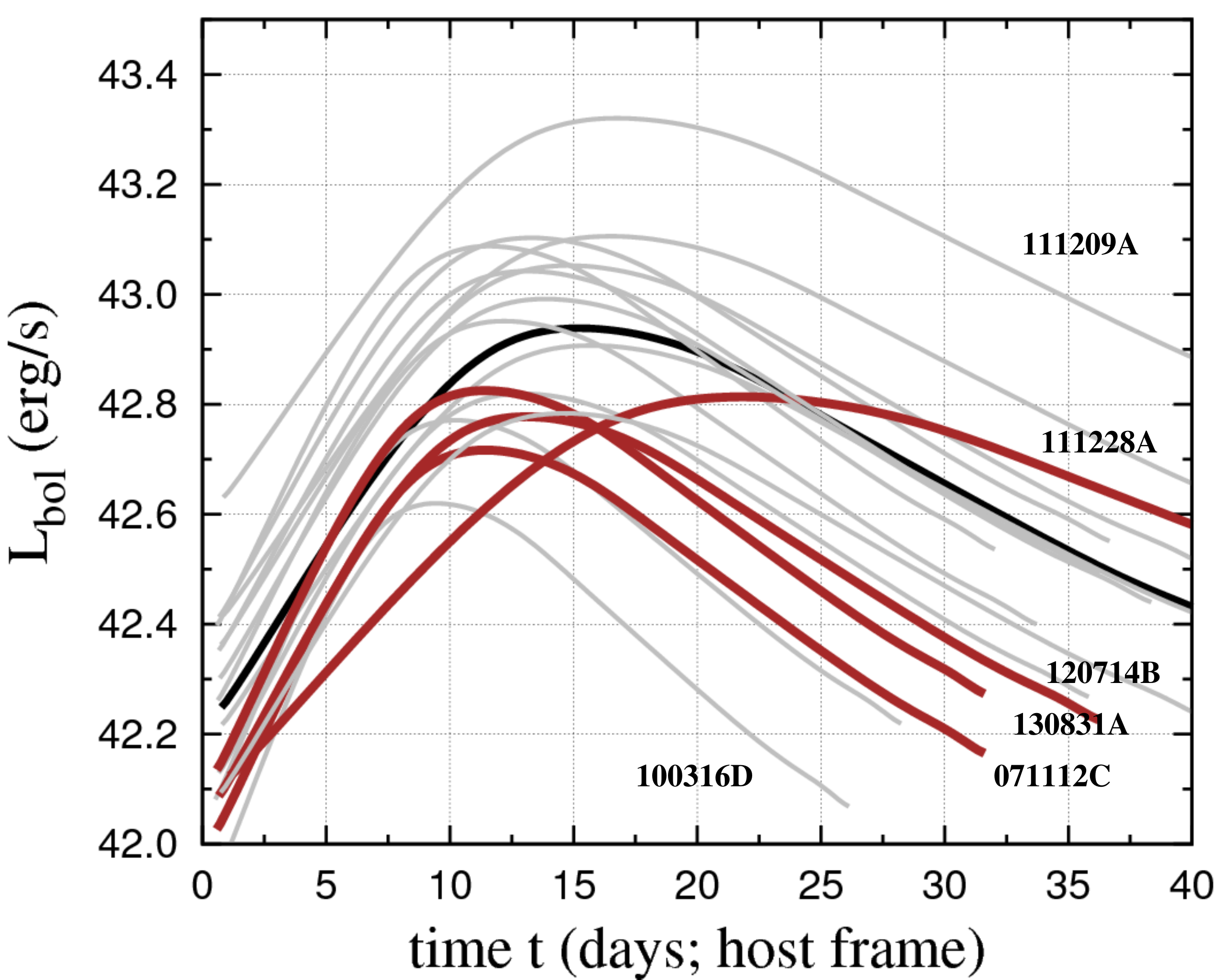}
\caption{Bolometric luminosity of the 17 GRB-SNe according to
  Table~\ref{skall}. 
  In addition we show the bolometric light curve of SN~1998bw (in black
  color), which we took from \cite{Greiner2015Natur} (their figure
  3) and performed a smoothed spline fit. All other SN light curves were then
  scaled accordingly using the corresponding bolometric values for the stretch
  and luminosity factor (Eq.~\ref{ot}).  The four GRB-SNe discussed here
  are drawn in brown color. Note that  this plot is a visualization of the
  $(s_{\rm bol}, k_{\rm bol})$  values in Table~\ref{skall}. It is not based
  on individual data points.}
\label{fig:lum}
\end{figure}

The $k_{\rm bol}, s_{\rm bol}$ pairs were then used to calculate 
bolometric light curves by using the corresponding light curve of SN~1998bw 
as a reference (\citealt{Greiner2015Natur}; Figure~\ref{fig:lum}). 

What is immediately apparent in Table~\ref{skall} is the outstanding large
stretch factor we have found  here for the SN associated with GRB 111228A. 
This solution has to be taken with care, however. As we have
noted in appendix~\ref{App.Sect.111228A}, due to the rather weak
SN bump in this case the fit is rather sensitive to the deduced afterglow
parameters. {For example,
fitting the \swift/XRT light curve from 2000~s onwards, and regardless
of whether we include the last X-ray
data point in the fit or not, the post-break
decay slope is much shallower than implied by the GROND data, 
closer to 1.4. Consequently, the reported $1\sigma$ error for this stretch
factor might therefore be even larger than given in Table~\ref{skall}.

Neglecting this burst as an outlier, for GRB-SNe the observed $s_{\rm
bol},k_{\rm bol}$ pairs suggest that a large $s_{\rm bol}$ implies a large
$k_{\rm bol}$ \citep[Fig.~\ref{ks};
see also][]{Cano2013a,Cano2014ApJ,Cano2016}.  The data
can formally be fitted with a function $k_{\rm bol}\sim s_{\rm bol}^{a}$,
$a=\textnormal{const.}$,  but the scatter in the data is large and the result of the fit
(i.e., the fit parameter $a$) sensitive to the exclusion of individual
events.  While hidden systematic errors could affect our determination of
the individual $k_{\rm bol}$ and $s_{\rm bol}$ values, this scatter could
also indicate that additional parameters come into play which are not
considered in this plot \citep{Klose2017}.

\begin{figure}[t!]
\includegraphics[width=1\columnwidth,angle=0]{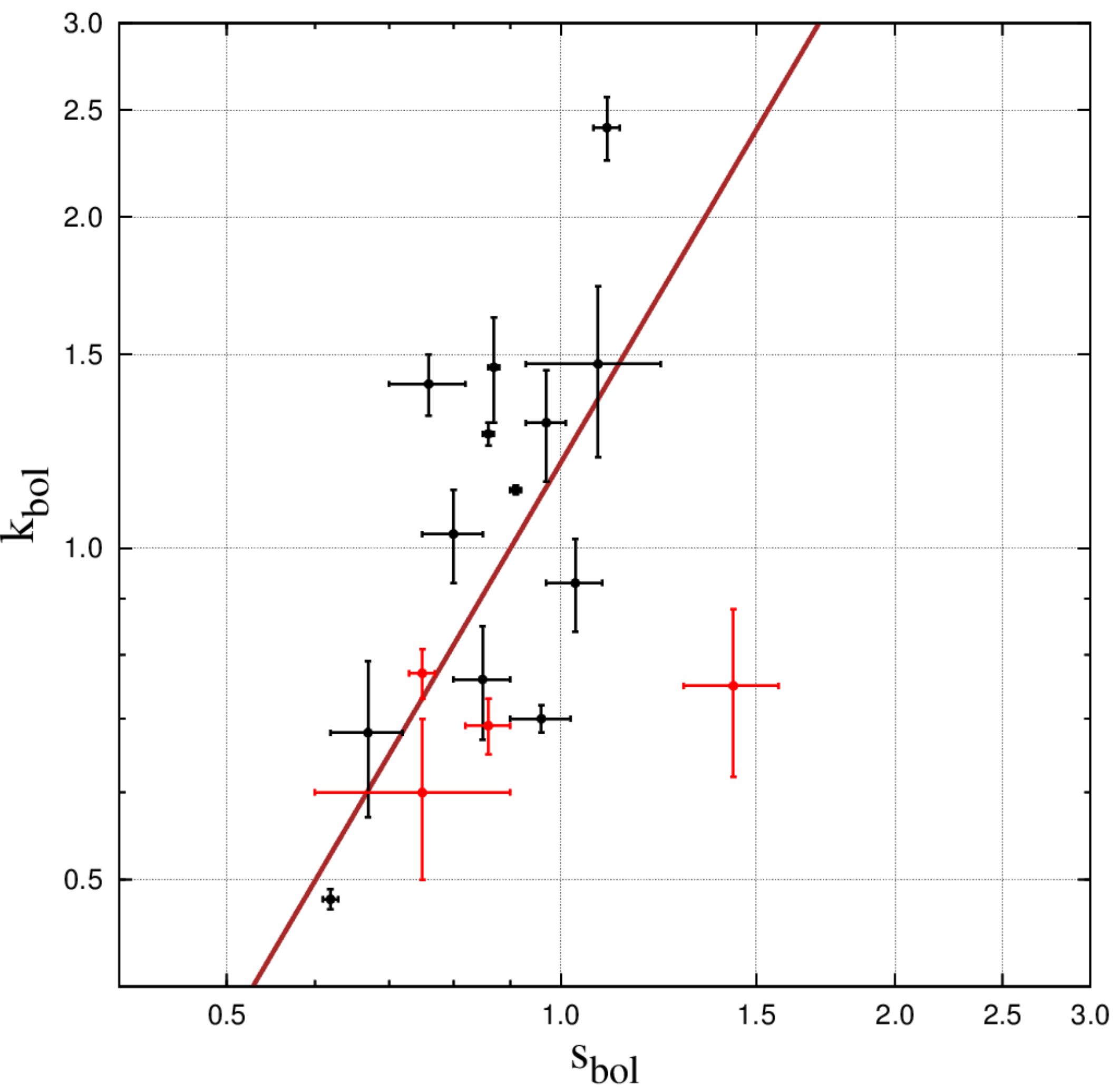}
\caption{Bolometric luminosity $k_{\rm bol}$ 
  vs. bolometric stretch factor $s_{\rm bol}$ for the 17 GRB-SNe  listed in
  Table~\ref{skall}. Red-colored data points refer to the four GRB-SNe
  discussed here, black-colored data points to the remaining 13 events.
  Note that SN~1998bw is not included here (by definition,
  $s_{\rm bol}^{\rm bw}=1, k_{\rm bol}^{\rm bw}=1$).
  The best fit is shown for an ansatz $k_{\rm bol}\sim s_{\rm bol}^{a}$,
  $a=$1.7 (excluding in the fit GRBs 111209A, 111228A and SN~1998bw). 
} 
\label{ks}
\end{figure}

\subsection{Supernova explosion parameters}

We used the Arnett model for type Ia SNe \citep{Arnett1982}, its slight
modification by \cite{Valenti+08}\footnote{for applications and explanations
of the Arnett-Valenti equations see, e.g.,
\cite{Olivares2015a,Wheeler2015,Toy2016ApJ}}, and the
\emph{Stritzinger-Leibundgut relation} \citep{Stritzinger2005}\footnote{see,
e.g., \cite{Takaki2013} and \cite{Prentice2016}} to derive the explosion
parameters (released mass of  radioactive nickel $M_{\rm Ni}$, ejected mass
$M_{\rm ej}$, kinetic energy of the ejecta $E_{\rm kin}$). We
normalize all values to SN~1998bw. As such we are insensitive to 
slight disagreements in the literature about 
the corresponding absolute values for SN~1998bw (e.g., the discussion in
\citealt{Dessart2017A&A}). Within this context
we have (\citealt{Klose2017}):
\begin{eqnarray}
M_{\rm Ni; \ norm} &=&  k_{\rm bol} \ \ \frac{\eta(t_{\rm peak}^{\rm bw})}{\eta(t_{\rm peak})}\,,  
\label{eq:Mni} \ \ \mbox{with}\\
\eta &=& 6.45 \, e^{-t_{\rm peak}/8.76} 
       + 1.45 \, e^{-t_{\rm peak}/111.27}\,,\\
\left(\frac{M^3_{\rm ej}}{E_{\rm kin}}\right)_{\rm norm}&=& 
s^4_{\rm bol}\ \left(\frac{\kappa^{\rm bw}}{\kappa}\right)^2\,, 
\label{eq:MejEkin}
\end{eqnarray} 
where $t_{\rm peak}$ is the peak time of the SN measured in days and 
$\kappa$ is the  volume and time-averaged matter opacity in the ejecta. 

\cite{Dessart2016MNRAS} have pointed out that the Arnett rule and its
underlying model assumptions are well established for type Ia SNe, but when
compared with numerical models for type II/Ib/Ic SNe the Arnett rule
overestimates the amount of released nickel mass by as much as 50\%. Moreover,
when asymmetric explosions  are considered, the derived explosion parameters
can change notably \citep{Dessart2017A&A}. We do not consider these issues
further but leave the reader with the warning that in the following error bars
on deduced SN parameters are basically simply mathematical in nature.

Using  the \emph{Stritzinger-Leibundgut relation}, in our sample 8 of the 17
GRB-SNe have $k_{\rm bol}>1$ (neglecting the $1\sigma$ error bars); 
they produced more
nickel than SN~1998bw (Fig.~\ref{MnickelCalc}).\footnote{This picture does not
change when we use the \emph{Arnett-Valenti relation} \citep{Valenti+08},
even though the sample of SNe that falls into this category is now slightly
different.} The ratio $M_{\rm Ni}/M_{\rm Ni}^{\rm bw}$ varies between about
0.3 and 2.6. It is highest for GRB 111209A/SN~2011kl (in agreement with
\citealt{Kann2016}; assuming the light curve was powered by $^{56}$Ni) 
and lowest for GRB 990712 and GRB/XRF 100316D/SN~2010bh. Excluding the
apparent outlier GRB 111209A, ninety percent (15) of the remaining 16 SNe lie
in a relatively narrow range between $M_{\rm Ni}/ M_{\rm Ni}^{\rm bw}=0.56$
and 1.23. Three of the four GRB-SNe discussed here (GRB 071112C, 120714B,
130831A) occupy the lower part of this distribution; they produced a rather
small  amount of radioactive nickel.

\begin{figure}[t!]
\includegraphics[width=1\columnwidth]{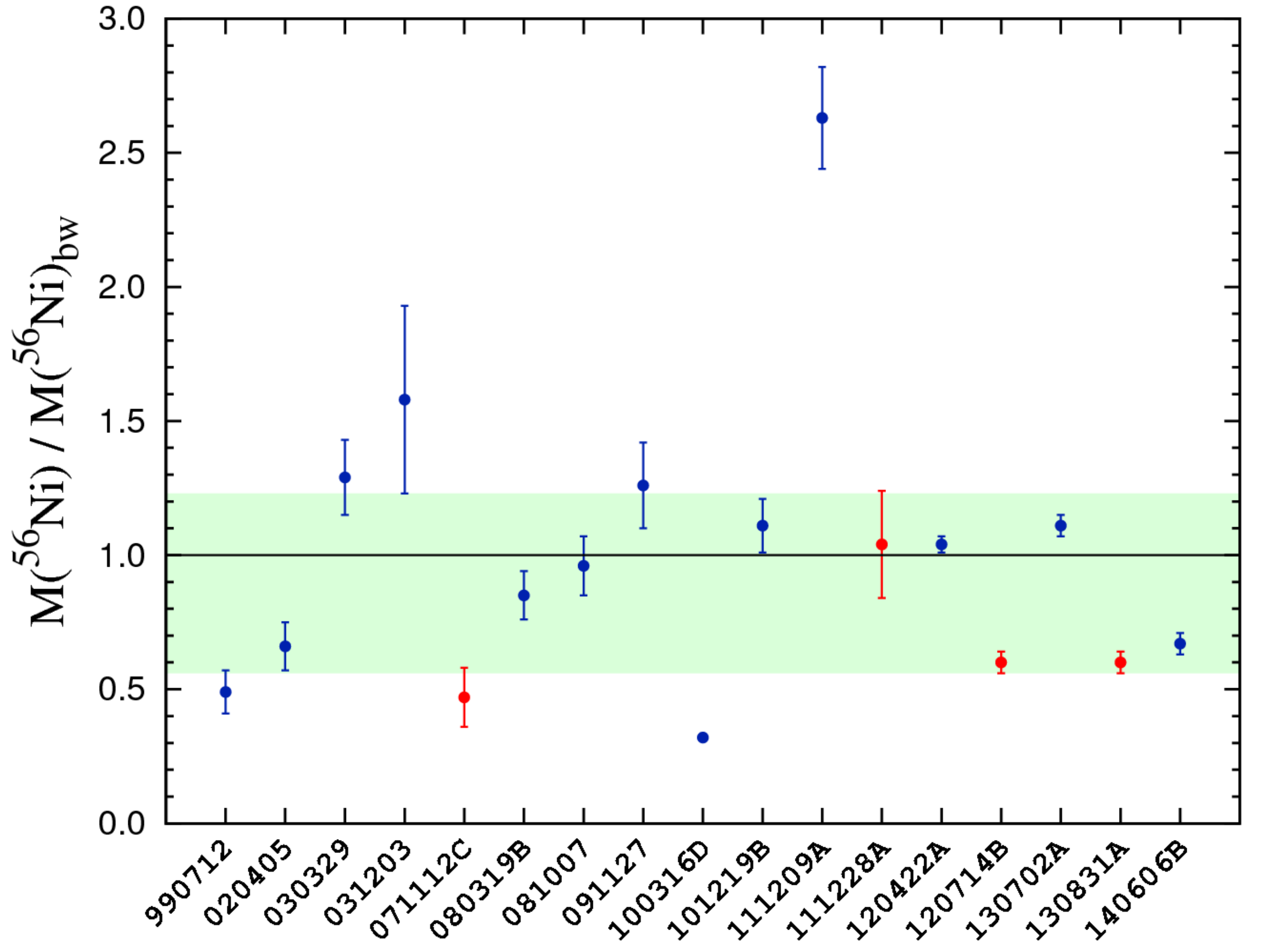}
\caption{The ejected mass of $^{56}$Ni for the 17 GRB-SNe listed in
  Table~\ref{skall} according to   the \emph{Stritzinger-Leibundgut
    relation}. In red are shown the results for the four GRB-SNe discussed
  here.  Excluding the apparent outlier GRB 111209A, ninety percent (15 SNe)
  of the remaining 16  SNe lie in the colored region (within their $1~\sigma$
  errors). The black horizontal line highlights the position of the template 
  SN~1998bw.} 
\label{MnickelCalc}
\end{figure}

The ultra-long GRB 111209A/SN~2011kl deserves a special note
\citep{Greiner2015Natur,Kann2016,Mazzali2016MNRAS} as it is an obvious outlier
in the distribution ($M_{\rm Ni}/M_{\rm Ni}^{\rm bw} \sim 2.6$).  We note
however,  if $M_{\rm Ni}^{\rm bw}=0.4~M_\odot$ (\citealt{Nakamura2001})
then the  absolute value we derive here is in agreement with the one reported
in \cite{Greiner2015Natur}. It has already been stated by these authors that
for SN~2011kl an additional energy source, most likely a newly-formed
magnetar, might have substantially affected its high peak luminosity 
(see also \citealt{Kann2017arXiv170600601K}).
Consequently, the approach we have used here overestimates the
amount of released radioactive $^{56}$Ni. 

Figure~\ref{fig:M3Ekin} shows for each SN the ratio  $M_{\rm ej}^3/E_{\rm
  kin}$, normalized to SN~1998bw (assuming $\kappa =
\kappa^{\rm bw}$).  Since this ratio does not depend on the 
photospheric expansion velocity,
it is more reliable than the value for its individual components $M_{\rm ej}$
and $E_{\rm kin}$. Not plotted in Fig.~\ref{fig:M3Ekin} is GRB 111228A since 
due to the relatively large stretch factor for this SN
$(M_{\rm ej}^3/E_{\rm kin})_{\rm norm}=
6.3\pm1.6$, which exceeds the parameter space occupied by the
other 16 GRB-SNe substantially. As noted beforehand, for this event the SN
bump is quite weak and the result of the fit relatively sensitive on the 
inclusion of individual data points. In other words, 
the error for $s_{\rm bol}$ is probably larger than what is quoted in
Table~\ref{skall}. Neglecting GRB 111228A as an outlier and
taking into account the $1~\sigma$ error bars, ninety  percent 
of all SNe (15 SNe) lie
between  $(M_{\rm ej}^3/E_{\rm kin})_{\rm norm}=0.23$ and 1.45.
 
\begin{figure}[t!]
\includegraphics[width=1\columnwidth]{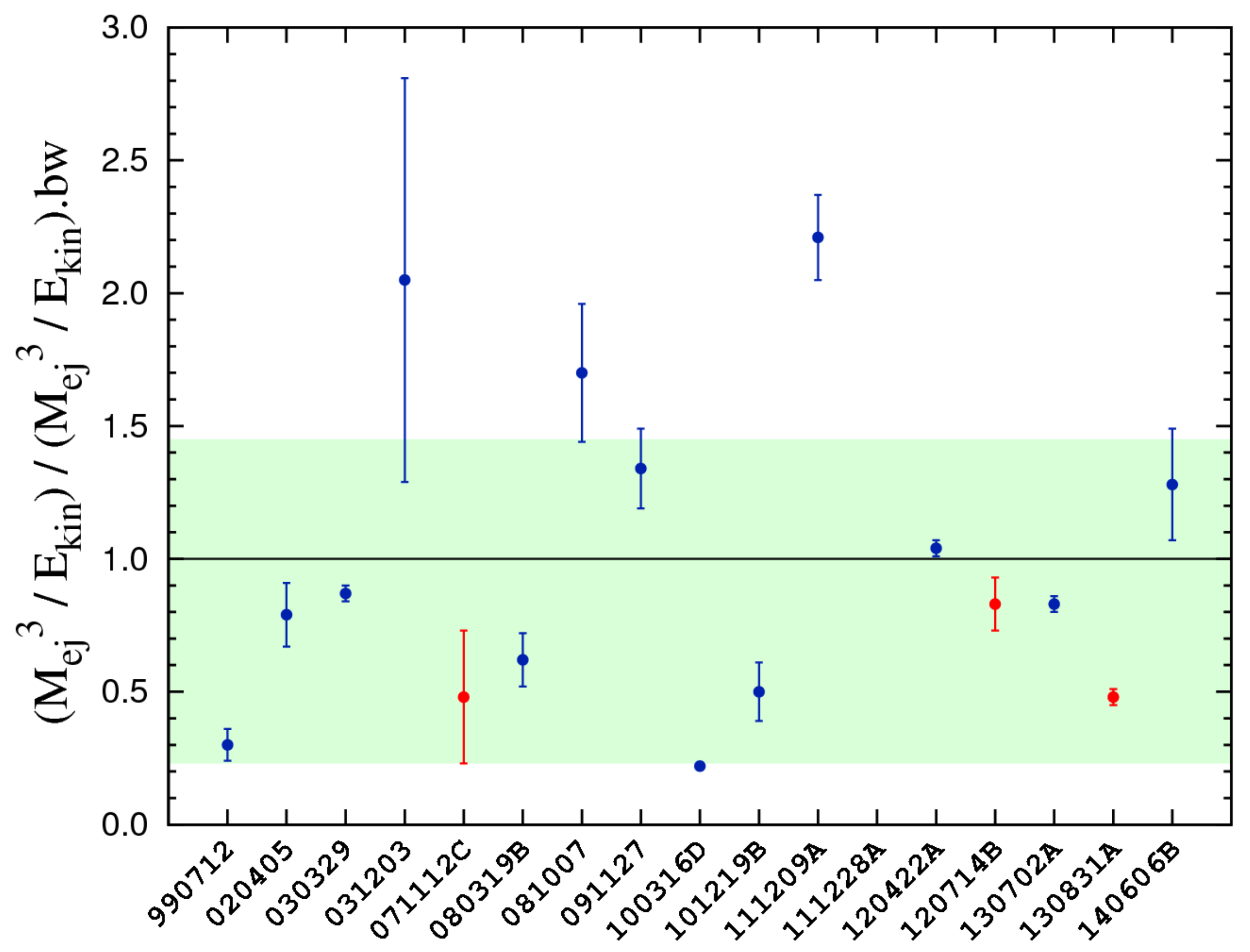}
\caption{The ratio $M_{\rm ej}^3/E_{\rm kin}$ for the SNe  listed in
  Table~\ref{skall} normalized to the corresponding value of SN~1998bw and
  adopting $\kappa = \kappa^{\rm bw}$. The burst GRB 111228A is not plotted
  here since its large value defines it  as an outlier in the distribution (see
  text).  Neglecting GRB 111228A, ninety percent (15 SNe) lie in the colored
  region. Symbols follow Fig.~\ref{MnickelCalc}. }
\label{fig:M3Ekin}
\end{figure}

A calculation of the ejected mass $M_{\rm ej}$ and the kinetic energy $E_{\rm
  kin}$ in the SN shell requires knowledge of the photospheric expansion
velocity $v_{\rm ph}$.  In our study, a snapshot of $v_{\rm ph}$ has been
obtained only for  GRB 120714B/SN~2012eb and 130831A/SN~2013fu. Given that
$v_{\rm ph}$ is a function of time and given that its value  usually depends
on the spectral feature used to measure it (e.g., \citealt{Cano2016}, their
figure 6), conclusions on  $M_{\rm ej}$ and $E_{\rm kin}$ are less secure even
if we are normalizing our results to the corresponding parameters of
SN~1998bw.

Using for SN~2012eb $v_{\rm ph; peak}=42\,000\pm5\,500$ km s$^{-1}$, we obtain
for the ejected mass $M_{\rm ej}/M_{\rm ej}^{\rm bw} = 2.2\pm0.5$ and a
kinetic energy of $E_{\rm kin}/E_{\rm kin}^{\rm bw} = 20.0\pm11.1$.  For
SN~2013fu ($v_{\rm ph; peak}=30\,300\pm6\,000$ km s$^{-1}$) the corresponding
numbers are $1.2\pm0.3$ and $5.7\pm4.7$, respectively.  The error is large, in
particular for the kinetic energy; it is dominated by the 
error of $v_{\rm ph; peak}$.  As noted beforehand, all values assume
$\kappa = \kappa^{\rm bw}$;  otherwise these numbers scale accordingly
(for a detailed discussion of this issue, see \citealt{Wheeler2015}).

Finally we note  that neither $k_{\rm bol}$ or $s_{\rm bol}$ revealed a
dependence on  redshift. As such there is no redshift-dependence of the
explosion parameters.  These data are fully consistent with the parameter
space occupied by the  so far known population of GRB-SNe, 
including those at low redshifts, confirming earlier
studies  (e.g., \citealt{Soderberg2006ApJ...636..391S,
  Tanvir2010ApJ...725..625T,Sparre2011ApJ...735L..24S}).

\begin{table}[t!]
\renewcommand{\tabcolsep}{3pt}
\caption{Summary of the afterglow energetics for an ISM and for a wind model.} 
\begin{center}
\begin{tabular}{l | rrrr}
\toprule
GRB     &  $\Theta^{\rm ISM}_{\rm jet}$ & log $E_{\rm corr}^{\rm ISM}$ & $\Theta^{\rm wind}_{\rm jet}$ & log $E_{\rm corr}^{\rm ISM}$ \\
        &  (deg) & (erg) & (deg) & (erg)\\
\midrule
\#1     &  $>$5.5$\pm$0.2  & $>$49.95$\pm$0.08  & $>$5.6$\pm$0.4  & $>$49.97$\pm$0.06\\
\#2     &   6.3$\pm$0.1  &  50.39$\pm$0.04  &  5.4$\pm$0.1  &  50.26$\pm$0.03\\
\#3     &  $>$9.5$\pm$0.3  & $>$48.90$\pm$0.09  & $>$14.5$\pm$1.0 & $>$49.26$\pm$0.06\\
\#4     &  $>$6.8$\pm$0.1  & $>$49.70$\pm$0.03  & $>$7.6$\pm$0.2  & $>$49.80$\pm$0.02\\
\bottomrule
\end{tabular}
\end{center}
\tablefoot{GRB numbers \#1 to \#4 refer to GRBs 071112C, 111228A, 120714B, 
and 130831A, respectively. 
Lower limits on the beaming angle and the corresponding 
beaming-corrected energy release refer to a break time of $t_b>$1 day.}
\label{Tab:summary4}
\end{table}

\subsection{Afterglow luminosities}\label{AGLum}

The parameters we have found for the decay slopes $\alpha$ of the afterglows
and their spectral slopes $\beta$ are summarized in Table~\ref{Tab:summary1}.
These parameters match what has been deduced for other long-duration
GRBs \citep[e.g.,][]{Zeh2006a,Kann2006ApJ,Kann2010}, $\alpha$ lies between about
0.5 and 1.6, $\beta$ between about 0.4 and 1.0.  Only in the case of GRB
111228A do we see evidence for a potential jet break around 1.7 days after the
burst. The observational constraints we can set on the jet-break times  for
the other three events do not characterize these bursts as special,
however (for a recent summary of measured jet half-opening angles see
\citealt{Tanvir2017}). 
It is possible that in these cases a break is hidden by the
rapidly rising SN components and, therefore, remain undetected. Keeping
this in mind, the results and constraints we obtain for the 
beaming-corrected isotropic equivalent energies $E_{\rm corr}$
and jet half-opening angles ($\Theta_{\rm jet}$;
Table~\ref{Tab:summary4}) do
not reveal atypical burst parameters. Based on the 
closure relations between temporal and spectral slopes 
\citep[$\alpha-\beta$ relations,
e.g.,][]{Zhang2004IJMPA,Gao2013NewAR..57..141G}, in no case can we rule 
out or favor a wind or an ISM model (Table~\ref{Tab:alphabeta}).

In order to characterize the luminosities of the afterglows, we compare them 
with the afterglow sample discussed in \cite{Kann2006ApJ,Kann2010,Kann2011ApJ}.
By using the derived intrinsic slopes of the afterglows as well as the
line-of-sight extinctions (which is significantly detected only for GRB
111228A), we applied the method described in \cite{Kann2006ApJ} to shift the
observed afterglow light curves to a common redshift of $z=1$.

\begin{figure*}[t!]
\includegraphics[width=\columnwidth,angle=0]{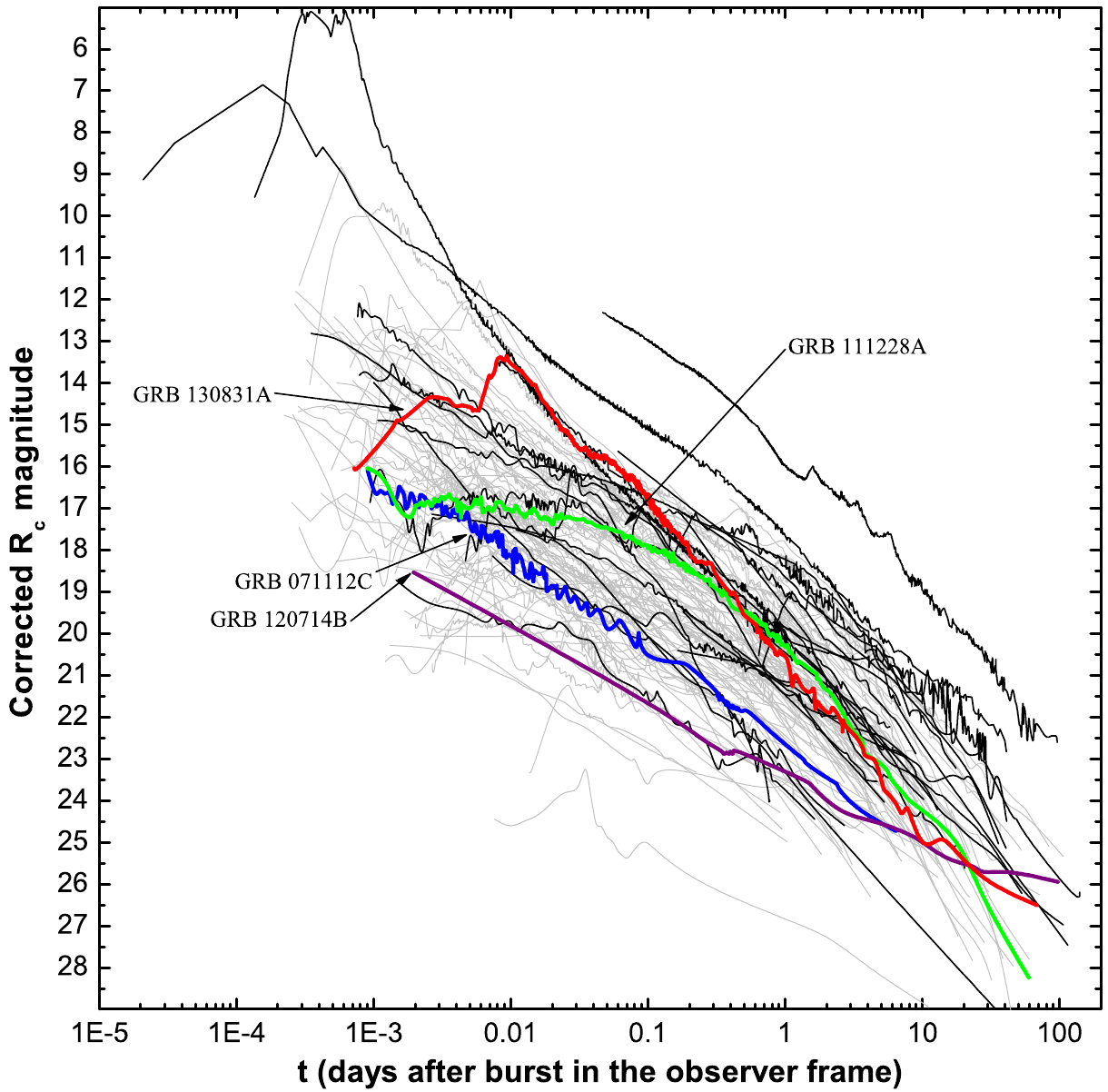}
\includegraphics[width=\columnwidth,angle=0]{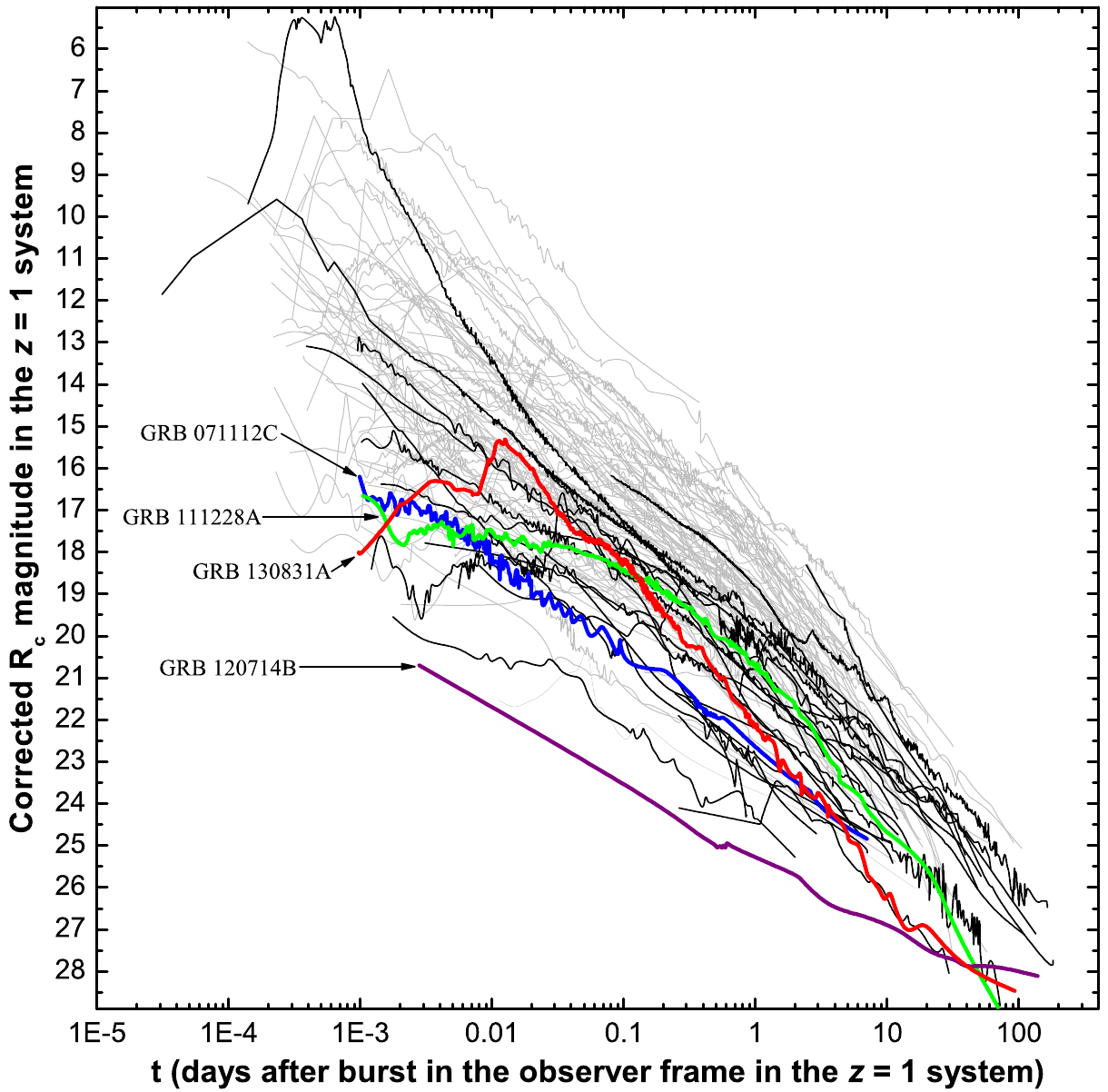}
\caption{\emph{Left:}\  The observed afterglows of (long) GRBs. These light
  curves are corrected for Galactic extinction and, where appropriate, for the
  contribution of the host galaxy and the late-time supernova signature. Light
  grey light curves are afterglows with no known SN contribution. Black lines
  are afterglows with detected SN signatures. Finally, the afterglows of the
  four GRBs discussed in this paper are color-coded and labeled. It can be
  seen that GRBs associated with SNe span a large observed magnitude range,
  from the brightest afterglows to among the faintest.  \emph{Right:}\  The
  afterglows, corrected for all line-of-sight extinction, and shifted to a
  common redshift $z=1$ (i.e., the afterglows are as they would be observed if
  they all lay at that redshift and were not affected by dust extinction).
  The SN-associated GRB afterglows are now seen to cluster in the lower part
  of the distribution; indeed, GRB 120714B has the least luminous afterglow
  within the entire sample except at very late times. For more details, see
  appendix~\ref{explained:kplots}.}
\label{kplot12}
\end{figure*}

We show the 
afterglow light curves in Fig.~\ref{kplot12} (left). These light
curves are in Vega magnitudes, in the $R_C$ or the $r^\prime$ band, corrected
for Galactic extinction, and, where appropriate, corrected for the
contribution of the host galaxy and the late-time SN signature (in detail see
appendix~\ref{explained:kplots}). Other than that, they are as observed. Light
grey afterglows represent the ``background sample'', these are overwhelmingly
those at $z>1$, though there are some cases where no SN search was undertaken
(to our knowledge), or the line-of-sight extinction toward the GRB site
prohibited a SN search despite the low redshift \citep[e.g., GRB 130925A,][]{Greiner2014AA}.
The black light curves represent the afterglows of
GRBs for which SN signatures have been detected, from late-time red bumps to
detailed spectral sequences. Note that several low-$z$ SN-GRBs, such as GRB
980425 \citep[e.g.,][]{Galama1998Natur,Clocchiatti2011AJ}, GRB 031203
\citep[e.g.,][]{Malesani2004a}, GRB/XRF 060218 \citep[e.g.,][]{Pian2006Nature,Ferrero2006a},
and GRB/XRF 100316D \citep[e.g.,][]{Cano2011,Olivares2012,Bufano2012ApJ} 
are not included in this sample, as they have no
discernable afterglow components (see \citealt{Kann2011ApJ} for a discussion on
upper limits on several of these events\footnote{in the case of GRB/XRF 100316D 
\cite{Olivares2012} report evidence for an afterglow component in two early-time
SEDs}). Finally, we plot the four GRB
afterglows discussed in this paper in separate colors and label them.

It can be seen that GRBs which are empirically associated with SNe span a wide
variety of behaviors and afterglow magnitudes. At very early times, there are
the extremely bright flashes of GRBs 080319B
\citep{Racusin2008Natur,Bloom2009ApJ} and 130427A \citep{Vestrand2014Sci}, but
also very faint light curves, like those of XRF 050416A \citep{Kann2010} and
GRB 120714B as presented in this paper. Afterglows of SN-associated GRBs can
also be very bright at late times, such as the exceptional case of GRB 030329
\citep[e.g.,][]{Lipkin2004ApJ,Kann2006ApJ}, and the similarly nearby GRB
130702A \citep{Singer2013ApJ,Singer2015ApJ,Toy2016ApJ,Volnova2017MNRAS}.
This is not unexpected, as there are strong biases involved. All
SN-GRBs lie at $z\lesssim1$ and any cases with large
line-of-sight extinction (such as GRB 130925A as mentioned above) will not be
followed up at late times. 

The light curves shifted to $z=1$ are shown in Fig.~\ref{kplot12} (right).
Interestingly,  the afterglows of SN-GRBs now reside in the lower part of the
luminosity distribution,  with the exceptions of the aforementioned bright
early flashes and a few further cases (e.g., the afterglow of GRB 991208 as
well as that of GRB 030329, which is generally not exceptionally luminous,
\citealt{Kann2006ApJ}).  Since all long GRBs should be associated with
late-time SN emission, this effect of a lower luminosity for SN-GRBs might be 
caused by an observational bias. In particular, only a low redshift enables
detailed observations of even intrinsically low-luminosity afterglows. One
such case is GRB 120714B, which is found to be the least luminous afterglow in
the entire sample for most of its observed timespan. At $t\sim0.03$ days  (in
the $z=1$ system) it was more than 12 mag less luminous than the brightest
afterglows at that time, at $t=1$ day it was still $\sim5$ mag fainter than
the median of the distribution; a clear advantage for the monitoring of the SN 
bump.

\subsection{Host-galaxy properties \label{sect:hosts} }

The host-galaxy magnitudes in our sample range between $r^\prime=23$ and 25
mag (AB system), the brightest is the one of GRB 120714B, which is also the
closest. On the GROND images all hosts appear basically featureless, their
morphology is not revealed. Additional archival Hubble Space
Telescope (HST) data in the case of GRB 071112C as well
as observations with the High Acuity Wide field $K$-band Imager
(HAWKI) mounted at the VLT (GRB 120714B) 
and deep imaging with FORS2 (GRB 130831A) do not show
any substructure either. An astrometry of these images shows that 
three of the four
GRB-SNe exploded within the inner  $1-2$ kpc projected distance from the
geometric center of their hosts. 

In order to measure the (global) star-formation rate (SFR) in the hosts, we
modeled their Galactic-extinction-corrected broad-band SEDs
(Table~\ref{Tab:summary3}) using \texttt{Le PHARE} \citep{Arnouts1999a,
  Ilbert2006a}\footnote{\url{http://www.cfht.hawaii.edu/~arnouts/LEPHARE}} 
(PHotometric Analysis for Redshift Estimations)
by applying a grid of galaxy templates based on
\citet{Bruzual2003a} stellar population-synthesis models with the Chabrier
initial mass function.

All four host galaxies (Fig.~\ref{fig:hosts}) turn out to be rather typical
members of the class of long-GRB hosts as they are found in the low-$z$
universe ($z=0.4-0.8$; look-back time between $\sim4.4$ and 7.1 Gyr;
Table~\ref{Tab:LePhare1short}).  They are low-luminosity ($M_B=-18.6$ to
$-17.7$), low-mass ($\log M/M_\odot= 8.4-8.7)$ hosts. Their global
star-formation rate is rather modest, on the order of 1~$M_\odot$ yr$^{-1}$.
Their specific SFR (sSFR) is (within the errors) rather normal when compared
to the $z<0.5$ GRB host-galaxy sample studied by \cite{Vergani2015a} and
\cite{Schulze2016}, log sSFR $\sim-8.7$ to $-8.4$ ($M_\odot$ yr$^{-1}$ per
$M_\odot$), suggesting that all four galaxies are undergoing episodes of
star-forming activity.  Their global parameters ($M_B, SFR,
M/M_\odot)$ lie close to the median found for the \emph{Swift}/BAT6 GRB
host-galaxy sample of long GRBs \citep{Vergani2015a}. 

Nevertheless,
some caution is required concerning the deduced SFR: (1) Due to the
lack of $HK_s$ detections for the hosts of GRBs 111228A and 130831A, in these
cases the stellar mass and sSFR obtained by  \texttt{Le PHARE}
should be taken with care.  (2) As
noted in Sect.~\ref{SN.Spectr}, the emission line strengths 
obtained (predicted) 
by \texttt{Le PHARE} for the hosts of GRBs 120714B and 130831A
might be too large. This could indicate that
for these hosts the SFR is notably smaller than what was obtained based
on the available broad-band photometry.

\section{Summary}

We have performed a detailed study 
of the GRB-SNe related to the long GRBs 071112C, 111228A,
120714B, and 130831A, covering a redshift range from  $z=0.4$ to 0.8. 
Partly in combination with public data, we were able
to follow their light curves in $g^\prime r^\prime i^\prime z^\prime $ over several weeks, 
with additional host-galaxy
observations more than one year after the  corresponding burst.  These events
belong to a rare number of fewer than 20 GRB-SNe for which multi-color light
curves in at least three  photometric bands are available. This allowed us to
derive (quasi-) bolometric light curve parameters  (stretch factor $s_{\rm
bol}$ and luminosity factor $k_{\rm bol}$) and, using the Arnett SN model, to
determine the SN explosion parameters $M_{\rm Ni}$ and $M_{\rm ej}^3/E_{\rm
kin}$ normalized to the corresponding parameters of SN~1998bw. In doing so, we
found that the four SNe studied here released between 0.5 and 1.0~M$_\odot$
of radioactive nickel, a parameter range which lies within the observed 
interval for GRB-SNe at low and medium redshifts ($z=0.1-0.9$). For
$M_{\rm ej}^3/E_{\rm kin}$ the parameter range is larger,
between 0.5 and 6.3. However, this large range
is mainly affected by the outlier GRB 111228A, whose
high value (6.3) could be a consequence of the relatively flat SN bump that
made the precise determination of its bolometric stretch factor difficult.
Omitting this burst, for the other three events we have found $0.5\kr M_{\rm
ej}^3/E_{\rm kin}\kr0.8$, a range which is consistent with the
parameter space occupied by the presently known GRB-SN ensemble with
multi-color light-curve data.

\begin{figure}[t!]
\includegraphics[width=\columnwidth]{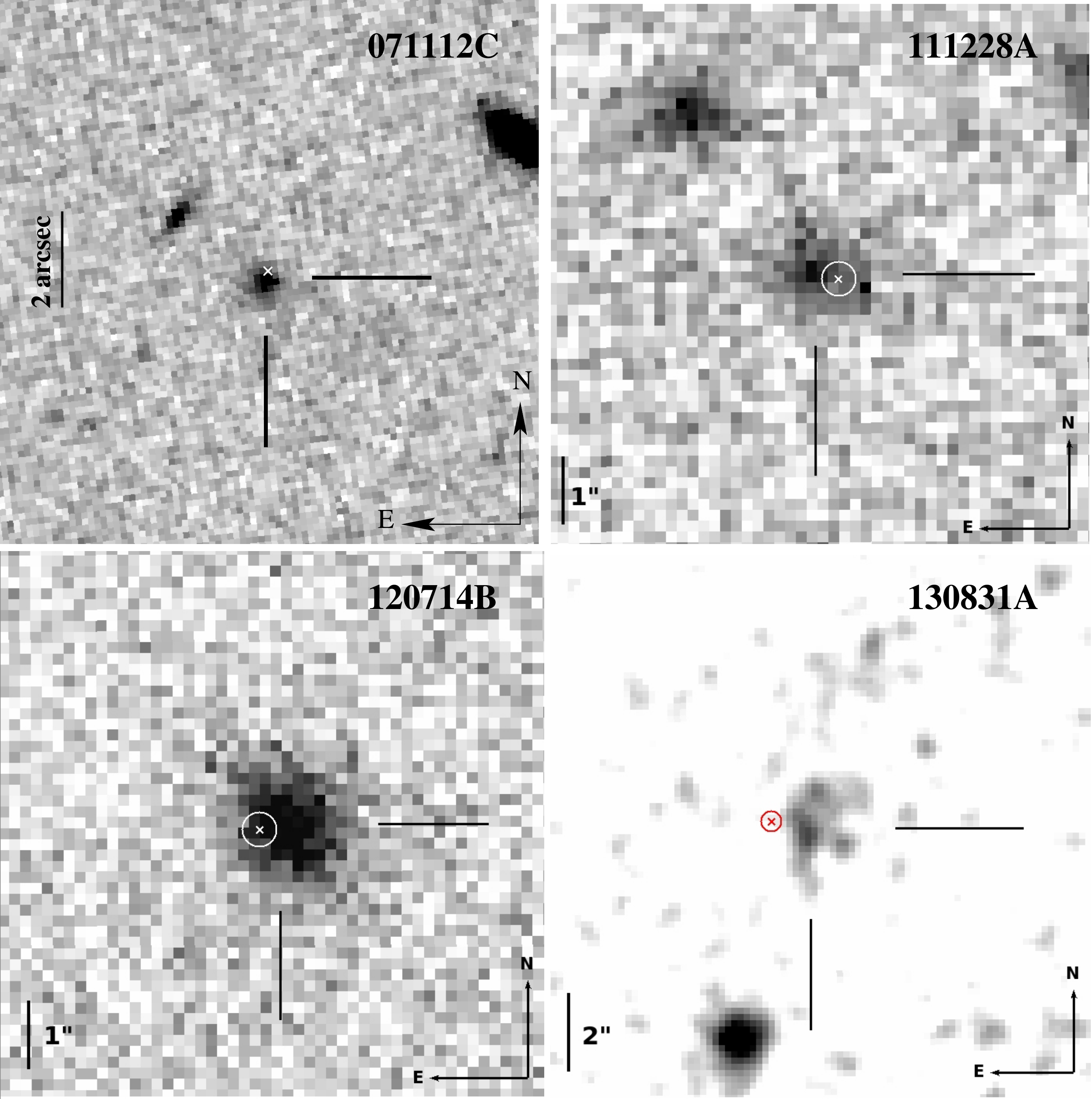}
\caption{The four GRB host galaxies. GRB
071112C: archived HST/WFC3 F160W image taken 3 years after the burst 
(HST proposal ID 12307, PI: A. J. Levan).
GRB 111228A: GROND $g^{\prime}r^{\prime}i^{\prime}z^{\prime}$ combined white-band.
GRB 120714B: GROND $g^{\prime}r^{\prime}i^{\prime}z^{\prime}$ combined white-band.
GRB 130831A: GROND $g^{\prime}r^{\prime}$ combined.}
\label{fig:hosts}
\end{figure}

Among our sample, GRB 120714B/SN~2012eb at
$z=0.40$ turned out to be of particular interest.  It belongs to the subclass
of intermediate-luminosity GRBs, has the least luminous optical afterglow
among all well-observed long-duration GRBs, and showed evidence for a
blackbody component in its optical transient at early times. In addition,
VLT/FORS2 spectroscopy revealed a photospheric expansion velocity of  $v_{\rm
ph}\sim$40\,000 km s$^{-1}$  close to SN maximum (based on the \ion{Si}{II}
feature). This is the highest expansion velocity found so far for a GRB-SN
close to maximum light. In contrast to this high value, for GRB
130831A/SN~2013fu   using the \ion{Fe}{II} $\lambda4570$ feature we measure
$v_{\rm ph}\sim$20\,000 km s$^{-1}$ near peak time, a value close to the lower
end of the observed distribution of $v_{\rm ph}$ for GRB-SNe measured via the
iron features.

The fact that GRB 120714B/SN~2012eb was an intermediate-luminosity burst
that showed a blackbody
component in its optical transient fits into the picture
(\citealt{Bromberg2011}) that there is a continuum between low- and
high-luminosity GRBs (\citealt{Schulze2014}; for a review see
also \citealt{Cano2016} and references therein). The case of SN~2012eb also
makes clear that a systematic search for such a blackbody component in GRB-SN
light curves in the optical bands (observer frame) might be a challenging
task. On the one hand it requires early-time data which reach into the UV
band in the GRB host frame, i.e., the higher the cosmological redshift the
better. On the other hand, it requires a bright thermal component and a
correspondingly faint afterglow and underlying host galaxy. 

We have shown that GRB-SNe at moderately high redshifts  can be well-studied
with 2-m class optical telescopes, good observing sites and instrumentation
provided. Based on ten years of GROND multi-color follow-up observing
campaigns at the MPG 2.2m telescope on ESO/La Silla, altogether nine GRB-SNe
have now been in detail investigated in the optical $g^\prime r^\prime
i^\prime z^\prime$ bands. Most of these were discovered with GROND. Even
though on average this is just one event per year, in total it represents 20\%
of the currently known GRB-SN sample.

\begin{acknowledgements}
S.K., Sebastian S., and A.N.G. acknowledge support by the Th\"uringer
Ministerium f\"ur Bildung, Wissenschaft und Kultur under FKZ 12010-514.  S.K.,
Sebastian S., A.N.G., and D.A.K. acknowledge support by grant DFG Kl
766/16-3. A.N.G., A.R., D.A.K., and A.U. are grateful for travel funding
support through MPE Garching.  A.R. acknowledges additional support by the
Jenaer Graduierten\-akademie, by TLS Tautenburg and DFG Kl/766
13-2. D.A.K. acknowledges support by TLS Tautenburg and MPE Garching, as well
as support from the Spanish research project AYA 2014-58381-P and Juan de la
Cierva Incorporaci\'on IJCI-2015-26153. 
P.S. and T.K. acknowledges support through the Sofja Kovalevskaja Award to
P.S. from the Alexander von Humboldt Foundation of Germany. 
R.F. acknowledges support from European Regional Development 
Fund-Project 'Engineering applications of
microworld physics' (No. CZ.02.1.01/0.0/0.0/16$_-$019/0000766).
Support for F.O. is provided by the Ministry of Economy, Development, 
and Tourism's Millennium
Science Initiative through grant IC120009, awarded to The 
Millennium Institute of Astrophysics, MAS. The Cosmic Dawn center
is funded by the DNRF. Part of
the funding for GROND (both hardware and personnel) was generously granted by
the Leibniz-Prize to G. Hasinger (DFG grant HA 1850/28-1). This work made use
of data supplied by the UK Swift science data center at the University of
Leicester. 
This work is based in part on observations made with the Spitzer Space
Telescope, which is operated by the Jet Propulsion Laboratory, California
Institute of Technology under a contract with NASA. Support for this work was
provided by NASA through an award issued by JPL/Caltech.
Some of the data presented herein were obtained at the W. M. Keck
Observatory, which is operated as a scientific partnership among the
California Institute of Technology, the University of California and the
National Aeronautics and Space Administration. The Observatory was made
possible by the generous financial support of the W. M. Keck Foundation. 
The authors wish to recognize and acknowledge the very significant cultural
role and reverence that the summit of Maunakea has always had within the
indigenous Hawaiian community.  We are most fortunate to have the opportunity
to conduct observations from this mountain. 
We thank the referee for a very constructive report. 
\end{acknowledgements}

\bibliographystyle{aa}
\bibliography{mypaper}

\begin{appendix}

\section{Details on the individual bursts \label{appgrbs}}
\subsection{GRB 071112C \label{app071112C}}

{\bf The burst and the afterglow data:}\ The burst was detected by 
the \swift~ Burst Alert Telescope (BAT; \citealt{Barthelmy2005SSRv..120})
on 2007 November 12 at 18:32:57 UT \citep{Perri2007a}. It consisted of a
single  peak (FRED-like: fast-rise, exponential decay) with a duration of
$T_{90} (15-350$ keV$)=15\pm2$~s (\citealt{KrimmGCN7081}). The optical
afterglow was discovered by \swift/UVOT (\citealt{Perri2007a}).  Based on a
spectrum of the afterglow with FORS2 a redshift of $z=0.823$ was found
\citep{Jakobsson2007a}. The properties of the burst and its early afterglow 
based on data available at the time are in detail summarized in \cite{Kann2010}.

The multi-channel imager
GROND was on target starting about 8~hr after the burst and continued visiting
the field for another seven epochs, spanning  a time span of altogether
28~days. The late-time GROND data (Table~\ref{071112Cmagdata}) reveal a
pronounced  SN bump in the $r^\prime i^\prime z^\prime $ bands with its peak
being about 2 weeks after the burst (Fig.~\ref{fig:SNe}). This is the most distant
GRB-SN in our sample.

At the time when the GRB went off,  the field of the burst was immediately
observable from Europe to East Asia, resulting in a well-covered early light
curve  of the optical transient.  Early  observations ($t>90$~s) are discussed
in \cite{Uehara2010a} and \cite{Huang2012a}. For the construction of the
light curve of the optical transient we took data from those two papers, as
well as \cite{Covino2013a}, \cite{Fynbo2009a}, the UVOT automatic analysis
page\footnote{\url{http://swift.gsfc.nasa.gov/uvot\_tdrss/296504/index.html}}, the
GCN Circulars (\citealt{Yuan2007a,Klotz2007a,Burenin2007a,Dintinjana2007a,
Oates2007a,Ishimura2007a,Greco2007a,Sposetti2007a,Yoshida2007a}, and
\citealt{Minezaki2007a}).  With the exception of a single measurement from
\cite{Covino2013a}, we are unaware of any other observations between one day
post-burst and late-time host-galaxy data.  Late host-galaxy data are taken
from \cite{Vergani2015a}. We note that there is a difference of an entire
magnitude between late-time host-galaxy magnitudes as given in
\cite{Huang2012a} and \cite{Vergani2015a}. Our GROND observations  yield an
intermediate value. We do not use the value from  \cite{Huang2012a}. We also
do not use late-time $i^\prime $-band data  from \cite{Covino2013a}, which is
significantly brighter than our  own data during the SN epoch as well as the
re-reduction in \cite{Vergani2015a}.

{\bf Afterglow properties:}\ In our analysis we included data from the
literature beginning 0.06 days after the GRB. The fit in $g^\prime
r^\prime i^\prime z^\prime $ then gives $\alpha=0.96\pm0.01$. When compared
with the results of the statistical study of afterglow light curves by
\cite{Zeh2006a} this suggests that for the considered time span  the fireball
was in the pre-break evolutionary phase. An identical fit result was reported
by \cite{Huang2012a}, who used optical data until about $t=1$ day to fit
their light curve. 

Assuming for the jet break time $t_b > 1$ day and using the observed isotropic
equivalent energy (Table~\ref{Tab:summary0}), this  constrains the jet
half-opening angle to  $\Theta_{\rm ISM} > 5.5 \pm 0.2$ deg for an ISM and
$\Theta_{\rm wind} > 5.6 \pm 0.4$ deg for a wind model, and  implies a
beaming-corrected energy release (in units of erg) of log $E_{\rm corr, ISM}$[erg]$ >
49.95\pm0.08$ and log $E_{\rm corr, wind} $[erg]$ > 49.97\pm0.06$, respectively.

Based on GROND $g^{\prime}r^{\prime}i^{\prime}z^{\prime}JHK_s$ data the SED of
the afterglow does not show evidence for host-galaxy extinction (Table~\ref{Tab:summary1},
Fig.~\ref{fig:071112C.AG2}). The observed decay slope ($\alpha= 0.96\pm0.01$)
and the obvious  lack of a break in the light curve until at least $2-3$ days after
the burst suggest that our $\alpha, \beta$ measurements refer to the
spherical expansion phase. The $\alpha-\beta$ relations
\citep{Zhang2004IJMPA} then show that during this phase the cooling frequency
was above the optical bands, $\nu_{\rm obs} < \nu_c$, in agreement with the
results reported by \cite{Huang2012a}. A wind model is preferred
compared to an ISM model, although the latter model is not ruled out
(Table~\ref{Tab:alphabeta}).

\begin{figure}[t!]
\includegraphics[width=\columnwidth]{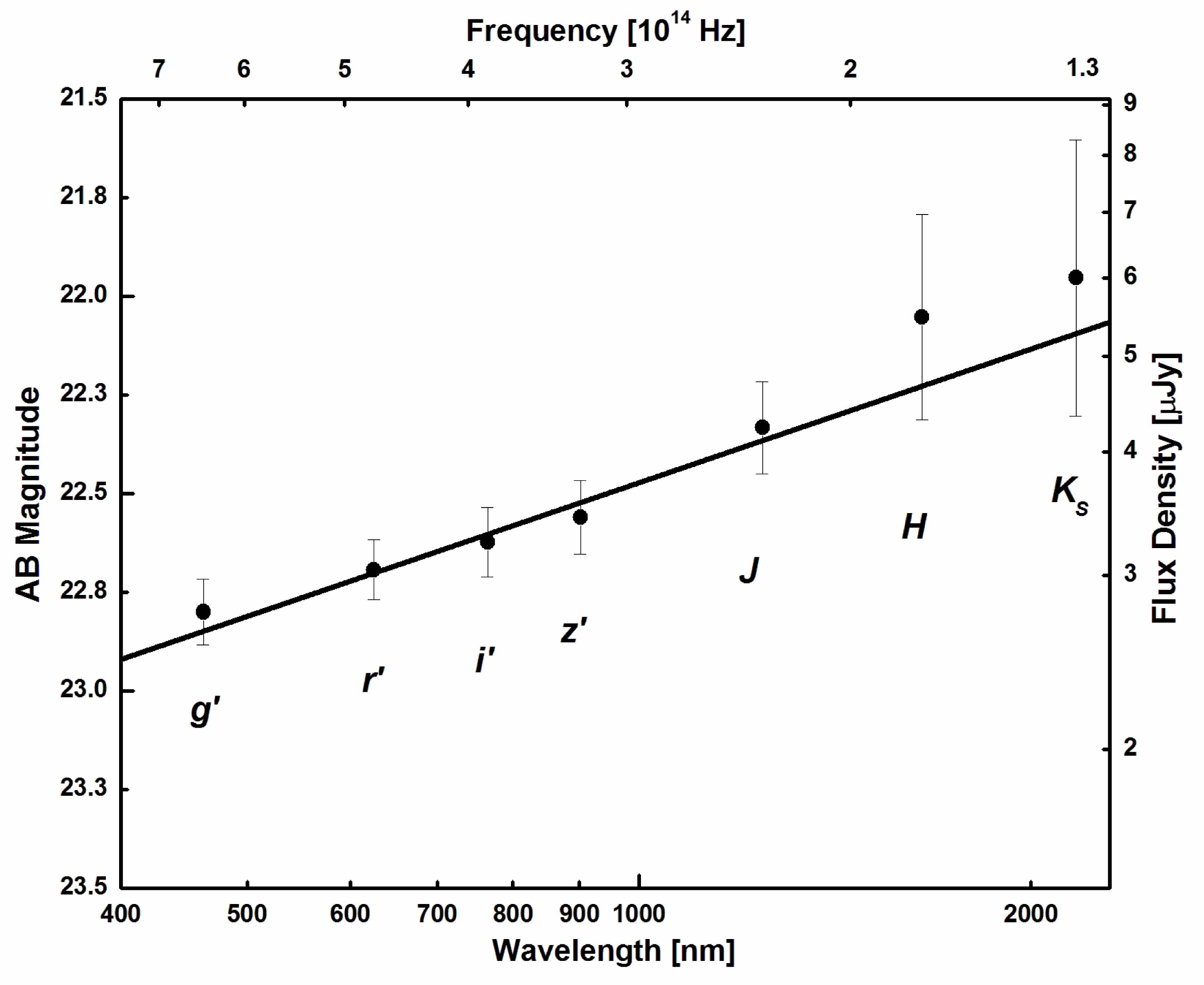}
\caption{SED of the afterglow of GRB 071112C based on GROND multi-color data
  and a joint fit of the $g^\prime r^\prime i^\prime z^\prime $-band light
  curves, assuming  a simple power-law decay. Magnitudes refer to $t=1$ day
  (corrected for Galactic extinction). The SED is well-described by a pure
  power-law, i.e., does not show evidence for host-galaxy extinction. }
\label{fig:071112C.AG2}
\end{figure}

{\bf Host galaxy properties:}\  The GRB host galaxy was observed with GROND
seven years after the burst. It turned out to be rather faint
(Table~\ref{Tab:summary1}). In order to measure the position  of the optical
transient with respect to its host galaxy, we made use of archival 
HST/Wide Field Camera 3 (WFC3)
F160W images taken 3 years after the burst (HST proposal ID 12307, PI:
A. J. Levan, see \citealt{Vergani2015a}). We astrometrized these images using the \texttt{Gaia} package and
compared them with the GROND images. We find that the optical transient was
placed $0\farcs26\pm0\farcs20$ ($2.1\pm1.6$ kpc) away from the brightness center
of its host, which on the HST image (centered at a wavelength of 8800~\AA\ in the rest frame)  does
not reveal any substructure (Fig.~\ref{fig:hosts}).  We note that on the
HST image a lenticular galaxy of roughly similar brightness and angular
extension lies only 2\farcs0 north-east of the GRB host, though no
redshift information is available for this galaxy so that a possible physical connection
between both galaxies cannot be claimed.

Using \texttt{Le PHARE} and fixing the redshift to  $z=0.823$,
 the SED of the host (Table~\ref{Tab:summary2};
including an HST/WFC3 F160W and a \emph{Spitzer}/IRAC1 detection) is best fit
by  the template of a galaxy 
with a modest SFR of $\sim1 M_\odot$
yr$^{-1}$ but a rather high specific SFR of log sSFR $\sim-8.6$
(Fig.~\ref{fig:071112C.2}; Table~\ref{Tab:LePhare1short}).  These results are
in rough agreement with those reported in \cite{Vergani2015a}.

\begin{figure}[t!]
\includegraphics[width=\columnwidth]{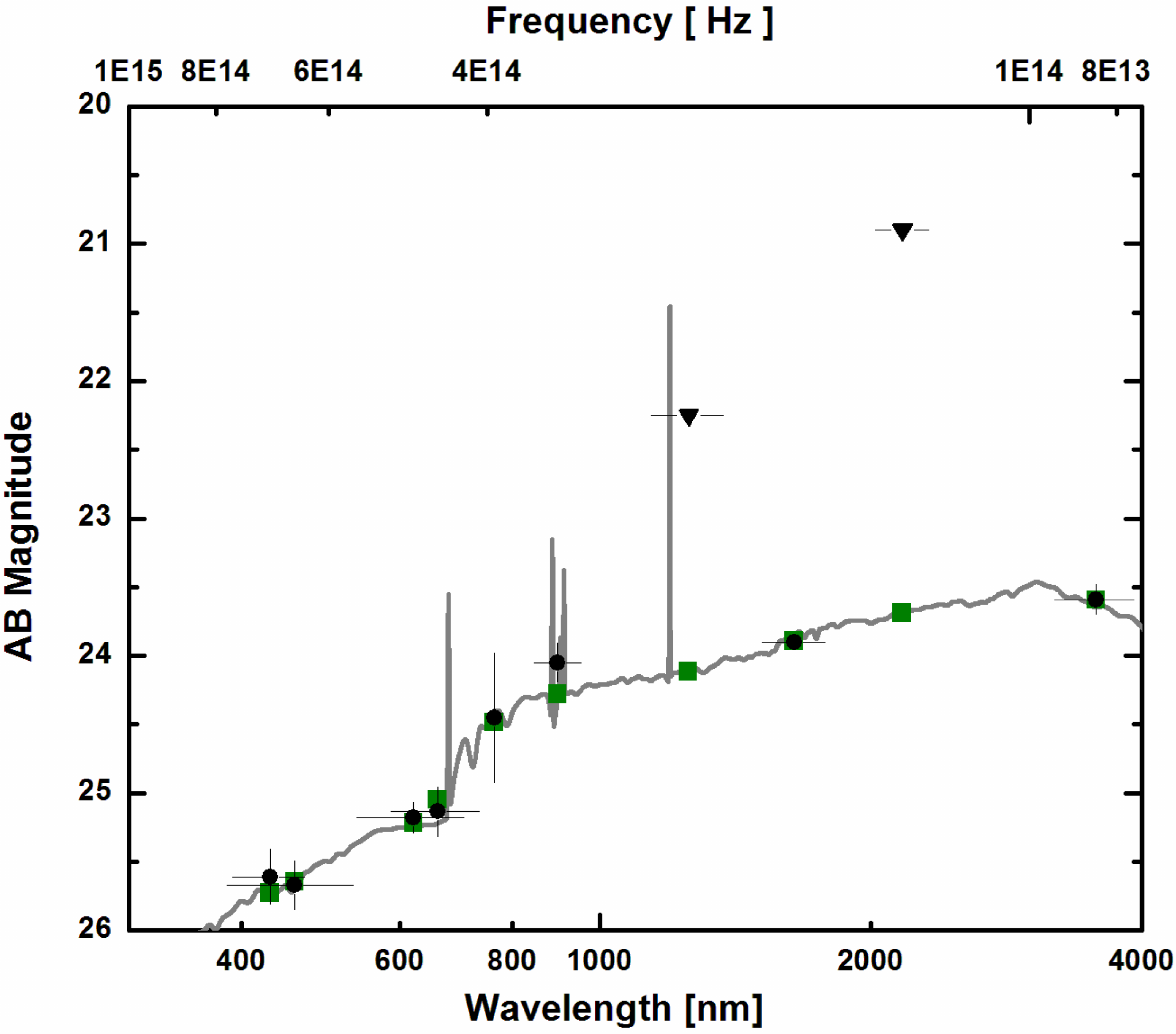}
\caption{SED of the host galaxy of GRB 071112C (corrected for Galactic
  extinction) based on GROND $g^\prime r^\prime i^\prime $-band data,  a GTC
  $z^\prime $, an HST/F160W ($H$ band), and a \emph{Spitzer}/IRAC1 
  detection as well as data taken with
  Keck/LRIS and Gemini-N/GMOS (Table~\ref{noGRONDmag}). In $JK_s$ we have only
  GROND upper limits (Table~\ref{071112Cmagdata}). The solid line displays the
  best template fit using \texttt{Le PHARE}, green data points are the
  template-predicted magnitudes.  }
\label{fig:071112C.2}
\end{figure}

{\bf Details on the SN fitting:}\  When performing the combined afterglow plus
SN fit (Eq.~\ref{ot}) in the $r^\prime $ band, we had to exclude data from 1
to 7 days post-burst. Including it yields an unrealistic SN component  with
$k\sim3.5$ and $s\sim0.3$, in strong contrast to the values derived in
$i^\prime $ and $z^\prime $ (and any known GRB-SN). Without these data points,
a result in accordance with those in the other photometric bands is derived
(Table~\ref{Tab:summary2}): The peak luminosity of the SN was $\sim0.7$ times
the luminosity of SN~1998bw, though our measurement error is in the order of
30\%. We find that
the SN developed notably faster than SN~1998bw (stretch factor
$s\sim0.75)$.  Our results indicate that the GRB-SN was similar in terms of
evolution and luminosity as GRB/XRF 060218/SN~2006aj and GRB/XRF
100316D/SN~2010bh, i.e., fainter and faster than the template SN~1998bw. 

\subsection{GRB 111228A} \label{App.Sect.111228A}

{\bf The burst and the afterglow data:}\ The burst triggered \swift/BAT on
2011 December 28 at 15:44:43 UT \citep{Ukwatta2011a}. It consisted of multiple
spikes and had a duration of $T_{90} (15-350$ keV$)=101.20\pm5.42$~s
(\citealt{Cummings12749}). It was also detected by the
\emph{Fermi}/Gamma-ray Burst Monitor (GBM;
\citealt{Meegan2009ApJ...702}) with a duration
$T_{90} (50-300$ keV)  of about 100~s (\citealt{Briggs12744}). Its optical
afterglow was bright enough to be found by \swift/UVOT
(\citealt{Ukwatta2011a}). Its spectroscopic redshift is $z=0.716$  (published
values vary between 0.713 and 0.716;
\citealt{Cucchiara2011a,Dittmann2011a,Palazzi2011GCN12765,Schulze2011a,
  Xu2011a}), with the most accurate one coming from X-shooter observations
\citep{Schulze2011a}.  The burst was also detected by \konus{ }\citep{Golenetskii2011a}.  A possible
detection of an SN signal in the light curve of the optical transient
34.5~days after the burst based on TNG imaging data  was reported by
\citet{DAvanzo2012a}. The early optical and X-ray afterglow is studied by
\cite{Xin2016ApJ} in detail. The authors conclude that the observed early plateau phase
of the afterglow light curve suggests  an energy injection period provided by
a freshly formed magnetar. Since their optical data do not go beyond
$\sim0.3$~Ms post-burst, no SN is discussed.
 
Using GROND, 
we started observations on December 29 at 04:53 UT, 13 hr after the trigger
\citep{Nic2011GCN12757}; additional data were obtained over further 11 epochs,
resulting in a good sampling of the optical light curve between about
0.5 and 90 days post burst.  This was finalized by a deep host-galaxy
observation 4 years later (Table~\ref{111228Amagdata}).

 We furthermore analyzed the UVOT data for this GRB, stretching from 0.00162
days (140 s) to 33.7 days after the trigger. The afterglow is detected until
5.6 days after the trigger, but neither the SN nor the host galaxy is detected
in any filter, even in deep stacks. The UVOT light curves are shown in
Fig.~\ref{fig:111228AUVOT} and the data is given in Table~\ref{111228AUVOTdata}.

\begin{figure}[t!]
\includegraphics[width=\columnwidth]{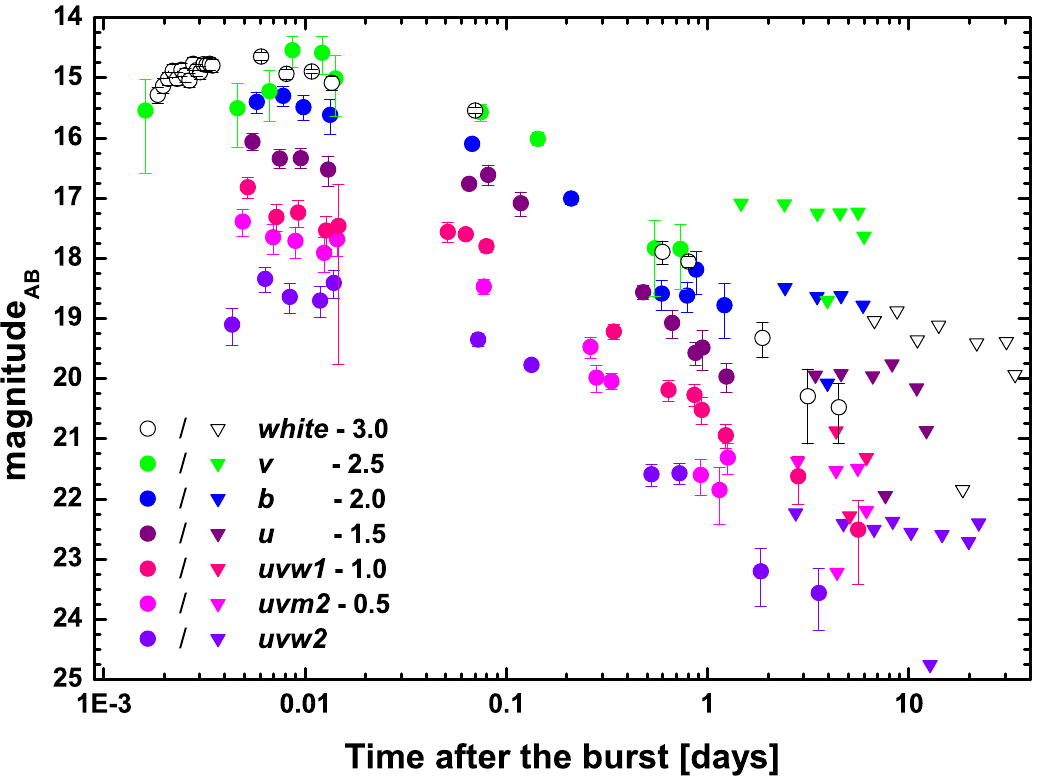}
\caption{UVOT observations of the afterglow of GRB 111228A. Circles with errors
are detections and upper limits are downward-pointed triangles. The different
bands are offset by steps of 0.5 magnitudes for clarity. Observations yielding
upper limits only have also been stacked to produce deeper upper limits, but
neither the SN nor the host galaxy are detected at late times.}
\label{fig:111228AUVOT}
\end{figure}

{\bf Afterglow properties:}\  Applying Eq.~(\ref{ot}), and using GROND data
only, for a joint fit in $g^\prime r^\prime i^\prime z^\prime JHK_s$ and using
a broken power-law, we find $\alpha_1=1.18\pm0.03$, $\alpha_2=1.79\pm0.05$,
$t_b=1.68\pm0.06$ days. We note that such a break is potentially also apparent
in the X-ray light curve (see the \swift/X-ray Telescope
(XRT) repository, \citealt{Evans2007a}; \citealt{Xin2016ApJ});
however, it would require to skip the very last X-ray data point.
The value we found for $\alpha_1$ agrees with the decay
slope reported by \cite{Xin2016ApJ}, $\alpha=1.12\pm0.10$, though their
results are based on a much less well-sampled data base at later times and,
therefore, these authors did not find evidence for a break in their optical
light curve.

The deduced light-curve parameters do substantially change if  early data
($t=0.02-0.28$ days)  from other telescopes (UVOT data and those of \citealt{Xin2016ApJ}) are taken
into account. In this case the light curve can still be fitted with a single
broken power law, but the pre-break decay slope flattens to $\alpha_1 =
0.06\pm0.04$.  Such a flat decay suggests an early plateau phase which
obviously smoothly developed into a normal afterglow decay phase. When this
transition happened is not apparent in the optical/NIR data.    Compared to
the GROND-only data the break time is at $t_b=0.41\pm0.07$ days and the
post-break decay parameter steepens slightly to  $\alpha_2=2.00\pm0.06$,
formally a more comfortable solution as it  does not exclude anymore the case
$p>2$. 

When only early GROND $g^{\prime}r^{\prime}i^{\prime}z^{\prime}JHK_s$ data are taken into account, the  SED of
the afterglow can be described by a power-law without evidence for host
extinction ($A_V^{\rm host}=0$ mag).  The joint fit of the GROND multi-color
data provides a spectral slope of $\beta= 0.88\pm0.03$, compared to
$\beta\sim0.75$ as it was deduced by \cite{Xin2016ApJ}. Taking into account
\emph{Swift}/UVOT data, such a fit however shows a deficit of flux in
the UV domain, which can be attributed to 0.16 mag of host-galaxy visual extinction
(adopting an SMC extinction law). This changes the spectral slope of the
afterglow to  $\beta= 0.69\pm0.07$ (Fig.~\ref{fig:111228A.AG2}), in agreement
with \cite{Xin2016ApJ}. 

\begin{figure}[t!]
\includegraphics[width=\columnwidth]{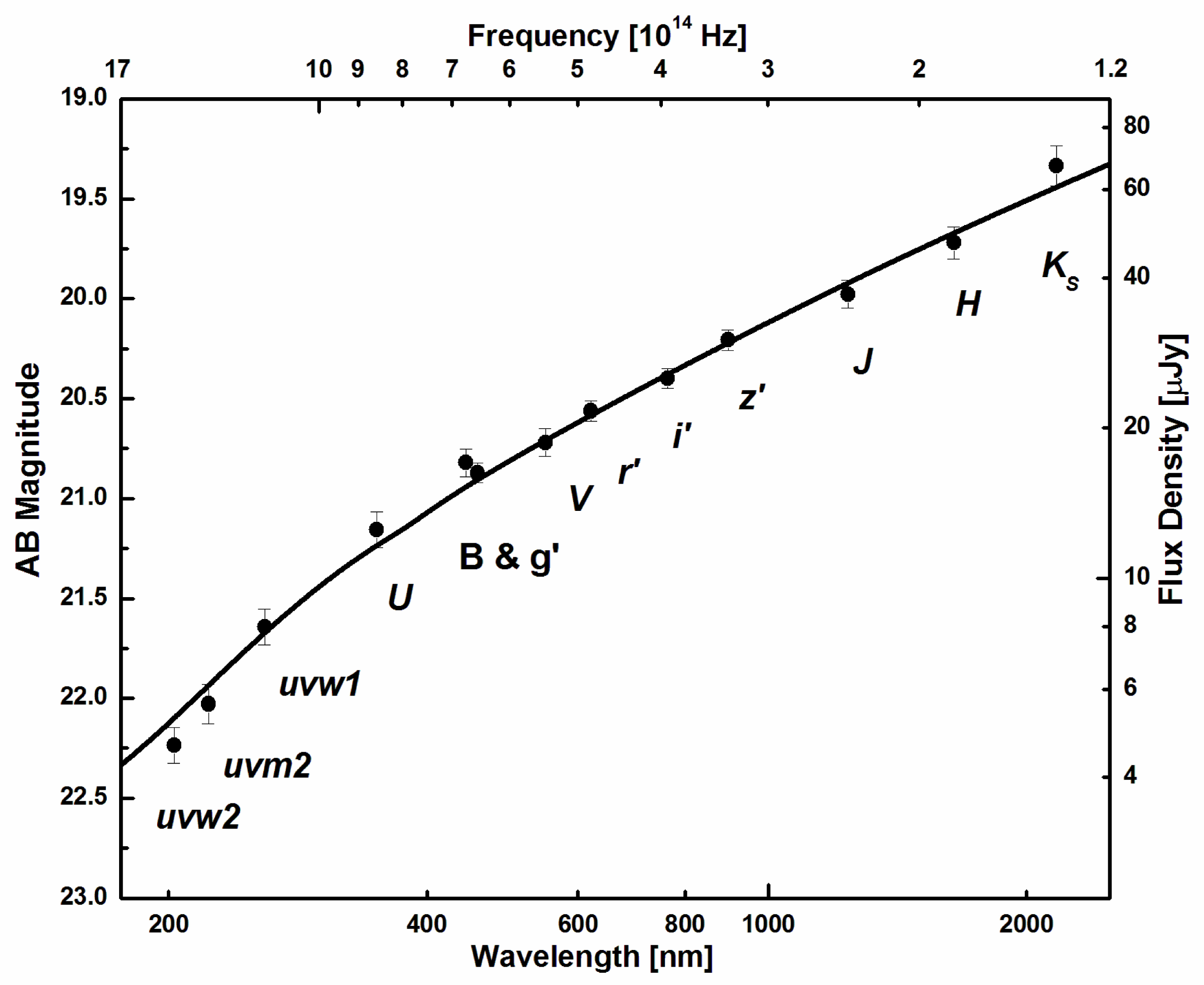}
\caption{SED of the afterglow of GRB 111228A based on GROND as well as
  \swift/UVOT data (Tables \ref{111228Amagdata},\ref{111228AUVOTdata}). Magnitudes refer to the time of the light
  curve break at $t=t_b=1$ day, i.e., are free of any contribution from an
  underlying SN component. Data are corrected for Galactic extinction. The SED
  suggests a host extinction of $A_V^{\rm host}=0.16$ mag.} 
\label{fig:111228A.AG2}
\end{figure}

Neither the GROND-only light-curve parameters (which imply $p<2$) nor the
light-curve parameters based on the extended data set  (which suggest $p>2$)
lead to a reasonable solution for the $\alpha-\beta$ relations
(Table~\ref{Tab:alphabeta}). Interpreting the (GROND-only) light curve break
at $t=1.68$ days as a jet break, and using the observed isotropic equivalent
energy (Table~\ref{Tab:summary0}), this implies a jet half-opening angle of
$\Theta_{\rm ISM} = 6.3 \pm 0.1$ deg for an ISM and $\Theta_{\rm wind} = 5.4
\pm 0.1$ deg for a wind model, and a beaming-corrected energy release (erg) of
log $E_{\rm corr, ISM} $[erg]$ = 50.39\pm0.04$ and log $E_{\rm corr, wind}  $[erg]$=
50.26\pm0.03$, respectively. 

{\bf Host galaxy properties:}\  We detect a galaxy at the location of the GRB in
$g^\prime r^\prime i^\prime z^\prime $ in our last-epoch GROND images taken
1499 days post burst (Fig.~\ref{fig:111228A.2}, Table~\ref{Tab:summary2}). On
a GROND white-band image the SN is located $0\farcs28\pm0\farcs20$ east of the
brightness center  of its host ($2.1\pm1.5$ kpc).  Using \texttt{Le PHARE}
and fixing the redshift to $z=0.716$, the SED of the host
(Table~\ref{Tab:summary2}) suggests a dusty galaxy ($E(B-V)=0.4$ mag).
Its  SFR, mass,
and specific SFR are similar to those of the host of GRB 071112C
(Fig.~\ref{fig:111228A.2}; Table~\ref{Tab:LePhare1short}). 

\begin{figure}[t!]
\includegraphics[width=\columnwidth]{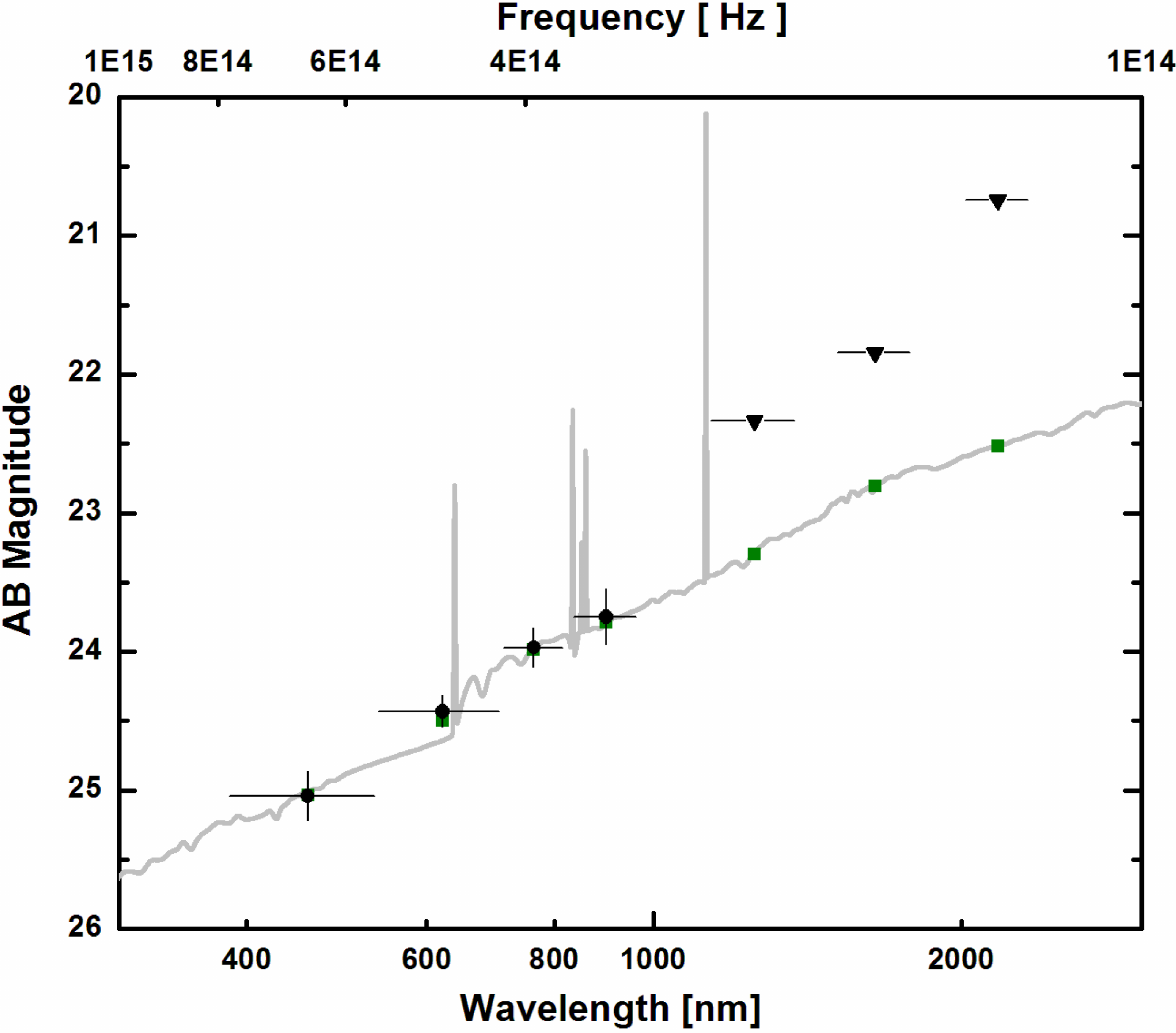}
\caption{SED of the host galaxy of GRB 111228A based on GROND data taken four
  years after the burst and the corresponding best galaxy-template SED found
  by \texttt{Le PHARE} (for the color coding see
  Fig.~\ref{fig:071112C.2}). Due to the lack of a NIR detection, the fit is
  rather ill-defined.}
\label{fig:111228A.2}
\end{figure}

{\bf Details on the SN fitting:}\  The GROND $r^\prime i^\prime$ light curves show a
slight but clearly visible bump around $t=20$ days (Fig.~\ref{fig:SNe}). It is
also marginally seen in $z^\prime $ but not apparent in the $g^\prime$ band.
When performing the joint fit, we therefore fixed $k_{g^\prime }=0$, i.e., we
assumed that there was no contribution from SN light in this photometric  band
(centered at a wavelength of 2650~\AA\ in the rest frame).  Leaving this $k$-value as a free
parameter did otherwise worsen the fit and finally, within the errors,
included a solution with $k=0$.  In doing so, for the SN parameters we
obtained $k$ values rather typical for most GRB-SNe. The stretch factor
however is rather large, $s\sim1.2-1.6$, implying that this SN evolved
slower than SN~1998bw (Table~\ref{Tab:summary2}). We note, however, that  due
to the rather weak SN bump the deduced SN peak time is rather  sensitive to
the light curve fit at early times. The error for $s$ is, as always, the pure
mathematical error of the fit but does not take into account this
problem. Therefore, the rather large stretch factor deduced for this event
should be taken with care.  

\subsection{GRB 120714B} \label{Appendix:120714B}
 
{\bf The burst and the afterglow data:}\ The burst was detected by \swift/BAT
on 2012 July 14 at 21:18:47 UT \citep{Saxton2012a}. It is dominated by a broad
peak with a duration $T_{90}$ (15-350 keV$)=159\pm34$~s
\citep{Cummings2012a}. The optical afterglow was discovered by \swift/UVOT
(\citealt{Marshall2012a}). Its spectroscopic redshift was soon found to be
$z=0.3984$ \citep{Fynbo2012a}, suggesting that an upcoming SN component might
be detectable. Given the redshift, the observed relatively low mean luminosity
of the burst (Table~\ref{Tab:summary1}) defines it as a member of the class of
intermediate-luminosity GRBs (\citealt{Schulze2014}). 

Using GROND, we started observing about 6 hr after the burst
(\citealt{Nic2012GCN13478}), though weather conditions were inclement at
that time (airmass $>2.2$, mean seeing 2\farcs4).  Because of the
very-well-detectable SN component in the light curve of the optical transient
(\citealt{Klose2012GCN13613,Klose2012a}), GROND observed the field for
altogether 17 epochs up to 449 days post burst  (Table~\ref{120714Bmagdata}).

{\bf Afterglow properties:}\  Since no SN is apparent in the $g^\prime$ band,
we used this band to measure the decay slope of the optical transient
(Table~\ref{Tab:summary1}).  Moreover, since no evidence for a break is
detectable in the $g^\prime$-band light curve, we adopted a single power-law
decay and find $\alpha=0.58\pm0.06$. Only  a lower limit on the time of a
potential jet break can be set: of $t_b\gtrsim10$ days.  No \emph{Swift}/X-ray
light-curve data are available to support this conclusion further.

Adopting a break time of $t_b>10$ day, and using the observed isotropic
equivalent energy (Table~\ref{Tab:summary0}), this gives for the jet
half-opening angle $\Theta_{\rm ISM}>9.5\pm0.3$ deg for an ISM and
$\Theta_{\rm wind}>14.5\pm1.0$ deg for a wind model. The corresponding
upper limits for the beaming-corrected energy release (erg) are log $E_{\rm
corr,ISM} $[erg]$>48.90\pm0.09$ and log $E_{\rm corr, wind} $[erg]$>49.26\pm0.06$,
respectively. 

As we have discussed in Sect.~\ref{Shocking}, our X-shooter spectroscopy
revealed that the early  first-epoch GROND data are affected by an additional
radiation component. Taking this into account provides for the afterglow
light a spectral slope of $\beta=0.7\pm0.4$. Better constraints on
$\beta$ can not be obtained, already at day two the rapid rise of the SN
component and the comparably rather bright host galaxy prevent a reliable
measurement of the afterglow SED. Using the $\alpha-\beta$ relations no
model is really ruled out (Table~\ref{Tab:alphabeta}).

{\bf Host galaxy properties:}\ In order to detect the GRB host galaxy, GROND
observed the field again 449 days post-burst. In addition, we obtained
HAWKI  $J$ and $K_s$-band images on Oct 13, 2013, one year after the SN
maximum (Program ID: 092.A-0231, PI: T. Kr\"uhler).

Stacking the GROND $g^\prime r^\prime i^\prime z^\prime $-band images into a
white-band frame reveals that the SN was located $0\farcs21\pm0\farcs20$
($1.2\pm1.1$ kpc) east of the center of its host (Fig.~\ref{fig:hosts}). 
The SED of the host is well-determined, with detections in $g^\prime r^\prime
i^\prime z^\prime$ by GROND and $JK_s$ with HAWKI. Using
\texttt{Le PHARE}, and fixing the redshift to $z=0.3984$, the
best template to describe the SED (Table~\ref{Tab:summary2}) is a dusty
($E(B-V)\sim0.3$ mag) galaxy 
with a rather modest SFR of about
$1~M_\odot$ yr$^{-1}$. Its mass in stars and specific SFR resemble the
properties of the host of GRB 071112C (Fig.~\ref{fig:120714B.2}; 
Table~\ref{Tab:LePhare1short}).

\begin{figure}[t!]
\includegraphics[width=\columnwidth]{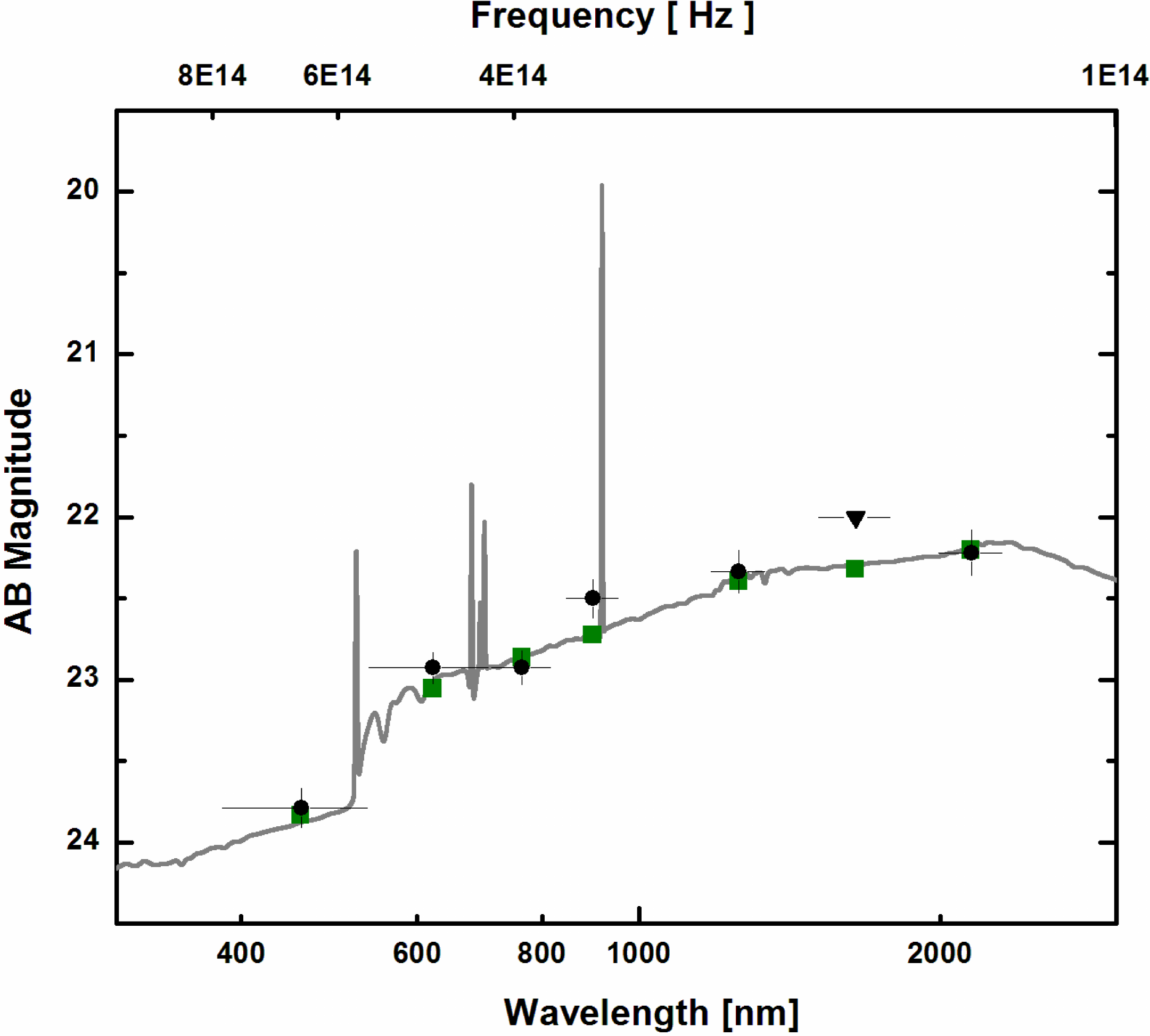}
\caption{SED of the host galaxy of GRB 120714B based on GROND $g^\prime r^\prime
  i^\prime z^\prime H$ and HAWKI $JK_s$-band data. The SED is best described
  by a galaxy undergoing an episode of intense star-forming activity (for the
  color coding see Fig.~\ref{fig:071112C.2}).}
\label{fig:120714B.2}
\end{figure}

{\bf Details on the SN fitting:}\  In the $r^\prime i^\prime z^\prime$-band
light curves a SN bump is clearly seen, while no SN signal is evident in the
$g^\prime$-band data (centered at a wavelength of 3250~\AA\ in the rest frame;
Fig.~\ref{fig:SNe}). This lack of SN flux at shorter wavelengths is also
evident in our SN spectrum  (Sect.~\ref{SN.Spectr}). Such a strong damping in
the host-UV band is reminiscent of the otherwise more luminous SN~2009nz
associated with GRB 091127 \citep{Olivares2015a,Kann2016}.  When performing the
joint fit in order to obtain the SN $(k,s)$ values, for the $g^\prime$ band
we fixed $k=0$.  In doing so, we find that in $r^\prime i^\prime z^\prime$
(corresponding to wavelengths of about 4480 to 6390~\AA\ in the GRB rest frame) the SN
reached between about 50 to 80\%  of the peak luminosity of SN~1998bw, while
in $i^\prime$ and $z^\prime$ it evolved about 10\% faster
(Table~\ref{Tab:summary2}).  Compared to GRB-SNe that have been analyzed with
the ($k,s$) methodology
(\citealt{Zeh2004a,Ferrero2006a,Thoene2011a,Cano2013a}), SN~2012eb is among
the faintest GRB-SNe detected so far. Especially, it is markedly fainter than
GRB 120422A/SN~2012bz which is, analogous to GRB 120714B, another
intermediate-luminosity GRB (\citealt{Schulze2014}, but see \citealt{Cano2016}). 

\subsection{GRB 130831A \label{Sect:130831A}}

{\bf The burst and the afterglow data:}\ GRB 130831A was detected by
\swift/BAT on 2013 August 31 at 13:04:16 UT \citet{Hagen2013a}. It is
dominated by a FRED-like profile followed by additional emission. Its duration
was $T_{90}$ (15-350 keV$)=32.5\pm2.5$~s (\citealt{Barthelmy15155}). The
optical afterglow was discovered by \swift/UVOT \citep{Hagen2013a}.  A
redshift $z=0.4791$ was soon reported by \cite{Cucchiara2013a} based on
observations with Gemini-North. The GRB was also detected by \konus\ at
13:04:22 UT and an isotropic energy release of $E_{\rm iso}=(4.6\pm0.2)
\,\times\,10^{51}$ erg was determined \citep{Golenetskii2013a}. 

First follow-up observations with GROND started at 03:33 UT (14.5 hrs after
the burst) and continued for 3.3 hr. Further data were obtained until 33.6
days after the burst, and a final host-galaxy observation was obtained 387
days after the GRB (in total 12 epochs, Table~\ref{130831Amagdata}). Observations were constrained by a
relatively narrow visibility window and affected by permanently high airmass
(1.9 at best). No GROND data could be obtained around the peak time of the
GRB-SN.

{\bf Afterglow properties:}\ The early optical
afterglow light curve (at $t\lesssim5$~ks) is complex and variable, and is
discussed in detail in \cite{DePasquale2015a}. We did not use this data for
the calculation of the light curve parameters.  Since there is no evidence for
a break in the optical light curve,  we adopted a single-power law decay,
which provides $\alpha=1.61\pm0.01$ and a spectral slope $\beta=1.00\pm0.05$
(Fig.~\ref{fig:SNe}), suggesting a post-break evolutionary phase. When
compared with the afterglow sample of \cite{Zeh2006a} its interpretation as a
pre-break decay slope is however not excluded. For example, the afterglow of
GRB 000926 had $\alpha_1=1.74\pm0.03$, $\alpha_2=2.45\pm0.05$.  

In order to characterize the underlying  SN component, we combined our data
with that of \cite{Cano2014a}, \cite{DePasquale2015a}, and
\cite{Khorunzhev2013a}. To avoid the early complex light-curve evolution, we
only used data from 0.39 days onward. We fitted the 
$g^{\prime} r^{\prime} R_C i^{\prime} I_C z^{\prime}$-band
data using a simple-power law, sharing the decay slope (we find no  evidence
for any chromatic evolution). The SN parameters $k$ and $s$ were left as free
parameters, individual in each band, as is the host galaxy magnitude. We also
tried a fit with the afterglow described by a broken power-law, but found that
the break time and post-break decay slope are unconstrained, while at the same
time the $k,s$ parameters hardly changed. Therefore, there is no evidence in
the afterglow of a break between 0.4 and $\sim3$ days, in  agreement with the
lack of a break in the late X-ray light curve (\citealt{DePasquale2015a}).

The SED of the optical/NIR afterglow is well-fit by a power-law
(Fig.~\ref{fig:130831A.AG2}). There is no evidence for host-galaxy extinction
along the line of sight, in agreement with \cite{Cano2014a} and
\cite{DePasquale2015a}. The $\alpha-\beta$ relations suggest that the deduced
afterglow parameters rule out a jet model with $\nu_{\rm obs}< \nu_c$ but
exclude neither a jet model with $\nu_{\rm obs}> \nu_c$ nor a spherical
wind/ISM model (Table~\ref{Tab:alphabeta}). If the jet-break time was at
$t_b>1$ day then the observed isotropic-equivalent energy release
(Table~\ref{Tab:summary0}), implies a jet half-opening angle of $\Theta_{\rm
  ISM}>6.8\pm0.1$ deg and a beaming-corrected energy (erg) of log $E_{\rm
  corr, ISM} $[erg]$>49.70\pm0.03$.  For a wind model these numbers are
$\Theta_{\rm wind} $[erg]$>7.6\pm0.2$ deg and log $E_{\rm corr, wind}>49.80\pm0.02$. 

Nevertheless, a very early jet break cannot be ruled out. A potential
example for such a case is GRB 061007, which was exceptionally bright
across the electromagnetic spectrum and whose afterglow showed a 
remarkably similar
temporal decay slope of $\alpha=1.65 \pm 0.01$ from early-on
(\citealt{Schady2007MNRAS.380.1041S}). 
It has been argued by \cite{Schady2007MNRAS.380.1041S}
that here the scenario
of a very narrow jet ($\Theta<0.8$ deg) can fit the observational data 
well.

\begin{figure}[t!]
\includegraphics[width=\columnwidth]{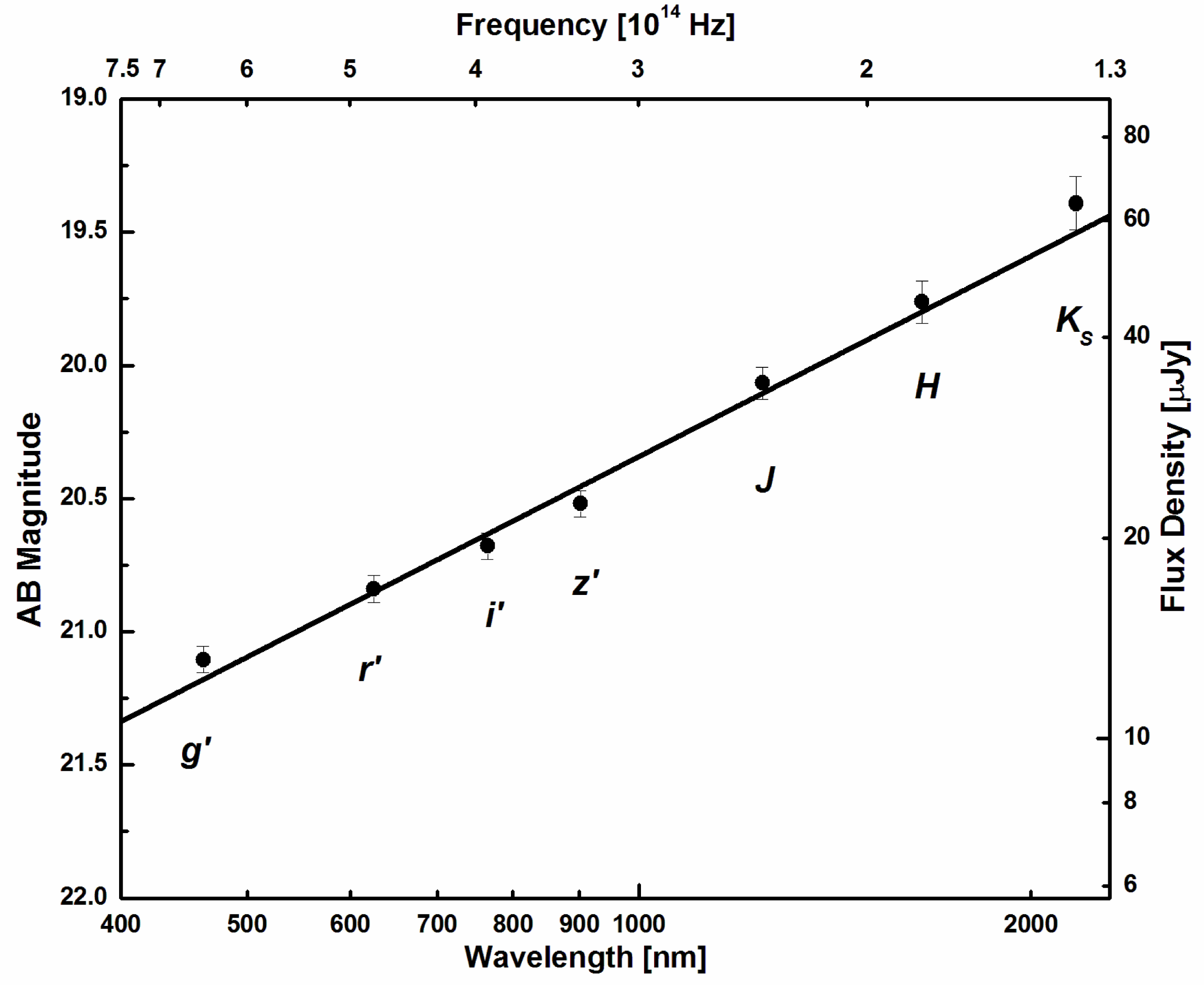}
\caption{GROND ($g^{\prime}r^{\prime}i^{\prime}z^{\prime}JHK_s$) SED of the
  afterglow of GRB 130831A. Magnitudes refer to $t=1$ day (corrected for
  Galactic extinction). There is no evidence for host-galaxy extinction.}
\label{fig:130831A.AG2}
\end{figure}

We finally note that \cite{DePasquale2015a} and
\cite{Zhang2016ApJ...823..156Z} proposed that the 
early X-ray and optical data of the afterglow suggest a continuous 
energy input from a magnetar. Our data cannot be used to evaluate this model
further.

{\bf Host galaxy properties:}\  We used the late-epoch GROND $g^\prime
r^\prime z^\prime $-band detections of the host galaxy one year after the
burst together with the results of our joint light-curve fit for the $i^\prime
$ band (Fig.~\ref{fig:130831A.2}), as well as a $J$-band detection by
Keck/MOSFIRE 
and a deep $3.6\mu$m upper limit from \emph{Spitzer}/IRAC (Table~\ref{noGRONDmag}) to construct the SED
(Table~\ref{Tab:summary2}). Applying \texttt{Le PHARE}, and fixing the
redshift to $z=0.4791$, the best template is a
dusty  galaxy ($E(B-V)=0.2$ mag) 
with a rather low SFR on the order of
0.3~M$_\odot$ yr$^{-1}$. Its mass and specific SFR are, within 
errors, rather normal for long-GRB hosts (Table~\ref{Tab:LePhare1short}).

\begin{figure}[t!]
\includegraphics[width=\columnwidth]{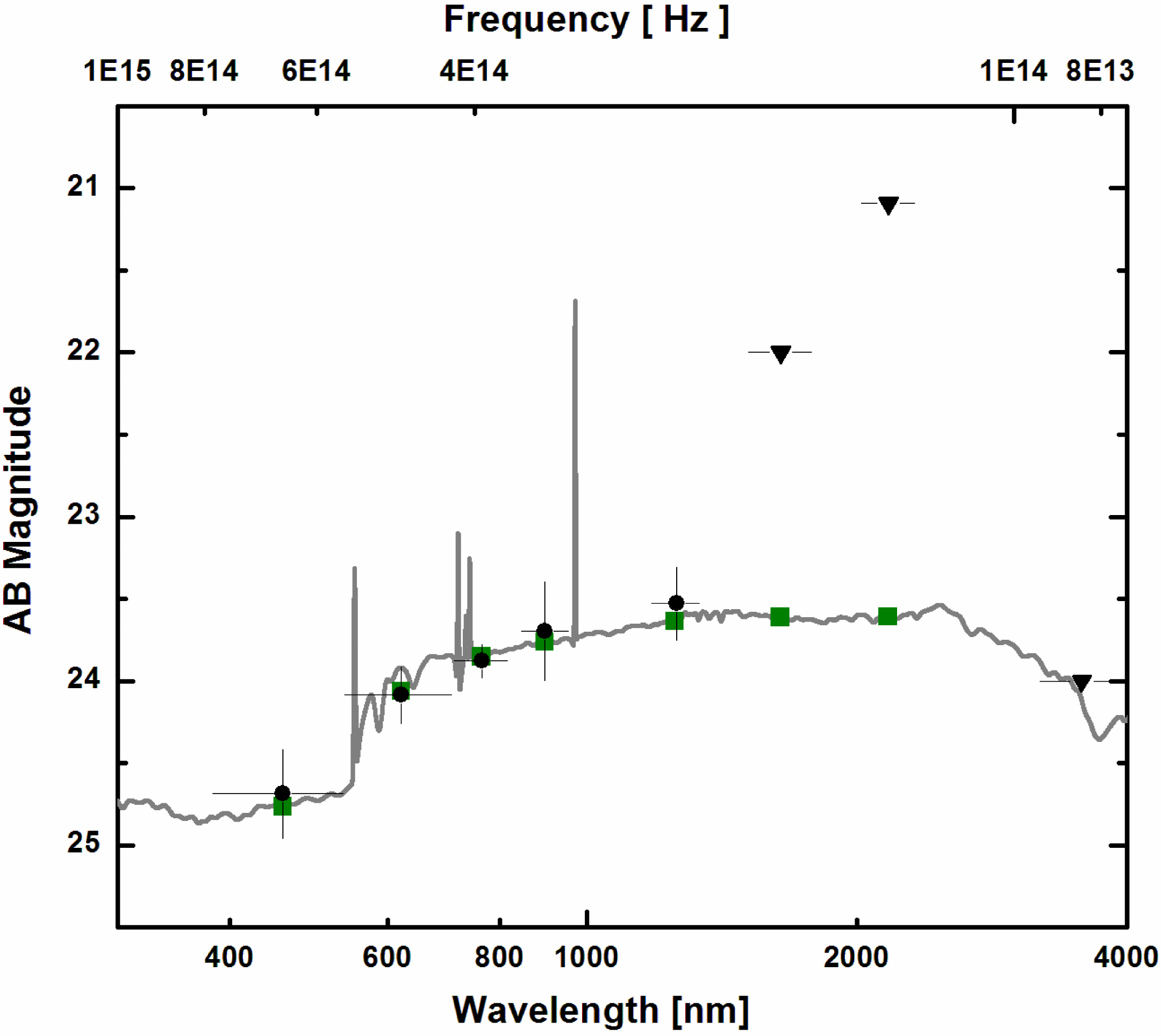}
\caption{SED of the host galaxy of GRB 130831A (the color coding follows
Fig.~\ref{fig:071112C.2}).} 
\label{fig:130831A.2}
\end{figure}

Using the GROND $g^\prime r^\prime$ combined image, we find that the SN was located
$0\farcs92\pm0\farcs20$ (projected distance $5.3\pm1.2$ kpc) east of the center
of its host galaxy (Fig.~\ref{fig:hosts}). 

{\bf Details on the SN fitting:}\  Due to some scatter in the data, the joint
fit is formally bad (Table~\ref{Tab:summary2}), but the SN is well-sampled in
all GROND optical bands. The steep afterglow decay slope
($\alpha=1.61\pm0.01$) is in agreement with the one derived by
\cite{Cano2014a} and \cite{DePasquale2015a}.  As we do not find any evidence
for rest-frame extinction, the $k$ values derived directly from  the fit for
the individual bands are not further corrected for host-galaxy extinction
(Table~\ref{Tab:summary2}). In general, we find that SN~2013fu evolves faster
than SN~1998bw ($s\sim0.6$ to 0.8) and is somewhat less luminous ($k\sim0.6$
to 1.0), though it shows signs of being photometrically different. It was
significantly fainter than SN~1998bw in $r^\prime$ and $z^\prime$ while of similar
luminosity in $g^\prime$ and $i^\prime$. These values are in agreement with those
derived by \cite{Cano2014a}, with  the exception of the $z^\prime$ band,
where our data point to a significantly fainter SN. \cite{Cano2014a} have
only a single data point during the SN phase in $z^\prime$, and this value is
not host-subtracted either, we therefore believe our derived value to be more trustworthy.

\begin{figure}[t!]
\includegraphics[width=\columnwidth]{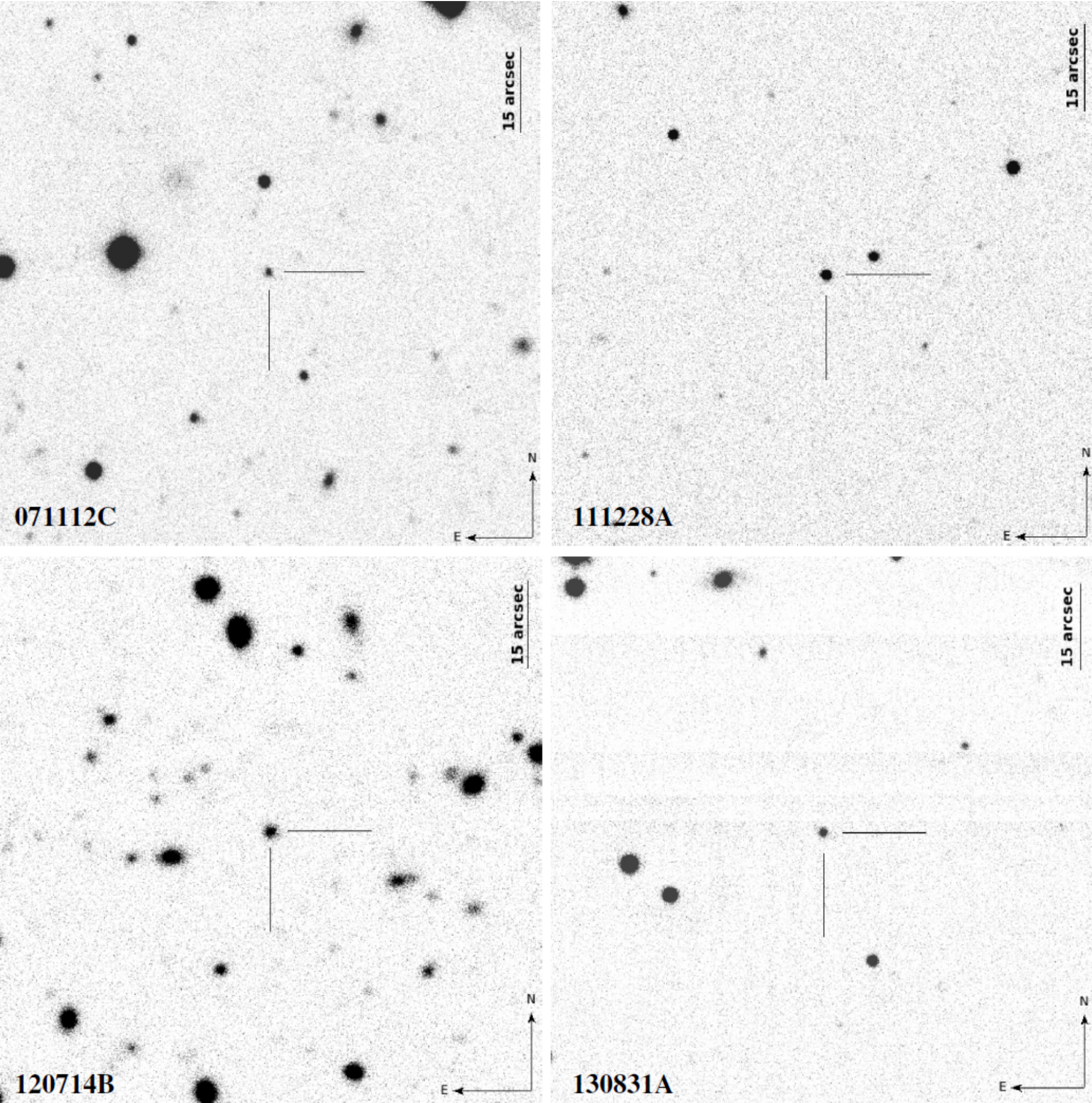}
\caption{Finding charts of the GRB fields. 
Shown here are the afterglows as imaged by GROND in the $r^\prime$ band.
071112C: 0.369 days after the burst,
111228A: 0.648 days, 120714B: 0.396 days, 130831A: 0.689 days.}
\label{fig:FCs}
\end{figure}

\section{Details on figure~\ref{kplot12} \label{explained:kplots}}

The data sources for the light curve of the afterglow of GRB 071112C are given
in \cite{Kann2010}.  In addition, we added the data presented in this work. We
shifted data from other bands to the $r^\prime$ band, subtracted the host-galaxy
contribution, and only data up to six days were considered. We used the data
presented in this paper as well as the data set of \cite{Xin2016ApJ} to
construct a compound light curve of the afterglow of GRB 111228A. Here, we used
late-time data in the $g^\prime$ band, for which we were unable to discern a SN
component, host-corrected and shifted them to the $r^\prime$ band. We also used
this method to derive the compound light curve of  the afterglow of GRB
120714B. At early times, we used two UVOT \emph{white}-filter
detections\footnote{\url{http://swift.gsfc.nasa.gov/uvot\_tdrss/526642/index.html}},
hereby, we assumed $white=R_C$ in Vega magnitudes. The host-subtracted light
curve is well-described by a single power-law decay over its entire
timespan. The extremely well time-resolved light curve of the afterglow of GRB
130831A was constructed from our data as well as the extensive data sets of
\cite{DePasquale2015a}, \cite{Cano2014a}, and \cite{Gorbovskoy2015GCN}. In
this case, the associated SN 2013fu is also detected in the $g^\prime$ band,
therefore we used the analytical results from our fit to subtract the SN
component and extend the pure afterglow light curve.

We extended the sample by adding light curves for further GRB-SNe studied by
\cite{Kann2016}. The data of XRF 020903 were host-subtracted already, and
we only extend the light curve to six days post-burst. In the case of GRB
120729A, the afterglow and the SN are well-separated (the afterglow breaks
early and decays steeply), so only data until 0.8 days were taken into
account. For GRB 130215A, the SN is only detected at low significance, we cut
the light curve off at 17 days; no host is detected in this case. In the cases
of GRBs 130702A and 140606B, we subtracted both the individual host-galaxy
magnitudes as well as the SN contributions from the $r^\prime$ and $i^\prime$
light curves, then merged them according to the colors derived from the
respective SEDs.

\cite{Kann2006ApJ} labeled the shift in magnitude $dRc$. For the GRBs in this
paper, we find: GRB 111228A: $dRc=+0.62^{+0.06}_{-0.07}$ mag; GRB 120714B:
$dRc=+2.17^{+0.05}_{-0.06}$ mag; GRB 130831A: $dRc=+1.97^{+0.01}_{-0.02}$ mag;
the value for GRB 071112C is taken from \cite{Kann2010}. 

\section{Summarizing tables and log of the observations}

\begin{table*}[!htbp]
\renewcommand{\tabcolsep}{5pt}
\caption{Summary of the measured supernova parameters.}
\begin{center}
\begin{tabular}{l | rrrr}
\toprule
GRB                     & 071112C          & 111228A         & 120714B         & 130831A \\
\midrule
$k_{g^\prime}$     	            & $\cdots$        & $\cdots$          & $\cdots$         & $0.87\pm0.24$ \\
$k_{r^\prime}$                   & $0.53\pm0.17$  & $0.63\pm0.14$    & $0.47\pm0.05$   & $0.61\pm0.04$ \\
$k_{i^\prime}$                   & $0.70\pm0.17$  & $0.73\pm0.15$    & $0.74\pm0.06$   & $0.99\pm0.05$ \\
$k_{z^\prime}$                   & $0.52\pm0.21$  & $0.85\pm0.27$    & $0.78\pm0.08$   & $0.58\pm0.12$ \\
$s_{g^\prime}$                   & $\cdots$        & $\cdots$          & $\cdots$         & $0.64\pm0.09$ \\
$s_{r^\prime}$                   & $0.75\pm0.19$  & $1.58\pm0.23$    & $0.66\pm0.04$   & $0.80\pm0.02$ \\
$s_{i^\prime}$                   & $0.78\pm0.23$  & $1.20\pm0.25$    & $0.91\pm0.06$   & $0.72\pm0.02$ \\
$s_{z^\prime}$                   & $0.71\pm0.35$  & $1.45\pm0.25$    & $0.95\pm0.09$   & $0.72\pm0.07$ \\[1mm]
$t_{g^\prime, obs, p}$ (days)    & $\cdots$         &$\cdots$           &  $\cdots$        & $13.7\pm2.5$  \\
$t_{r^\prime, obs, p}$           & $23.1\pm 6.6$   & $45.8\pm 9.1$    &  $15.6\pm2.3$   & $20.0\pm2.8$  \\
$t_{i^\prime, obs, p}$           & $26.2\pm 8.7$   & $37.9\pm 9.8$    &  $23.4\pm4.9$   & $19.6\pm3.0$  \\
$t_{z^\prime, obs, p}$           & $24.7\pm12.8$   & $47.5\pm11.3$    &  $25.4\pm5.9$   & $20.3\pm3.8$  \\
$t_{g^\prime, host, p}$          & $\cdots$         & $\cdots$          & $\cdots$         &  $9.3\pm1.7$  \\
$t_{r^\prime, host, p}$          & $12.7\pm 3.6$   & $26.7\pm 5.3$    &  $11.2\pm1.7$   & $13.5\pm1.9$  \\
$t_{i^\prime, host, p}$          & $14.4\pm 4.8$   & $22.1\pm 5.7$    &  $16.7\pm2.8$   & $13.2\pm2.0$  \\
$t_{z^\prime, host, p}$          & $13.6\pm 7.0$   & $27.7\pm 6.6$    &  $18.1\pm3.4$   & $13.8\pm2.6$  \\[1mm]
$\log L_{\rm bol}$ (erg s$^{-1}$)  &$42.73\pm0.08$  &$42.83\pm0.09$    &$42.79\pm0.05$   &$42.84\pm0.05$  \\
$t_{p}$ (days)   & $11.9\pm2.4$   & $22.7\pm2.2$     & $13.6\pm0.7$     & $11.9\pm0.3$  \\[1mm]
$v_{\rm exp; obs}$ (km s$^{-1}$) & $\cdots$        & $\cdots$     & $43\,000\pm5\,300$ & $20\,400\pm3\,500$ \\
$v_{\rm exp; peak}$         & $\cdots$        & $\cdots$          & $42\,000\pm5\,500$ & $30\,300\pm6\,000$ \\
\bottomrule
\end{tabular}
\end{center}
\tablefoot{The first eight rows summarize the measured $(s,k)$ values in the
  GROND $g^{\prime}r^{\prime}i^{\prime}z^{\prime}$ bands. The following rows contain the corresponding SN peak
  times in the observer as well as in the host galaxy frame, $t_p(\mbox{SN})
  = t_p(\mbox{SN~1998bw}) \,\times\,s.$ We used the light-curve peak
  times for SN~1998bw as they follow from our SN fitting procedure
  \citep{Zeh2004a}: $t_p(g^\prime )=14.5\pm1.7$, $t_p(r^\prime)=16.9\pm2.3$,
  $t_p(i^\prime )=18.4\pm2.8$, $t_p(z^\prime )=19.1\pm3.1$ days. The
  following two rows provide the  bolometric luminosity,  $L_{\rm bol} =
  k_{\rm bol}\, L_{\rm bol}^{\rm bw}$, as well as the bolometric light-curve
  peak time (given by $t_p^{bw}\,\times\,s_{\rm bol}$, with 
  $t_p^{\rm bw}=15.86\pm0.18$ days; \citealt{Prentice2016}).
  The two last rows list the  photospheric
  expansion velocity as it was measured and its calculated value for the time
  of the  peak of the bolometric light curve.} 
\label{Tab:summary2}
\end{table*}
\begin{table*}[!htbp]
\renewcommand{\tabcolsep}{10pt}
\caption{Summary of the afterglow light-curve fits.}
\begin{center}
\begin{tabular}{ll | rrrr}
\toprule
& GRB                     & 071112C          & 111228A         & 120714B         & 130831A \\
\midrule
{\bf Afterglow} 
&RA (J2000)             & 02:36:50.955     & 10:00:16.032    & 23:41:38.076     & 23:54:29.882\\  
&Dec. (J2000)           & $+$28:22:16.70   & $+$18:17:51.94  & $-$46:11:01.72   & $+$29:25:46.11\\[2mm]
&$\alpha_1$             & $0.96\pm0.01$    & $1.18\pm0.03$   & $0.58\pm0.06$    & $1.61\pm0.01$ \\
&$\alpha_2$             & $\cdots$          &$1.79\pm0.05$    & $\cdots$          & $\cdots$        \\
&$t_{\rm b}$ (days)     & $\cdots$          &$1.68\pm0.06$    & $\cdots$          & $\cdots$        \\
&$n$                    & $\cdots$          &   10     & $\cdots$          & $\cdots$        \\
&$\chi^2$/d.o.f.        & 1.08           &    0.78         & 0.87              & 2.50            \\
\midrule
{\bf Afterglow SED} &
$g^\prime$ (corr. mag)        & $22.80\pm 0.08$ & $20.87\pm 0.05$  & $\cdots$  & $21.11\pm0.05$ \\
&$r^\prime$                   & $22.69\pm 0.08$ & $20.56\pm 0.05$  & $\cdots$  & $20.84\pm0.05$ \\
&$i^\prime$                   & $22.62\pm 0.09$ & $20.40\pm 0.05$  & $\cdots$  & $20.68\pm0.05$ \\
&$z^\prime$                   & $22.56\pm 0.09$ & $20.21\pm 0.05$  & $\cdots$  & $20.52\pm0.05$ \\
&$J$                    & $22.33\pm 0.12$ & $19.98\pm 0.07$  & $\cdots$  & $20.07\pm0.06$ \\
&$H$                    & $22.05\pm 0.26$ & $19.72\pm 0.08$  & $\cdots$  & $19.76\pm0.08$ \\
&$K_s$                  & $21.95\pm 0.35$ & $19.33\pm 0.10$  & $\cdots$  & $19.39\pm0.10$ \\[1mm]
&$E_{\rm (B-V), Gal}$ (mag)  & 0.12           & 0.03           & 0.01              & 0.04            \\
&$\beta_{\rm obs}$      & $0.44\pm0.11$  & $0.69\pm0.07$    & $0.7\pm0.4$    & $1.00\pm0.05$ \\
&$A_V^{\rm host}$ (mag) 	& 0              & $0.16\pm0.04$    & 0          & 0 \\
&$\chi^2$/d.o.f.        & 0.23             & 0.86            & 6.5          & 0.66 \\
\bottomrule
\end{tabular}
\end{center}
\tablefoot{
The afterglow magnitudes forming the SEDs are corrected for the Galactic reddening, $E_{\rm
  (B-V), Gal}$, along the line of sight. In the case of GRB 120714B the
afterglow SED was only roughly determined based on VLT/X-shooter data taken
0.353 days after the burst (see Sect.~\ref{Shocking}).  In this
case, for the SN light curve fits $A_V^{\rm host}=0$ mag is assumed, not measured. In
order to be consistent with the  host-galaxy measurements (see
Table~\ref{Tab:summary3}), in the case of GRB 130831A afterglow coordinates
were measured on the VLT acquisition images. Otherwise GROND images were
used.} 
\label{Tab:summary1}
\end{table*}
\begin{table*}[!htbp]
\renewcommand{\tabcolsep}{7pt}
\caption{Summary of the $\alpha-\beta$ relations.}
\begin{center}
\begin{tabular}{l | cc cc }
\toprule
Model / GRB             & 071112C     & 111228A & 120714B  & 130831A \\
                        & $\beta_{\rm obs}=0.44\pm0.11$ 
                        & $\beta_{\rm obs}=0.69\pm0.07$
                        & $\beta_{\rm obs}=0.7\pm0.4$ 
                        & $\beta_{\rm obs}=1.00\pm0.05$ \\
                        & $\beta_{\rm theo}$ 
                        & $\beta_{\rm theo} (p>2)$, $\beta_{\rm theo} (p<2)$ 
                        & $\beta_{\rm theo}$ 
                        & $\beta_{\rm theo}$ \\
\midrule
ISM,  iso, $\nu < \nu_c$ &  0.64$\pm$0.01 &{ }{ }0.04$\pm$0.03,  1.65$\pm$0.08 &  0.39$\pm$0.04 &   1.07$\pm$0.01   \\
ISM,  iso, $\nu > \nu_c$ &  0.97$\pm$0.01 &{ }{ }0.37$\pm$0.03,  1.48$\pm$0.08 &  0.72$\pm$0.04 &   1.41$\pm$0.01   \\
wind, iso, $\nu < \nu_c$ &  0.31$\pm$0.01&$-$0.29$\pm$0.03,  0.22$\pm$0.12 &  0.05$\pm$0.04 &   0.74$\pm$0.01   \\
wind, iso, $\nu > \nu_c$ &  0.97$\pm$0.01 &{ }{ }0.37$\pm$0.03,  1.72$\pm$0.12 &  0.72$\pm$0.04 &   1.41$\pm$0.01   \\[1mm]
ISM,  jet, $\nu < \nu_c$ &  --            &{ }{ }0.50$\pm$0.03,  0.08$\pm$0.10 &  --            &   0.31$\pm$0.01   \\
ISM,  jet, $\nu > \nu_c$ &  --            &{ }{ }1.00$\pm$0.03,  0.58$\pm$0.10 &  --            &   0.81$\pm$0.01   \\
wind, jet, $\nu < \nu_c$ &  --            &{ }{ }0.50$\pm$0.03,  0.08$\pm$0.10 &  --            &   0.31$\pm$0.01   \\
wind, jet, $\nu > \nu_c$ &  --            &{ }{ }1.00$\pm$0.03,  0.58$\pm$0.10 &  --            &   0.81$\pm$0.01   \\
\bottomrule
\end{tabular}
\end{center}
\tablefoot{Predicted spectral slope $\beta_{\rm theo}$  for the afterglows
  adopting for the power-law index $p$ of the electron distribution function a
  value $p>2$  (GRB 071112C, 111228A, 120714B, 130831A) or $p<2$ (GRB 111228A;
  \citealt{Zhang2004IJMPA}). Values for $\alpha_1$ and $\alpha_2$ are taken from
  Table~\ref{Tab:summary1}. \emph{Individual
    events:}\ GRB 071112C: The observed decay slope $\alpha$ implies that the
  afterglow is in the spherical expansion phase. GRB 111228A: Shown here are the
  results for the extended data set (which suggests $p>2$) and for the
  GROND-only data (which imply $p<2$). GRB 120714B: The observed decay slope
	$\alpha$ implies that the afterglow is in the spherical expansion phase. GRB
	130831A: The observed decay slope could either be pre- or post-break.
	Therefore, both options are calculated.}
\label{Tab:alphabeta}
\end{table*}
\begin{table*}[!htbp]
\renewcommand{\tabcolsep}{10pt}
\caption{Summary of the host-galaxy data.}
\begin{center}
\begin{tabular}{l | rrrr}
\toprule
GRB                     & 071112C          & 111228A         & 120714B         & 130831A \\
\midrule

RA (J2000)              & 02:36:50.965     & 10:00:16.051    & 23:41:38.057    & 23:54:29.831\\  
Dec.                    & $+$28:22:16.48   & $+$18:17:52.01  & $-$46:11:01.64  & $+$29:25:45.73\\[1mm]
$g^\prime$              & $25.67\pm0.06$   & $25.04\pm0.17$  & $23.80\pm0.13$  & $24.69\pm0.27$ \\
$r^\prime$              & $25.18\pm0.11$   & $24.44\pm0.12$  & $22.93\pm0.10$  & $24.08\pm0.17$ \\
$i^\prime$              & $24.45\pm0.47$   & $23.98\pm0.14$  & $22.70\pm0.10$  & $23.88\pm0.10$ \\
$z^\prime$              & $24.05\pm0.14$   & $23.75\pm0.20$  & $22.50\pm0.12$  & $23.69\pm0.30$ \\
$J$                     & $>22.3$          & $>22.3$         & $22.34\pm0.13$  & $23.49\pm0.22$    \\
$H$                     & {\bf $23.91\pm0.06$}   & $>21.8$         & $>21.8$         & $>22.0$     \\
$K_s$                   & $>21.0$          & $>20.7$         & $22.22 \pm0.14$ & $>21.1$     \\[1mm]
offset SN               & $0\farcs26\pm0\farcs20$&$0\farcs28\pm0\farcs20$&$0\farcs21\pm0\farcs20$&$0\farcs92\pm0\farcs20$          \\
offset SN (kpc)         & $2.1\pm1.6$      & $2.1 \pm 1.5$   & $1.2 \pm 1.1$  & $5.3\pm1.2$    \\
\bottomrule
\end{tabular}
\end{center}
\tablefoot{
{Magnitudes are given in the AB system and corrected for Galactic extinction according to \cite{Schlafly2011a}.
{\it Notes on individual objects:} GRB 071112C: $z^\prime$ and $H$-band magnitudes
were taken from \cite{Vergani2015a} (the near-infrared band data point 
is actually based on an observation with the HST F160W filter), 
$g^\prime$- and $r^\prime$-band magnitudes are based on Gemini-North/GMOS observations.
GRB 120714B: $JK_s$-band magnitudes stem from our VLT/HAWKI observations.
GRB 130831A: the $J$-band magnitude 
is based on Keck/MOSFIRE observations (Table~\ref{noGRONDmag}), the $i^\prime$-band
magnitude was derived based on our light-curve fit.}
See Table~\ref{noGRONDmag}
for additional host-galaxy detections and upper limits in other filters. All other
magnitudes are based on late-epoch observations with GROND. In the case of GRB 
130831A the host-galaxy coordinates were measured on the VLT acquisition images, otherwise
GROND images were used.} 
\label{Tab:summary3}
\end{table*}
\begin{table*}
\caption{Results of the stellar population synthesis fits using 
\texttt{Le PHARE} and adopting a starburst model.}
\begin{center}
\begin{tabular}{c|cccccc}
\toprule
GRB     & 
$M_B$   & 
$E(B-V)_{\rm host}$ & 
$\log SFR $  & 
$\log M/M_\odot$ & 
$\log sSFR$ & 
$\chi^2$ \\
        & (mag)   & (mag)&$(M_\odot\,{\rm yr}^{-1})$ && (${\rm yr}^{-1}$) &\\
\midrule
 071112C & $-$18.49 & 0.15 & $\,\,\,\,  0.22_{-0.44}^{+0.60}$ & $8.67_{-0.23}^{+0.16}$ & $-8.44_{-0.61}^{+0.86}$ & 2.11\\[1mm]
 111228A & $-$18.57 & 0.40 & $\,\,\,\,  0.33_{-0.61}^{+0.63}$ & $8.72_{-0.24}^{+0.24}$ & $-8.41_{-0.65}^{+0.68}$ & 0.01\\[1mm]
 120714B & $-$18.33 & 0.08 & $-0.32_{-0.29}^{+0.42}$ & $8.72_{-0.17}^{+0.16}$ & $-9.06_{-0.38}^{+0.58}$ & 0.01\\[1mm]
 130831A & $-$17.72 & 0.02 & $-0.25_{-0.56}^{+0.75}$ & $8.43_{-0.32}^{+0.47}$ & $-8.66_{-0.83}^{+0.82}$ & 0.55\\
\midrule
\multicolumn{6}{c}{Median values of the unbiased \swift/BAT6 GRB host
galaxy sample at $z<1$ from \citet{Vergani2015a}}\\
\midrule
  & $\cdots$ & $0.00^{+0.08}_{-0.00}$ & $-0.07^{+0.59}_{-0.30}$ & $8.84^{+0.25}_{-0.60}$ & $-8.72^{+0.31}_{-0.26}$ & $\cdots$ \\
\midrule
\multicolumn{6}{c}{Median values of the $z<0.5$ GRB host
galaxy sample from \citet{Schulze2016}}\\
\midrule
  & $-18.33\pm0.41$ & $\cdots$ & $-0.19\pm0.13$ & $8.83\pm0.20$ & $-9.15\pm0.11$ & $\cdots$ \\
\bottomrule
\end{tabular}
\end{center}
\tablefoot{
The integrated host properties were derived by fitting the spectral
energy distribution of the host galaxies with
stellar-population-synthesis models by \citet{Bruzual2003a} in
\texttt{Le PHARE}. The SEDs
of the BAT6 sub-sample were fitted using the same method. The errors
of the median values indicate the distance to the 16 and 84\%-iles.}
\label{Tab:LePhare1short}
\end{table*}


\begin{table*}
\caption{GRB 071112C: GROND AB magnitudes and upper limits of the optical transient.}
\begin{center}
\begin{tabular}{r|ccccccc}
\toprule
$t - t_0$ & $g^\prime$ & $r^\prime$ & $i^\prime$ & $z^\prime$ & $J$ & $H$ & $K_s$  \\
\midrule
0.322	      &	$22.54\pm0.14$	&	$22.20\pm0.08$	&	$21.95\pm0.16$	&	$21.84\pm0.24$	&		--	          	&		--		          &		--		\\
0.330	      &	$22.63\pm0.06$	&	$22.21\pm0.06$	&	$21.97\pm0.11$	&	$21.88\pm0.16$	&		--		          &		--	          	&		--		\\
0.347	      &	$22.57\pm0.03$	&	$22.31\pm0.03$	&	$22.13\pm0.06$	&	$21.89\pm0.07$	&		--	          	&		--		          &		--		\\
0.369	      &	$22.64\pm0.03$	&	$22.44\pm0.04$	&	$22.17\pm0.06$	&	$22.08\pm0.10$	&		--		          &		--		          &		--		\\
0.392	      &	$22.72\pm0.03$	&	$22.44\pm0.03$	&	$22.30\pm0.05$	&	$22.13\pm0.08$	&	$21.66\pm0.13$	  &	$21.44\pm0.26$	  &	$21.31\pm0.35$	\\
0.415	      &	$22.73\pm0.03$	&	$22.49\pm0.03$	&	$22.33\pm0.06$	&	$22.07\pm0.07$	&		--		          &		--		          &		--		\\
0.438	      &	$22.76\pm0.03$	&	$22.48\pm0.04$	&	$22.32\pm0.07$	&	$22.17\pm0.08$	&		--		          &		--	           	&		--		\\
0.454	      &	$22.80\pm0.11$	&	$22.55\pm0.08$	&	$22.30\pm0.11$	&	$22.16\pm0.18$	&		--		          &		--		          &		--		\\
1.392	      &	$23.90\pm0.04$	&	$23.63\pm0.06$	&	$23.34\pm0.07$	&	$23.24\pm0.14$	&	$>22.35$		      &	$>21.82$	      	&	$>21.02$		\\
2.402	      &	$24.73\pm0.10$ 	&	$24.05\pm0.10$  &	$23.79\pm0.14$	&	$23.65\pm0.23$	&	$>22.34$	      	&	$>21.62$	      	&	$>20.93$		\\
3.387	      &	$25.20\pm0.27$	&	$24.48\pm0.18$	&	$23.77\pm0.21$	&	$>23.53$        &	$>22.30$	      	&	$>21.30$	      	&	$>20.94$		\\
6.409	      &	$25.05\pm0.40$ 	&	$24.33\pm0.22$	&	$24.28\pm0.29$	&	$>24.05$	      &	$>22.17$	      	&	$>21.46$	       	&	$>20.70$		\\
8.365	      & $>24.23$		    &	$>24.30$  	    &	$24.38\pm0.47$	&	$23.82\pm0.35$	&	$>22.17$	      	&	$>21.32$	      	&	$>20.58$		\\
14.369	    &	$>24.46$	  	  &	$24.36\pm0.31$	&	$>24.01$	      &	$>23.69$		    &	$>21.88$	      	&	$>21.09$	      	&	$>20.73$		\\
17.343	    &	$25.50\pm0.23$	&	$24.73\pm0.13$	&	$24.07\pm0.20$  &	$23.54\pm0.21$	&	$>22.24$	      	&	$>21.32$	      	&	$>20.51$		\\
21.357    	&	$>25.36$		    &	$24.72\pm0.24$	&	$23.91\pm0.18$	&	$23.45\pm0.25$	&	$>22.11$	      	&	$>21.35$	      	&	$>20.29$		\\
22.317    	&	$>25.61$		    &	$25.03\pm0.46$	&	$23.85\pm0.19$	&	$23.76\pm0.33$	&	$>22.20$		      &	$>21.30$	      	&	$>20.78$		\\
24.327    	&	$25.58\pm0.31$	&	$24.63\pm0.19$	&	$24.22\pm0.23$	&	$24.03\pm0.38$	&	$>22.35$	      	&	$>21.51$	      	&	$>20.86$		\\
28.333    	&	$25.49\pm0.25$	&	$25.05\pm0.22$	&	$24.33\pm0.26$	&	$23.72\pm0.25$	&	$>22.14$	      	&	$>21.55$	       	&	$>20.74$		\\
2534.453   	&	$25.51\pm0.43$	&	$25.13\pm0.28$	&	$24.69\pm0.47$	&	$>23.68$				&	$>21.90$	      	&	$>21.37$	       	&	$>20.54$		\\
\bottomrule
\end{tabular}
\tablefoot{
$t-t_0$ is the time in units of days after the burst 
($t_0=12$ November 2007, 18:32:57 UT; \citealt{Perri2007a}). 
The data are not corrected for Galactic extinction. }
\label{071112Cmagdata}
\end{center}
\end{table*}

\begin{table*}
\caption{GRB 111228A: GROND AB magnitudes and upper limits of the optical transient.}
\begin{center}
\begin{tabular}{r|ccccccc}
\toprule
$t - t_0$ & $g^\prime$ & $r^\prime$ & $i^\prime$ & $z^\prime$ & $J$ & $H$ & $K_s$  \\
\midrule
0.552	      &		--	          	&	19.80	$\pm$ 0.02	&		--	          	&		--	          	&		--		          &		--		          &		--		\\
0.556	      &	20.09	$\pm$ 0.02	&	19.81	$\pm$ 0.02	&	19.59	$\pm$ 0.03	&	19.54	$\pm$ 0.052	&	19.27	$\pm$ 0.12	&	19.32	$\pm$ 0.23	&	18.55	$\pm$ 	0.21	\\
0.561      	&	20.09	$\pm$ 0.02	&	19.81	$\pm$ 0.01	&	19.63	$\pm$ 0.03	&	19.45	$\pm$ 0.045	&	19.14 $\pm$ 0.10	&	18.63	$\pm$ 0.10	&	18.49	$\pm$ 	0.22	\\
0.568	      &	20.15	$\pm$ 0.02	&	19.82 $\pm$ 0.01	&	19.63	$\pm$ 0.02	&	19.40	$\pm$ 0.023	&	19.23	$\pm$ 0.06	&	18.84	$\pm$ 0.07	&	18.84	$\pm$ 	0.22	\\
0.577      	&		--	           	&	19.85	$\pm$ 0.01	&	19.70	$\pm$ 0.02	&	19.43	$\pm$ 0.027	&	19.19	$\pm$ 0.05	&	18.93 $\pm$ 0.07	&	18.64	$\pm$ 	0.20	\\
0.586      	&	20.20	$\pm$ 0.01	&	19.86	$\pm$ 0.01	&	19.67	$\pm$ 0.02	&	19.46	$\pm$ 0.020	&	19.15	$\pm$ 0.06	&	18.88	$\pm$ 0.07	&	18.54	$\pm$ 	0.15	\\
0.595      	&	20.20	$\pm$ 0.02	&	19.89	$\pm$ 0.01	&	19.70	$\pm$ 0.02	&	19.47	$\pm$ 0.019	&	19.21	$\pm$ 0.05	&	19.28	$\pm$ 0.09	&	18.76	$\pm$ 	0.16	\\
0.604	      &		--		          &	19.91 $\pm$	0.01	&	19.70	$\pm$ 0.02	&	19.49	$\pm$ 0.022	&	19.19	$\pm$ 0.05	&	18.96	$\pm$ 0.07	&	18.60	$\pm$ 	0.16	\\
0.613      	&	20.23	$\pm$ 0.01	&	19.92	$\pm$ 0.01	&	19.72	$\pm$ 0.02	&	19.51	$\pm$ 0.022	&	19.23	$\pm$ 0.06	&	19.05	$\pm$ 0.08	&	19.00	$\pm$ 	0.21	\\
0.622      	&	20.25	$\pm$ 0.01	&	19.92	$\pm$ 0.01	&	19.73	$\pm$ 0.01	&	19.52	$\pm$ 0.022	&	19.21	$\pm$ 0.06	&	18.94	$\pm$ 0.07	&	18.55	$\pm$ 	0.14	\\
0.631	      &	20.28	$\pm$ 0.01	&	19.95	$\pm$ 0.01	&	19.77	$\pm$ 0.02	&	19.56	$\pm$ 0.023	&	19.25	$\pm$ 0.06	&	19.08	$\pm$ 0.09	&	18.60	$\pm$ 	0.14	\\
0.640      	&	20.28	$\pm$ 0.01	&	19.98	$\pm$ 0.01	&	19.78	$\pm$ 0.01	&	19.60	$\pm$ 0.022	&	19.32	$\pm$ 0.05	&	19.05	$\pm$ 0.07	&	18.52	$\pm$ 	0.14	\\
0.648	      &	20.31	$\pm$ 0.01	&	19.99	$\pm$ 0.01	&	19.77	$\pm$ 0.02	&	19.56	$\pm$ 0.021	&	19.40	$\pm$ 0.06	&	19.12	$\pm$ 0.08	&	18.85	$\pm$ 	0.17	\\
0.657	      &	20.30	$\pm$ 0.01	&	19.99	$\pm$ 0.01	&	19.79	$\pm$ 0.02	&	19.62	$\pm$ 0.021	&	19.29	$\pm$ 0.06	&	19.10	$\pm$ 0.08	&	18.70	$\pm$ 	0.15	\\
0.666      	&	20.36	$\pm$ 0.01	&	20.01	$\pm$ 0.01	&	19.85	$\pm$ 0.01	&	19.61	$\pm$ 0.023	&	19.34	$\pm$ 0.06	&	19.03	$\pm$ 0.07	&	18.66	$\pm$ 	0.15	\\
0.675      	&	20.33	$\pm$ 0.01	&	20.04	$\pm$ 0.01	&	19.81	$\pm$ 0.02	&	19.62	$\pm$ 0.021	&	19.37	$\pm$ 0.06	&	19.14	$\pm$ 0.08	&	18.55	$\pm$ 	0.13	\\
0.684      	&	20.37	$\pm$ 0.01	&	20.05	$\pm$ 0.01	&	19.85	$\pm$ 0.02	&	19.65	$\pm$ 0.018	&	19.47	$\pm$ 0.06	&	19.16	$\pm$ 0.07	&	18.47	$\pm$ 	0.12	\\
0.693	      &	20.37	$\pm$ 0.01	&	20.05	$\pm$ 0.01	&	19.87	$\pm$ 0.02	&	19.68	$\pm$ 0.022	&	19.48	$\pm$ 0.06  &	19.16	$\pm$ 0.07	&	18.79	$\pm$ 	0.16	\\
0.702      	&	20.36	$\pm$ 0.02	&	20.06	$\pm$ 0.01	&	19.88	$\pm$ 0.02	&	19.68	$\pm$ 0.023	&	19.49	$\pm$ 0.06	&	19.22	$\pm$ 0.07	&	18.91	$\pm$ 	0.16	\\
0.709	      &		--	           	&		--		          &		--		          &		--	           	&	19.50	$\pm$ 0.09	&	19.33	$\pm$ 0.11	&	18.82	$\pm$ 	0.19	\\
0.714      	&		--	           	&		--	           	&		--		          &		--		          &	19.47	$\pm$ 0.09	&	19.26	$\pm$ 0.12	&	19.03	$\pm$ 	0.26	\\
0.719      	&		--	           	&		--	           	&		--		          &		--		          &	19.59	$\pm$ 0.10	&	19.16	$\pm$ 0.10	&	18.75	$\pm$ 	0.18	\\
0.724	      &		--		          &		--		          &		--	           	&		--	           	&	19.36	$\pm$ 0.10	&	19.15	$\pm$ 0.10	&	18.72	$\pm$ 	0.18	\\
0.728      	&		--		          &		--	          	&		--	           	&		--	           	&	19.58	$\pm$ 0.18	&	19.24	$\pm$ 0.11	&	18.97	$\pm$ 	0.23	\\
1.700       &	21.63	$\pm$ 0.05	&	21.32	$\pm$ 0.03	&	21.09	$\pm$ 0.06	&	20.81	$\pm$ 0.07	&	20.73	$\pm$ 0.18	&	20.51	$\pm$ 0.23	&	19.96	$\pm$ 	0.31	\\
2.626	      &	22.37	$\pm$ 0.05	&	22.01	$\pm$ 0.04	&	21.72	$\pm$ 0.06	&	21.59	$\pm$ 0.07	&	$>$	21.60	       	&	$>$	21.15		      &	$>$	19.93		\\
3.707	      &	23.25	$\pm$ 0.27	&	22.65	$\pm$ 0.12	&	22.01	$\pm$ 0.18	&	22.41	$\pm$ 0.26	&	$>$	21.27	       	&	$>$	20.76		      &	$>$	20.14		\\
5.686	      &	23.73	$\pm$ 0.09	&	23.20	$\pm$ 0.06	&	22.69	$\pm$ 0.09	&	22.51 $\pm$ 0.09	&	$>$	21.65		      &	$>$	21.15	       	&	$>$	20.09		\\
7.675	      &	24.08	$\pm$ 0.16	&	23.68	$\pm$ 0.11	&	23.02	$\pm$ 0.15	&	22.95	$\pm$ 0.17	&	$>$	21.64	      	&	$>$	21.16		      &	$>$	20.04		\\
18.654	    &	24.54	$\pm$ 0.30	&	24.08	$\pm$ 0.16	&	23.27	$\pm$ 0.13	&	22.95	$\pm$ 0.10	&	$>$	22.35	      	&	$>$	21.80		      &	$>$	20.89		\\
25.669	    &	24.96	$\pm$ 0.23	&	24.17	$\pm$ 0.12	&	23.61	$\pm$ 0.18	&	23.29	$\pm$ 0.16	&	$>$	22.26	      	&	$>$	21.78	      	&	$>$	20.91		\\
59.538	    &	25.12	$\pm$ 0.26	&	24.33	$\pm$ 0.13	&	23.72	$\pm$ 0.21	&	23.40	$\pm$ 0.21	&	$>$	22.23	      	&	$>$	21.82	      	&	$>$	20.52		\\
89.434	    &	25.50	$\pm$ 0.31	&	24.40	$\pm$ 0.18	&	24.01	$\pm$ 0.22	&	23.29	$\pm$ 0.15	&	$>$	22.28      		&	$>$	21.68	      	&	$>$	20.60		\\
1499.564	  &	25.16	$\pm$ 0.17	&	24.52	$\pm$ 0.12	&	24.04	$\pm$ 0.14	&	23.80	$\pm$ 0.20	&	$>$	22.34      		&	$>$	21.84	      	&	$>$	20.74		\\
\bottomrule
\end{tabular}
\tablefoot{$t-t_0$ is the time in units of days after the burst 
($t_0=28$ December 2011, 15:44:43 UT; \citealt{Ukwatta2011a}).
The data are not corrected for Galactic extinction. }
\label{111228Amagdata}
\end{center}
\end{table*}

\begin{table*}
\caption{GRB 111228A: UVOT AB mags (not corrected for Galactic extinction) 
and upper limits.}
\renewcommand{\tabcolsep}{4.0pt}
\begin{center}
\begin{tabular}{rrll|rrll|rrll}
\toprule
$t - t_0$ & exp & mag & filter & $t - t_0$ & exp & mag & filter & $t - t_0$ & exp & mag & filter \\
\midrule
0.00433	&	19.5	& $	19.35	^{+	0.35	}_{-	0.26	}$ &	$uvw2$	&	5.58389	&	2477.5	& $	23.73	^{+	0.91	}_{-	0.49	}$ &	$uvw1$	&	0.00185	&	10.0	& $	18.44	^{+	0.14	}_{-	0.13	}$ &	$white$	\\
0.00634	&	19.4	& $	18.59	^{+	0.23	}_{-	0.19	}$ &	$uvw2$	&	6.16284	&	1605.3	& $ >	22.53					$ &	$uvw1$	&	0.00197	&	10.0	& $	18.29	^{+	0.13	}_{-	0.12	}$ &	$white$	\\
0.00839	&	19.4	& $	18.89	^{+	0.27	}_{-	0.22	}$ &	$uvw2$	&	0.00547	&	19.4	& $	17.72	^{+	0.15	}_{-	0.13	}$ &	$u$	&	0.00208	&	10.0	& $	18.17	^{+	0.12	}_{-	0.11	}$ &	$white$	\\
0.01190	&	19.5	& $	18.95	^{+	0.29	}_{-	0.23	}$ &	$uvw2$	&	0.00749	&	19.5	& $	18.00	^{+	0.17	}_{-	0.15	}$ &	$u$	&	0.00220	&	10.0	& $	18.04	^{+	0.11	}_{-	0.10	}$ &	$white$	\\
0.01390	&	19.5	& $	18.66	^{+	0.25	}_{-	0.20	}$ &	$uvw2$	&	0.00952	&	19.4	& $	17.99	^{+	0.18	}_{-	0.16	}$ &	$u$	&	0.00231	&	10.0	& $	18.17	^{+	0.12	}_{-	0.11	}$ &	$white$	\\
0.07245	&	196.6	& $	19.60	^{+	0.12	}_{-	0.10	}$ &	$uvw2$	&	0.01303	&	16.6	& $	18.18	^{+	0.28	}_{-	0.22	}$ &	$u$	&	0.00243	&	10.0	& $	18.02	^{+	0.11	}_{-	0.10	}$ &	$white$	\\
0.13358	&	885.6	& $	20.02	^{+	0.07	}_{-	0.06	}$ &	$uvw2$	&	0.06532	&	196.6	& $	18.42	^{+	0.07	}_{-	0.06	}$ &	$u$	&	0.00255	&	10.0	& $	18.11	^{+	0.12	}_{-	0.11	}$ &	$white$	\\
0.52521	&	804.9	& $	21.84	^{+	0.20	}_{-	0.17	}$ &	$uvw2$	&	0.08107	&	47.7	& $	18.27	^{+	0.18	}_{-	0.16	}$ &	$u$	&	0.00266	&	10.0	& $	18.20	^{+	0.12	}_{-	0.11	}$ &	$white$	\\
0.72515	&	885.6	& $	21.82	^{+	0.19	}_{-	0.16	}$ &	$uvw2$	&	0.11799	&	30.1	& $	18.74	^{+	0.22	}_{-	0.18	}$ &	$u$	&	0.00278	&	10.0	& $	17.92	^{+	0.11	}_{-	0.10	}$ &	$white$	\\
1.83617	&	1209.7	& $	23.45	^{+	0.59	}_{-	0.38	}$ &	$uvw2$	&	0.47818	&	746.5	& $	20.22	^{+	0.12	}_{-	0.11	}$ &	$u$	&	0.00289	&	10.0	& $	18.03	^{+	0.11	}_{-	0.10	}$ &	$white$	\\
2.73793	&	1123.3	& $ >	22.48					$ &	$uvw2$	&	0.66585	&	275.5	& $	20.73	^{+	0.26	}_{-	0.21	}$ &	$u$	&	0.00301	&	10.0	& $	18.06	^{+	0.12	}_{-	0.10	}$ &	$white$	\\
3.56617	&	1674.5	& $	23.81	^{+	0.63	}_{-	0.40	}$ &	$uvw2$	&	0.87038	&	885.1	& $	21.23	^{+	0.21	}_{-	0.17	}$ &	$u$	&	0.00312	&	10.0	& $	17.93	^{+	0.11	}_{-	0.10	}$ &	$white$	\\
4.73622	&	2169.3	& $ >	22.66					$ &	$uvw2$	&	0.94385	&	292.4	& $	21.14	^{+	0.38	}_{-	0.28	}$ &	$u$	&	0.00324	&	10.0	& $	17.94	^{+	0.11	}_{-	0.10	}$ &	$white$	\\
6.72761	&	5232.4	& $ >	22.75					$ &	$uvw2$	&	1.23909	&	1036.5	& $	21.62	^{+	0.27	}_{-	0.21	}$ &	$u$	&	0.00336	&	10.0	& $	17.92	^{+	0.11	}_{-	0.10	}$ &	$white$	\\
8.33080	&	2974.9	& $ >	22.62					$ &	$uvw2$	&	3.43175	&	1658.5	& $ >	21.60					$ &	$u$	&	0.00347	&	9.7	& $	17.95	^{+	0.11	}_{-	0.10	}$ &	$white$	\\
10.2592	&	7029.7	& $ >	22.80					$ &	$uvw2$	&	4.60168	&	1999.8	& $ >	21.57					$ &	$u$	&	0.00605	&	19.4	& $	17.80	^{+	0.07	}_{-	0.07	}$ &	$white$	\\
12.7314	&	49449.0	& $ >	25.00					$ &	$uvw2*$	&	6.60718	&	3520.6	& $ >	21.61					$ &	$u$	&	0.00807	&	19.5	& $	18.09	^{+	0.09	}_{-	0.08	}$ &	$white$	\\
14.6397	&	3521.2	& $ >	22.84					$ &	$uvw2$	&	7.63888	&	15759.0	& $ >	23.60					$ &	$u*$	&	0.01081	&	147.4	& $	18.05	^{+	0.04	}_{-	0.04	}$ &	$white$	\\
19.8672	&	25310.8	& $ >	22.96					$ &	$uvw2$	&	8.26114	&	2625.5	& $ >	21.41					$ &	$u$	&	0.01361	&	19.4	& $	18.24	^{+	0.12	}_{-	0.11	}$ &	$white$	\\
22.2613	&	1252.6	& $ >	22.65					$ &	$uvw2$	&	10.9366	&	5174.4	& $ >	21.81					$ &	$u$	&	0.07006	&	196.6	& $	18.70	^{+	0.05	}_{-	0.05	}$ &	$white$	\\
0.00491	&	19.5	& $	18.17	^{+	0.25	}_{-	0.20	}$ &	$uvm2$	&	12.1980	&	3912.1	& $ >	22.52					$ &	$u$	&	0.59775	&	164.3	& $	21.05	^{+	0.21	}_{-	0.18	}$ &	$white$	\\
0.00693	&	19.5	& $	18.43	^{+	0.28	}_{-	0.22	}$ &	$uvm2$	&	0.00576	&	19.4	& $	17.52	^{+	0.19	}_{-	0.16	}$ &	$b$	&	0.80219	&	817.4	& $	21.21	^{+	0.10	}_{-	0.09	}$ &	$white$	\\
0.00896	&	19.4	& $	18.49	^{+	0.30	}_{-	0.23	}$ &	$uvm2$	&	0.00779	&	19.5	& $	17.42	^{+	0.18	}_{-	0.15	}$ &	$b$	&	1.87291	&	885.1	& $	22.48	^{+	0.33	}_{-	0.25	}$ &	$white$	\\
0.01246	&	19.5	& $	18.69	^{+	0.33	}_{-	0.25	}$ &	$uvm2$	&	0.00981	&	19.4	& $	17.61	^{+	0.22	}_{-	0.19	}$ &	$b$	&	3.14088	&	1236.5	& $	23.45	^{+	0.79	}_{-	0.45	}$ &	$white$	\\
0.01446	&	19.5	& $	18.46	^{+	0.29	}_{-	0.23	}$ &	$uvm2$	&	0.01332	&	19.4	& $	17.74	^{+	0.32	}_{-	0.25	}$ &	$b$	&	4.47754	&	2421.9	& $	23.63	^{+	0.61	}_{-	0.39	}$ &	$white$	\\
0.07719	&	196.6	& $	19.25	^{+	0.13	}_{-	0.11	}$ &	$uvm2$	&	0.06769	&	196.6	& $	18.22	^{+	0.08	}_{-	0.08	}$ &	$b$	&	6.74437	&	4351.5	& $ >	22.19					$ &	$white$	\\
0.26183	&	248.8	& $	20.25	^{+	0.19	}_{-	0.16	}$ &	$uvm2$	&	0.21029	&	822.3	& $	19.13	^{+	0.09	}_{-	0.08	}$ &	$b$	&	8.75036	&	2143.0	& $ >	22.03					$ &	$white$	\\
0.28021	&	312.3	& $	20.76	^{+	0.25	}_{-	0.20	}$ &	$uvm2$	&	0.59146	&	885.1	& $	20.71	^{+	0.28	}_{-	0.22	}$ &	$b$	&	11.0237	&	8490.5	& $ >	22.51					$ &	$white$	\\
0.33289	&	885.6	& $	20.82	^{+	0.14	}_{-	0.12	}$ &	$uvm2$	&	0.79203	&	885.2	& $	20.74	^{+	0.28	}_{-	0.22	}$ &	$b$	&	14.0660	&	2872.2	& $ >	22.27					$ &	$white$	\\
0.92634	&	885.6	& $	22.38	^{+	0.35	}_{-	0.26	}$ &	$uvm2$	&	0.87714	&	244.3	& $	20.31	^{+	0.42	}_{-	0.30	}$ &	$b$	&	18.5185	&	61512.0	& $ >	25.00					$ &	$white*$	\\
1.14712	&	682.1	& $	22.63	^{+	0.58	}_{-	0.37	}$ &	$uvm2$	&	1.21404	&	664.6	& $	20.90	^{+	0.55	}_{-	0.36	}$ &	$b$	&	21.6931	&	15200.1	& $ >	22.56					$ &	$white$	\\
1.26110	&	885.6	& $	22.09	^{+	0.29	}_{-	0.23	}$ &	$uvm2$	&	2.43093	&	1292.4	& $ >	20.61					$ &	$b$	&	30.5139	&	18317.7	& $ >	22.54					$ &	$white$	\\
2.81012	&	1688.1	& $ >	22.15					$ &	$uvm2$	&	3.50282	&	1251.3	& $ >	20.76					$ &	$b$	&	33.7246	&	13313.4	& $ >	23.09					$ &	$white$	\\
4.34176	&	2187.1	& $ >	22.31					$ &	$uvm2$	&	3.93518	&	5332.0	& $ >	22.20					$ &	$b*$	&		&		&						&		\\
4.39814	&	6573.0	& $ >	24.00					$ &	$uvm2*$	&	4.60650	&	1999.8	& $ >	20.74					$ &	$b$	&		&		&						&		\\
5.57821	&	2638.5	& $ >	22.27					$ &	$uvm2$	&	5.92886	&	1577.8	& $ >	20.90					$ &	$b$	&		&		&						&		\\
6.16084	&	1666.1	& $ >	22.97					$ &	$uvm2$	&	0.00161	&	10.6	& $	18.14	^{+	1.04	}_{-	0.52	}$ &	$v$	&		&		&						&		\\
0.00519	&	19.4	& $	18.04	^{+	0.19	}_{-	0.16	}$ &	$uvw1$	&	0.00462	&	19.5	& $	18.10	^{+	0.66	}_{-	0.41	}$ &	$v$	&		&		&						&		\\
0.00721	&	19.4	& $	18.53	^{+	0.25	}_{-	0.21	}$ &	$uvw1$	&	0.00665	&	19.5	& $	17.82	^{+	0.50	}_{-	0.34	}$ &	$v$	&		&		&						&		\\
0.00924	&	19.4	& $	18.46	^{+	0.25	}_{-	0.20	}$ &	$uvw1$	&	0.00867	&	19.5	& $	17.14	^{+	0.29	}_{-	0.23	}$ &	$v$	&		&		&						&		\\
0.01275	&	19.5	& $	18.76	^{+	0.31	}_{-	0.24	}$ &	$uvw1$	&	0.01218	&	19.5	& $	17.18	^{+	0.37	}_{-	0.27	}$ &	$v$	&		&		&						&		\\
0.01464	&	1.7	& $	18.68	^{+	2.31	}_{-	0.69	}$ &	$uvw1$	&	0.01418	&	19.4	& $	17.61	^{+	0.63	}_{-	0.39	}$ &	$v$	&		&		&						&		\\
0.05120	&	49.6	& $	18.78	^{+	0.17	}_{-	0.15	}$ &	$uvw1$	&	0.07482	&	196.6	& $	18.17	^{+	0.15	}_{-	0.13	}$ &	$v$	&		&		&						&		\\
0.06295	&	196.6	& $	18.82	^{+	0.09	}_{-	0.08	}$ &	$uvw1$	&	0.14359	&	792.7	& $	18.61	^{+	0.12	}_{-	0.11	}$ &	$v$	&		&		&						&		\\
0.07957	&	196.6	& $	19.02	^{+	0.11	}_{-	0.10	}$ &	$uvw1$	&	0.54557	&	656.2	& $	20.43	^{+	0.81	}_{-	0.46	}$ &	$v$	&		&		&						&		\\
0.34185	&	622.0	& $	20.44	^{+	0.13	}_{-	0.12	}$ &	$uvw1$	&	0.73368	&	545.8	& $	20.44	^{+	0.67	}_{-	0.41	}$ &	$v$	&		&		&						&		\\
0.64009	&	900.8	& $	21.41	^{+	0.19	}_{-	0.16	}$ &	$uvw1$	&	1.46837	&	1180.2	& $ >	19.68					$ &	$v$	&		&		&						&		\\
0.85984	&	885.6	& $	21.49	^{+	0.20	}_{-	0.17	}$ &	$uvw1$	&	2.40912	&	891.7	& $ >	19.69					$ &	$v$	&		&		&						&		\\
0.93684	&	885.6	& $	21.74	^{+	0.24	}_{-	0.20	}$ &	$uvw1$	&	3.51254	&	1157.7	& $ >	19.85					$ &	$v$	&		&		&						&		\\
1.23288	&	1771.2	& $	22.16	^{+	0.22	}_{-	0.18	}$ &	$uvw1$	&	3.93518	&	4958.0	& $ >	21.30					$ &	$v*$	&		&		&						&		\\
2.81533	&	1544.0	& $	22.84	^{+	0.48	}_{-	0.33	}$ &	$uvw1$	&	4.54936	&	1494.0	& $ >	19.84					$ &	$v$	&		&		&						&		\\
4.34832	&	2008.8	& $ >	22.09					$ &	$uvw1$	&	5.58572	&	1119.1	& $ >	19.83					$ &	$v$	&		&		&						&		\\
5.09259	&	4582.0	& $ >	23.50					$ &	$uvw1*$	&	5.98716	&	739.8	& $ >	20.23					$ &	$v$	&		&		&						&		\\
\bottomrule
\end{tabular}
\tablefoot{Midtimes are derived logarithmically,
  $t=10^{([log(t_1-t_0)+log(t_2-t_0)]/2)}$, hereby $t_{1.2}$ are the 
  start and stop times, $t_0$ is the \emph{Swift} trigger time. Upper
  limits marked with a $*$ represent stacks of all observations in each
  specific filter that yielded only upper limits, to obtain deeper limits on
  the host galaxy.}
\label{111228AUVOTdata}
\end{center}
\end{table*}

\begin{table*}
\caption{GRB 120714B: GROND AB magnitudes and upper limits of the optical transient.}
\begin{center}
\begin{tabular}{r|ccccccc}
\toprule
$t - t_0$ & $g^\prime$ & $r^\prime$ & $i^\prime$ & $z^\prime$ & $J$ & $H$ & $K_s$  \\
\midrule
0.265	      &	22.43 $\pm$	0.07	&	22.09	$\pm$	0.07	&	22.33	$\pm$	0.16	&	21.65	$\pm$	0.13	&	$>$	21.04		      &	$>$	20.51	        &	$>$	19.94		\\
0.354	      &	22.49	$\pm$	0.06	&	22.19	$\pm$	0.05	&	22.18	$\pm$	0.08	&	21.78	$\pm$	0.08	&		--		          &	--	            	&	--			\\
0.375      	&	22.44	$\pm$	0.05	&	22.21	$\pm$	0.05	&	22.27	$\pm$	0.10	&	21.97	$\pm$	0.10	&		--		          &	--	            	&	--			\\
0.396	      &	22.40	$\pm$	0.05	&	22.23 $\pm$	0.05	&	22.12	$\pm$	0.09	&	21.86	$\pm$	0.09	&	21.75	$\pm$	0.17	&	$>$	21.72	        &	$>$	21.32		\\
0.417      	&	22.39	$\pm$	0.04	&	22.24	$\pm$	0.05	&	22.11	$\pm$	0.09	&	--			          &		--		          &	--		            &	--			\\
0.438      	&	22.44	$\pm$	0.05	&	22.18	$\pm$	0.05	&	22.32	$\pm$	0.11	&	21.90	$\pm$	0.10	&		--		          &	--		            &	--			\\
1.480      	&	22.94	$\pm$	0.08	&	22.48	$\pm$	0.07	&	22.35	$\pm$	0.15	&	22.36	$\pm$	0.12	&	$>$	22.10		      &	$>$	21.53	        &	$>$	20.95		\\
2.488      	&	23.33	$\pm$	0.10	&	22.49	$\pm$	0.08	&	22.52	$\pm$	0.13	&	22.28	$\pm$	0.13	&	$>$	21.96		      &	$>$	21.43  	      &	$>$	20.08		\\
4.511      	&	23.40	$\pm$	0.20	&	22.47	$\pm$	0.10	&	--		           	&	22.06	$\pm$	0.14	&	$>$	21.86	        &	$>$	21.31	        &	$>$	19.50	\\
7.391	      &	23.49	$\pm$	0.09	&	22.28	$\pm$	0.06	&	22.15	$\pm$	0.10	&	22.00	$\pm$	0.11	&	$>$	21.75	      	&	$>$	20.95	        &	$>$	20.28		\\
9.471      	&	23.55	$\pm$	0.10	&	22.28	$\pm$	0.05	&	22.02 $\pm$	0.09	&	22.18	$\pm$	0.09	&	$>$	22.37	      	&	$>$	21.77	        &	$>$	20.39		\\
12.469    	&	--		          	&	22.24	$\pm$	0.05	&	21.95	$\pm$	0.09	&	21.91	$\pm$	0.10	&	$>$	22.27	      	&	$>$	21.63	        &	$>$	20.09		\\
15.511    	&	23.68	$\pm$	0.14	&	22.31	$\pm$	0.07	&	21.89	$\pm$	0.10	&	21.82	$\pm$	0.12	&	$>$	21.59	      	&	$>$	20.91	        &	$>$	20.27		\\
24.432    	&	23.70	$\pm$	0.20	&	22.53	$\pm$	0.06	&	21.86	$\pm$	0.06	&	21.81	$\pm$	0.08	&	$>$	21.97	      	&	$>$	21.43	        &	$>$	20.55		\\
26.380    	&	23.74	$\pm$	0.12	&	22.49	$\pm$	0.06	&	21.88	$\pm$	0.07	&	21.74	$\pm$	0.07	&	$>$	22.04	      	&	$>$	21.42	        &	$>$	20.24		\\
41.391     	&	23.72	$\pm$	0.10	&	22.68	$\pm$	0.06	&	22.20	$\pm$	0.09	&	22.10	$\pm$	0.11	&	$>$	22.27	      	&	$>$	21.59        	&	$>$	20.77		\\
97.244     	&	23.77	$\pm$	0.10	&	22.86	$\pm$	0.08	&	22.55	$\pm$	0.10	&	22.57	$\pm$	0.11	&	$>$	22.39	      	&	$>$	21.77	        &	$>$	20.79		\\
144.166    	&	23.88	$\pm$	0.13	&	22.80	$\pm$	0.09	&	22.86	$\pm$	0.14	&	22.44	$\pm$	0.14	&	$>$	21.97		      &	$>$	21.50         &	$>$	20.66		\\
449.274    	&	23.83	$\pm$	0.13	&	22.95	$\pm$	0.10	&	22.72	$\pm$	0.10	&	22.52	$\pm$	0.12	&	$>$	21.75	      	&	$>$	21.12	        &		--		\\
\bottomrule
\end{tabular}
\tablefoot{$t-t_0$ is the time in units of days after the burst 
($t_0=14$ July 2012,  21:18:47 UT; \citealt{Saxton2012a}).
The data are not corrected for Galactic extinction.}
\label{120714Bmagdata}
\end{center}
\end{table*}

\begin{table*}
\caption{GRB 130831A: GROND AB magnitudes and upper limits of the optical transient.}
\begin{center}
\begin{tabular}{r|ccccccc}
\toprule
$t - t_0$ & $g^\prime$ & $r^\prime$ & $i^\prime$ & $z^\prime$ & $J$ & $H$ & $K_s$  \\
\midrule
0.618	      &	20.30	$\pm$	0.12	&	20.09	$\pm$	0.13	&	19.85	$\pm$	0.14	&	19.58	$\pm$	0.16	&	$>$	19.46	      	&	$>$	18.59	      	&	$>$	18.16		\\
0.639	      &	20.55	$\pm$	0.08	&	20.17	$\pm$	0.10	&	19.92	$\pm$	0.10	&	19.65	$\pm$	0.09	&	$>$	20.00       	&	$>$	19.22	      	&		--		\\
0.665	      &	20.52	$\pm$	0.07	&	20.19	$\pm$	0.08	&	19.93	$\pm$	0.12	&	19.74	$\pm$	0.08	&	19.40	$\pm$	0.13	&	18.96	$\pm$	0.19	&		--		\\
0.689	      &	20.60	$\pm$	0.03	&	20.35	$\pm$	0.04	&	20.10	$\pm$	0.04	&	19.89	$\pm$	0.05	&	19.50	$\pm$	0.08	&	19.25	$\pm$	0.12	&	18.55	$\pm$	0.12	\\
0.714	      &	20.73	$\pm$	0.02	&	20.35	$\pm$	0.03	&	20.16	$\pm$	0.03	&	19.97	$\pm$	0.04	&	19.48	$\pm$	0.06	&	19.15	$\pm$	0.09	&	19.11	$\pm$	0.14	\\
0.736	      &		--		          &	20.43	$\pm$	0.06	&	20.16	$\pm$	0.08	&	19.96	$\pm$	0.10	&	19.68	$\pm$	0.20	&	$>$	19.49	      	&	$>$	18.54		\\
0.762	      &		--	          	&		--          		&		--		&		--		                    &	$>$	19.78		      &		--	          	&		--		\\
1.645	      &	21.98	$\pm$	0.07	&	21.80	$\pm$	0.08	&	21.76	$\pm$	0.12	&	21.56	$\pm$	0.14	&	$>$	21.32		      &	$>$	20.50	      	&		--		\\
1.689      	&	22.01	$\pm$	0.05	&	21.85	$\pm$	0.05	&	21.81	$\pm$	0.10	&	21.66	$\pm$	0.12	&	$>$	21.63		      &	$>$	20.98		      &		--		\\
1.736      	&	22.05 $\pm$	0.05	&	21.80	$\pm$	0.05	&	21.85	$\pm$	0.08	&	21.82	$\pm$	0.13	&	$>$	21.61		      &	$>$	20.92	      	&		--		\\
2.624	      &	22.59	$\pm$	0.05	&	22.38	$\pm$	0.06	&	22.25	$\pm$	0.11	&	22.42	$\pm$	0.15	&	$>$	21.80	      	&	$>$	20.73	      	&	$>$	20.23		\\
3.724	      &	23.00	$\pm$	0.14	&	22.59	$\pm$	0.11	&	22.91	$\pm$	0.33	&	22.46	$\pm$	0.22	&	$>$	21.25	      	&	$>$	20.28	      	&		--		\\
7.712      	&	23.36	$\pm$	0.11	&	22.95	$\pm$	0.09	&	22.62	$\pm$	0.12	&	22.88	$\pm$	0.20	&	$>$	22.04	      	&	$>$	21.07	      	&	$>$	19.94		\\
11.663    	&	23.56	$\pm$	0.16	&	23.05	$\pm$	0.13	&	22.72	$\pm$	0.18	&	22.80	$\pm$	0.15	&	$>$	22.20	      	&	$>$	21.07	      	&	$>$	20.52		\\
24.606    	&	24.56	$\pm$	0.29	&	23.22	$\pm$	0.13	&	22.75	$\pm$	0.15	&	22.98	$\pm$	0.19	&	$>$	22.07	      	&	$>$	20.86		      &		--		\\
29.598    	&	$>$	24.60		      &	23.58	$\pm$	0.21	&	23.18	$\pm$	0.19	&	23.09	$\pm$	0.19	&	$>$	21.99	      	&	$>$	21.11	      	&	$>$	20.49		\\
33.622    	&	24.81	$\pm$	0.22	&	23.67	$\pm$	0.16	&	23.48	$\pm$	0.17	&	23.12	$\pm$	0.12	&	$>$	22.71	      	&	$>$	22.00	      	&	$>$	21.09		\\
386.643	    &	24.86	$\pm$	0.27	&	24.21	$\pm$	0.17	&	$>$	24.25		      &	23.76	$\pm$	0.30  &		--		          &		--		          &		--		\\
\bottomrule
\end{tabular}
\tablefoot{$t-t_0$ is the time in units of days after the burst 
($t_0=31$ August 2013, 13:04:16 UT; \citealt{Hagen2013a}).
The data are not corrected for Galactic extinction.}
\label{130831Amagdata}
\end{center}
\end{table*}

\begin{table*}[ht]
\renewcommand{\tabcolsep}{12pt}
\caption{Additional data not based on GROND observations.}
\begin{center}
\begin{tabular}{c|llll}
\toprule
GRB    & band   & time (days) & AB magnitude & telescope    \\
\midrule
 071112C & $B$       & 2189  & $25.99\pm0.12$ & Keck/LRIS       \\
         & $g^\prime$& 1763  & $26.06\pm0.06$ & Gemini North/GMOS \\ 
         & $r^\prime$&  290  & $25.47\pm0.11$ & Gemini North/GMOS  \\
         & $R_c$     & 2189  & $25.43\pm0.18$ & Keck/LRIS   \\
         & 3.6$\mu$m & 1826  & $23.59\pm0.11$ & Spitzer/IRAC \\[1mm]
 111228A & $R_c$     & 29.33 & $24.08\pm0.16$ & TNG/DoLoRes     \\
         & $R_c$     & 34.51 & $24.12\pm0.13$ & TNG/DoLoRes      \\
         & $R_c$     & 76.23 & $24.53\pm0.22$ & TNG/DoLoRes      \\
         & $I_c$     & 34.44 & $23.33\pm0.10$ & TNG/DoLoRes       \\
         & $I_c$     & 76.30 & $24.10\pm0.14$ & TNG/DoLoRes      \\[1mm]
 120714B & $J$       &  455  & $22.34\pm0.13$ & VLT/HAWKI     \\
         & $K_s$     &  455  & $22.22\pm0.14$ & VLT/HAWKI     \\[1mm]
 130831A & $J$       &  289  & $23.53\pm0.22$ & Keck/MOSFIRE    \\
         & 3.6$\mu$m &  756  & $>24.0$          & Spitzer/IRAC\\
\bottomrule
\end{tabular}
\end{center}
\tablefoot{Given magnitudes are not corrected for Galactic extinction.
The time refers to how many days after the corresponding burst
the data were taken.} 
\label{noGRONDmag}
\end{table*}

\end{appendix}

\end{document}